\documentclass[pre,twocolumn,preprintnumbers,amsmath,amssymb,floatfix,longbibliography,superscriptaddress]{revtex4-2}
\usepackage{graphicx}
\usepackage{epstopdf}
\usepackage{psfrag}
\usepackage{hyperref}
\usepackage{multirow}
\usepackage[sort&compress]{natbib}
\usepackage{amssymb}
\usepackage{amsmath}
\usepackage{amsthm}
\usepackage{bbold}
\usepackage{bbm}
\usepackage{enumerate}
\usepackage{textcomp}
\usepackage{verbatim}
\usepackage{soul}
\usepackage{ulem}
\usepackage{float}
\usepackage{leftindex}
\usepackage{array}
\usepackage{mathrsfs}
\DeclareMathOperator{\sech}{sech}
\usepackage{orcidlink}

\newcommand{\vare}{\varepsilon}

\newcommand{\rmi}{{\rm i}}

\newcommand{\tr}[1]{{\rm Tr}\left(#1\right)}
\newcommand{\bk}{{\boldsymbol{k}}}

\newcommand{\ket}[1]{| #1 \rangle}

\newcommand{\bra}[1]{\langle #1 |}

\newcommand{\braket}[2]{\langle #1 | #2 \rangle}
\newcommand{\ketbra}[2]{|#1\rangle\langle#2|}
\newcommand{\Mel}[3]{\langle #1| #2 | #3 \rangle}

\newcommand{\ii}{{\rm i}}
\newcommand{\beq}{\begin{equation}\begin{aligned}}
\newcommand{\eeq}{\end{aligned}\end{equation}}

\newcommand{\bd}[1]{\boldsymbol{#1}}
\newcommand{\uc}{{\rm{u}}}

\newcommand{\bh}{{\boldsymbol{h}}}
\newcommand{\0}{{\boldsymbol{0}}}
\newcommand{\G}{{\mathcal{G}}}

\renewcommand{\vec}[1]{\boldsymbol{#1}}

\usepackage{xcolor}
\usepackage{cancel}

\begin{document}

\hypersetup{pdftitle={title}}
\title{Walking on Archimedean Lattices: Insights from Bloch Band Theory}

\author{Davidson Noby Joseph\,\orcidlink{0009-0008-2420-0951}}
\affiliation{Department of Physics, University of Alberta, Edmonton, Alberta, Canada}
\affiliation{Theoretical Physics Institute, University of Alberta, Edmonton, Alberta, Canada}

\author{Igor Boettcher\,\orcidlink{0000-0002-1634-4022}}
 \affiliation{Department of Physics, University of Alberta, Edmonton, Alberta, Canada}
 \affiliation{Theoretical Physics Institute, University of Alberta, Edmonton, Alberta, Canada}
 \affiliation{Quantum Horizons Alberta, University of Alberta, Edmonton, Alberta, Canada}

\begin{abstract}
 Returning walks on a lattice are sequences of moves that start at a given lattice site and return to the same site after $n$ steps. Determining the total number of returning walks of a given length $n$ is a typical graph-theoretical problem with connections to lattice models in statistical and condensed matter physics. We derive analytical expressions for the returning walk numbers on the eleven two-dimensional Archimedean lattices by developing a connection to the theory of Bloch energy bands. We benchmark our results through an alternative method that relies on computing the moments of adjacency matrices of large graphs, whose construction we explain explicitly. As condensed matter physics applications, we use our formulas to compute the density of states of tight-binding models on the Archimedean lattices and analytically determine the asymptotics of the return probability. While the Archimedean lattices provide a sufficiently rich structure and are chosen here for concreteness, our techniques can be generalized straightforwardly to other two- or higher-dimensional Euclidean lattices. 
\end{abstract}

\maketitle

\section{Introduction}

Lattices have always played a pivotal role in science, technology, and human culture: from artistic mosaic and weaving patterns, through early applications in cartography and sphere packing problems, to modern uses in finite-element numerics and lattice gauge theory. While in applications the lattice is often viewed as an approximate description of the continuum, it is particularly in condensed matter physics that the specific features of certain lattices are of interest. For instance, the electronic structure of graphene is intimately tied to the carbon atoms forming a periodic honeycomb lattice \cite{Wallace1947,Boehm1962, Boem1986,Saito1992,Shioyama2001,Reich2002,Geim2004,Geim2007,Mecklenburg2011,Yang2018}, quantum spin liquid ground states are predicted for spin models on specific lattices \cite{SRIRAMSHASTRY19811069,Kitaev2006,Song2016,Kos2017,Takagi2019,Verresen2022,PhysRevB.110.014414}, or recently negatively curved  spaces have been emulated in experiments by utilizing so-called hyperbolic lattices \cite{kollar2019hyperbolic,kollar2019line,PhysRevX.10.011009,Yu2020,PhysRevD.102.034511,PhysRevA.102.032208,Jahn2020,maciejko2020hyperbolic,brower2021lattice,Zhu:2021,Maciejko2022,Stegmaier2022,zhang2022observation,Lenggenhager2022,Boettcher2022,LenggenhagerJsp2023,PhysRevLett.133.061603,PhysRevB.111.L121108}. Such studies opened a rich and dynamic interdisciplinary field of research that connects mathematical graph theory with physics experiments.

A common graph-theoretic problem with many applications in physics is the counting of walks on a lattice \cite{graphsBook}. For this, consider first an infinitely extended one-dimensional grid of points, where each site (or vertex) has two neighbors. Let $S_n$ denote the number of walks to return from a given site to the same site in $n$ steps. Obviously, this quantity does not depend on which specific site was chosen, and $S_n$ vanishes for odd $n$. Since at each step a walker may go left or right, we find $S_{2n}=\binom{2n}{n}$, which is the sequence of central binomial coefficients in Pascal's triangle. Extending this exercise to walking on a square lattice in the two-dimensional plane, since steps in the horizontal and vertical direction are independent, we arrive at
\begin{align}
 \label{intro1} S_{2n} = \binom{2n}{n}^2,
\end{align}
with the natural generalization to higher-dimensional cubic lattices.

To see a connection between the integers $S_{2n}$ and condensed matter physics, consider for $E\in[-4,4]$ the infinite sum
\begin{align}
 \label{intro2} D(E) &= \frac{1}{\pi}\lim_{\eta\to 0}\mathfrak{I}\sum_{n=0}^\infty \frac{S_{2n}}{(E-\rmi \eta)^{2n+1}} =\frac{1}{2\pi^2} K\Bigl(1-\frac{E^2}{16}\Bigr),
\end{align}
where $K(k^2)$ is the complete elliptic integral of the first kind. The function $D(E)$ constitutes the density of states (DOS) of single-particle excitations with energy $E$ on the square lattice. It is a central object in the description of layered electronic materials and can be measured experimentally through a variety of techniques. Even if not all terms in the sequence $\{S_n\}_{n\geq 1}$ are known, the first few can be used to estimate $D(E)$ through the continued fraction method \cite{Haydock1972,Haydock1975}, which has recently been applied successfully to otherwise intractable hyperbolic lattices \cite{Mosseri2023,PhysRevLett.133.146601}. The number of walks on graphs can also be used to define discrete path integrals for lattice quantum models \cite{MnevPath,MnevArxiv}. Furthermore, the return probability, which is $S_n$ normalized by the total number of walks of length $n$, play a role for transient behavior of statistical models or Anderson localization in disordered solids \cite{Polya1921,Montroll,HUGHES1986443,DEHARO2004201,Manai2015,PhysRevA.97.062105,Tikhonov2021,RevModPhys.95.031003,Chen2024aug}.

Having established that the returning walks numbers $S_n$ appear in physics, we may wonder how to efficiently compute them. The challenge here resides in the fact that not all lattices allow for a simple combinatorial treatment as in the case of the square lattice. It is, in fact, the connection to electronic band theory in physics that yields a powerful tool to compute the $S_n$. We might expect such a result, since $D(E)$ in Eq. (\ref{intro2}) can be written as an integral over the Brillouin zone involving the energy-momentum-dispersion $\vare(\boldsymbol{k})$. Indeed, we have the remarkable formula
\begin{align}
\label{intro3} S_{2n} = \int_{\boldsymbol{k}}  \vare(\boldsymbol{k})^{2n}= \binom{2n}{n}^2
\end{align}
for the square lattice, where $\vare(\bk)=-2(\cos k_1+\cos k_2)$ is the expression for the Bloch energy bands on the square lattice (in units of the hopping amplitude) and
\begin{align}
 \label{intro4} \int_{\boldsymbol{k}} = \int_0^{2\pi}\frac{\mbox{d}k_1}{2\pi}\int_0^{2\pi}\frac{\mbox{d}k_2}{2\pi}
\end{align}
is the integral over crystal momenta $\bk=(k_1,k_2)\in[0,2\pi]^2$.

Equation (\ref{intro3}) generalizes to other periodic lattices, where a combinatorial determination of $S_n$ seems out of reach, while determining the Bloch energies $\vare(\boldsymbol{k})$ is possible. As an example, consider the two-dimensional honeycomb lattice, which has two  sites in the unit cell and two energy bands given by $\vare_\pm(\boldsymbol{k})=\pm|1+e^{\rmi k_1}+e^{\rmi k_2}|$. We then have
\begin{align}
 \label{intro5} S_{2n} = \frac{1}{2}\int_{\boldsymbol{k}}[\vare_+(\boldsymbol{k})^{2n}+\vare_-(\boldsymbol{k})^{2n}].
\end{align} 
Again, $S_n$ vanishes for odd $n$ and, after some algebra explained later in this work, we recover the known formula 
\begin{align}
 \label{intro6} S_{2n} = \sum_{\ell=0}^n \binom{2\ell}{\ell}\binom{n}{\ell}^2,
\end{align} 
given for instance in Vidakovic's ``All Roads Lead to Rome--Even in the Honeycomb World'' \cite{AllRoads}.

In this work, we continue on the path laid out by Eqs. (\ref{intro3}) and (\ref{intro5}) by generalizing them to periodic tessellations of the two-dimensional plane with $N_{\rm u}$ sites in the unit cell. In this case, we find that for periodic tilings where each vertex has the same coordination number, we have
\begin{align}
 \label{intro7} S_n = \frac{(-1)^n}{N_{\rm u}}\sum_{\alpha=1}^{N_{\rm u}} \int_{\boldsymbol{k}} \vare_\alpha(\boldsymbol{k})^n,
\end{align}
where $\vare_{\alpha}(\boldsymbol{k})$ are the Bloch energy bands with $\alpha=1,\dots,N_{\rm u}$ the band index. In the above examples, $N_{\rm u}=1$ for the square lattice and $N_{\rm u}=2$ for the honeycomb lattice. 

In practice, Eq. (\ref{intro7}) is not always useful for $N_{\rm u}>2$, since the determination of the eigenvalues $\vare_{\alpha}(\boldsymbol{k})$ requires the diagonalization of the $N_{\rm u}\times N_{\rm u}$ Bloch Hamiltonian matrix $H(\bk)=-A(\bk)$ introduced below. However, we show that computing traces of powers of this matrix, namely
\begin{align}
 \label{intro7b} S_n=\frac{1}{N_{\rm u}}\int_{\vec{k}} \mbox{Tr}[A(\vec{k})^n],
\end{align}
presents a way to always efficiently determine the $S_n$. For lattices where the number of returning walks depends on the start site $i_{\rm u}$, we derive the more general formula
\begin{equation}
\label{intro7c}
    S_{n}^{(i_\uc)}= \int_{\bk} [A(\bk)^n]_{i_\uc i_\uc},
\end{equation}
which yields Eq. (\ref{intro7b}) for vertex-transitive lattices upon summing over $i_{\rm u}$. We further introduce a convenient generating function for $S_n$ that can be computed readily for lattices that are periodic tessellations by a finite unit cell on a Bravais lattice. 

Formulas like Eq. (\ref{intro7b}), including a variant that also counts the area enclosed by the random walk, have been applied to counting walks before, see for instance Refs. \cite{Montroll,OUVRY2020115174,PhysRevE.105.014112,OuvryArea,PhysRevE.108.054104,Gan:2024rzb}. A universal spectrum moment theorem, which implies Eq. (\ref{intro7b}) and also applies to non-Hermitian systems, was derived in Ref. \cite{PhysRevLett.133.216401}. However it appears that the formalism has not been applied to lattices with larger unit cells ($N_{\rm u}>2$) such as the Archimedean lattices, or where the dependence on the start site $i_{\rm u}$ is nontrivial as in Eq. (\ref{intro7c}). The derivation of Eq. (\ref{intro7c}) in a general context, and Eq. (\ref{intro7b}) alongside its application to the eleven Archimedean lattices constitutes the main result of this work.

\begin{figure}[t]
    \includegraphics[width=\linewidth]{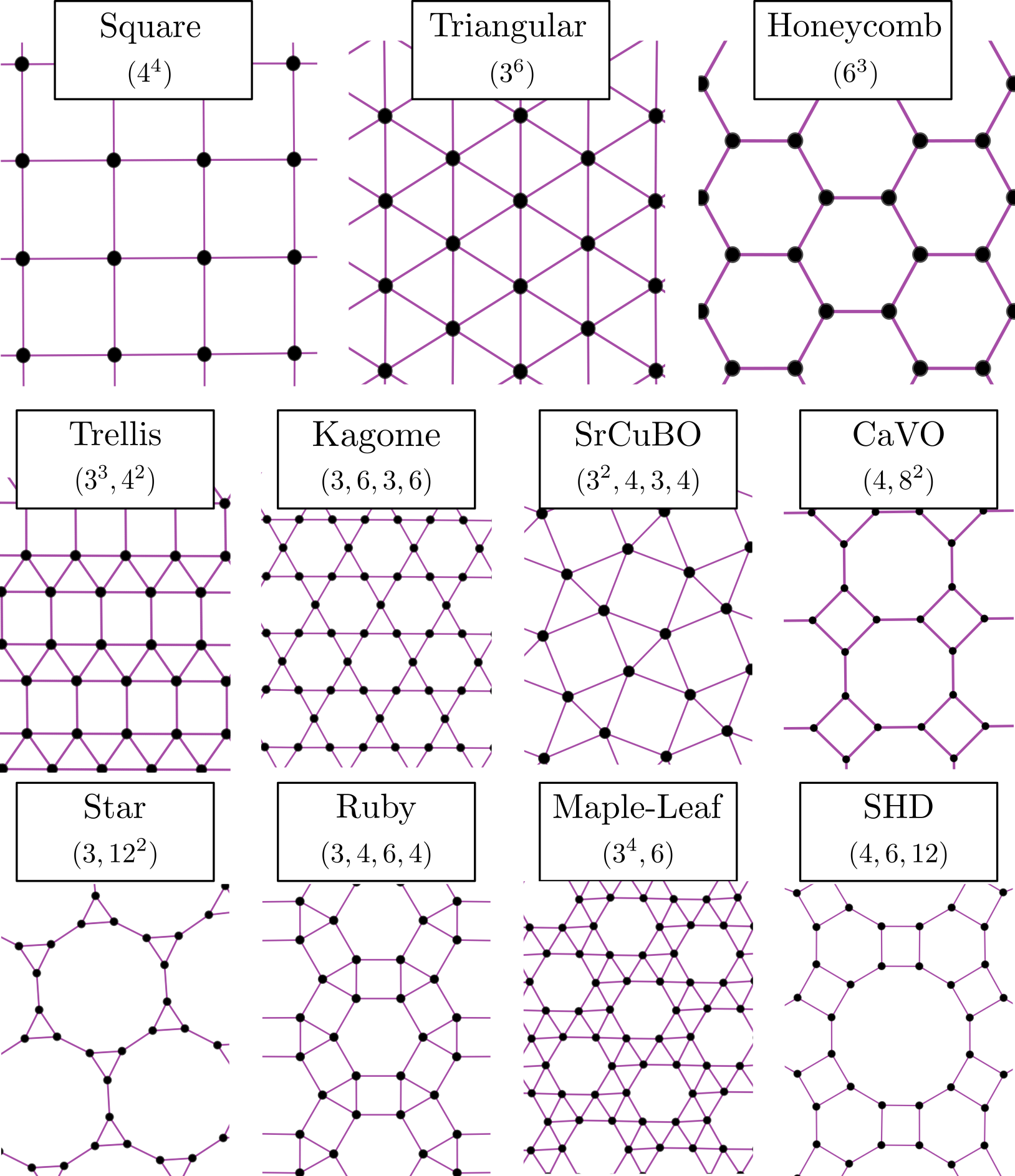}
    \label{AL}
    \caption{We show the eleven two-dimensional Archimedean lattices. Some frequently studied members of the family include the Square, Triangular, Honeycomb, and Kagome lattices, while the names of the other lattices follow common conventions in the literature, often named after chemical compounds related to the lattice \cite{Farnell,Betts1995}. 
    Note that the Ruby lattice \cite{PhysRevB.84.155116,Verresen2022,PhysRevE.109.045305} is also referred to as Bounce lattice \cite{Farnell}. The numbers in brackets indicate the Gr\"{u}nbaum--Shephard notation: For this, note that every vertex or site is surrounded by the same sequence of regular polygons such that, for instance, for the Square and Triangular lattice, at each vertex $4$ squares ($4^4$) or $6$ triangles ($3^6$) meet, respectively. Also note that every site has the same coordination number, defined as the number of nearest neighbors of any vertex.}
\label{FigArch}
\end{figure}

\renewcommand{\arraystretch}{1.5}
\begin{table*}[t!]
\begin{tabular}{|c|l|c|}
\hline
 \multicolumn{3}{|c|}{Bipartite Archimedean lattices}\\
\hline\hline
 Square 
 &  \ $S_{2n}=1, 4, 36, 400, 4900, 63504, 853776, 11778624, 165636900, \dots$ & $S_{2n}= \binom{2n}{n}^2$ \\
\hline
 \ Honeycomb \ 
 &  \	$S_{2n} =1, 3, 15, 93, 639, 4653, 35169, 272835, 2157759, 17319837, \dots$	&	\ $S_{2n}^{(\rm H)}=\sum_{\ell=0}^n \binom{2\ell}{\ell}\binom{n}{\ell}^2$ \ 	\\
\hline
 CaVO 
 &  \	$S_{2n} =1,3, 17, 111, 773, 5623, 42269, 325923, 2563645, 20486055,\dots$	&	$\mathcal{G}(z)$: Eq. (\ref{uv42})	\\
\hline
 SHD	
 &  \	$S_{2n}=1,3, 17, 113, 809, 6063, 46835, 369407, 2957769, 23952947,\dots$	&	$\mathcal{G}(z)$: Eqs. (\ref{uv58}), (\ref{uv61})	\\
\hline
\multicolumn{3}{c}{} \\
\hline
 \multicolumn{3}{|c|}{Non-bipartite Archimedean lattices} \\
\hline\hline
Triangular 
& \	$S_n = 1, 0, 6, 12, 90, 360, 2040, 10080, 54810, 290640, 1588356, \dots$	&	$S_n=\sum_{\ell=0}^n \binom{n}{\ell}(-3)^{n-\ell} S_{2\ell}^{(\rm H)}$	\\
\hline
Kagome 
& \	$S_n = 1, 0, 4, 4, 28, 60, 264, 784, 3004, 10204, 37824, 135784, 502784,\dots$	&	\ $S_n=\frac{2}{3}\sum_{\ell=0}^{[n/2]}\binom{n}{2\ell}S_{2\ell}^{(\rm H)}+\frac{1}{3}(-2)^n$ \	\\
\hline
Trellis 
& \	$S_n = 1, 0, 5, 6, 57, 140, 863, 2940, 15297, 61296, 297945, 1294810,\dots$	&	$S_n$: Eq. (\ref{uv32})	\\
\hline
Star  
& \	$S_n =1, 0, 3, 2, 15, 20, 89, 168, 591, 1346, 4223, 10648, 31621, 84266, \dots $	&	$S_n$: Eq. (\ref{uv35})	\\
\hline
 \ SrCuBO 
 & \	$S_n = 1, 0, 5, 6, 53, 140, 797, 2898, 14317, 59784, 282425, 1255430,\dots$	&	 $\mathcal{G}(z)$: Eq. (\ref{uv50})	\\
\hline
Ruby 
& \	$S_n = 1, 0, 4, 2, 32, 40, 314, 616, 3488, 8864, 42144, 125312, 540250,\dots$	&	 $\mathcal{G}(z)$: Eqs. (\ref{uv56}), (\ref{uv61}) \\
\hline
Maple-Leaf 
& \	$S_n = 1, 0, 5, 8, 57, 180, 907, 3612, 16777, 72896, 334265,1508584,\dots$	&	$\mathcal{G}(z)$: Eqs. (\ref{uv57}), (\ref{uv61})	\\
\hline
\end{tabular}
    \caption{The first few returning walk numbers $S_n$ for all eleven Archimedean lattices $S_n$ are shown, with $n=0,1,2,\dots$. Four of the lattices are bipartite and therefore require an even number of steps to return back to the starting site. For the cases when explicit formulae of the returning walk numbers could not be found, we list the generating function $\mathcal{G}(z)=\sum_{n\geq 0}S_n z^n$, which allows us to efficiently compute $S_n$ as the coefficient of $z^n$.}
\label{TabWalks}
\end{table*}
\renewcommand{\arraystretch}{1}

In the following, we focus on the eleven Archimedean lattices in  two dimensions \cite{codello2010exact,Grunbaum2016-rs,Farnell,CrastodeLima2019,IsingArch}, shown in Fig. \ref{FigArch}, with the associated numbers of returning walks $S_n$ computed in this work summarized in Table \ref{TabWalks}.  First described by Kepler in 1619, the Archimedean lattices contain the most frequently considered planar lattices, namely the Square, Triangular, Honeycomb, and Kagome lattice. (The latter recently came into the focus of condensed matter physicists because of its unique geometry and flat band physics in the context of frustrated systems and superconductivity \cite{Ortiz2021,Kato2024,Pratt2025,Cai2025}.) We choose this set of lattices here because it is rich enough to highlight many features of our approach, exemplifies how the procedure would extend to other periodic tilings, but at the same time can be treated in a largely analytical fashion with closed-form expressions.

The Archimedean lattices are tessellations of the plane by regular polygons. They have the important property of vertex transitivity, meaning roughly that each lattice site is equivalent. More precisely, each site has the same coordination number $q$ (number of nearest neighbors) and the lattice is homogeneous in the sense that each site is surrounded by the same sequence of regular polygons, making these lattices ``semi-regular''. This is reflected in the Gr\"{u}nbaum--Shephard notation \cite{Grunbaum2016-rs}, where, for instance, $(3,6,3,6)$  for the Kagome lattice means that each site is surrounded by a triangle, followed by a hexagon, triangle, and another hexagon. The Archimedean lattices, however, are not isotropic, in the sense that the direction in which these polygons appear depends on the site chosen. This homogeneity without isotropy is shared with hyperbolic $\{p,q\}$ tessellations of the plane \cite{Boettcher2022} and we expect that some features of Archimedean lattices studied here carry over to more general non-Euclidean tessellations and hyperbolic lattices.

\section{Bloch band theory}\label{SecBloch}

In this section, we develop the Bloch band theory for the eleven Archimedean lattices. This topic is relevant in itself for understanding the electronic properties of materials and metamaterials on such lattices that are described by tight-binding Hamiltonians of the form
\begin{align}
 \label{bloch1} \hat{\mathcal{H}} = -t \sum_{\langle i,j\rangle} (\hat{c}_i^\dagger \hat{c}_j+\text{h.c.}),
\end{align}
where $\hat{c}_i$ is the annihilation operator for a particle on site $i$ of the lattice, $t>0$ is the hopping amplitude, and the sum is over nearest-neighbor sites. On two-dimensional periodic tilings of the plane such as the Archimedean lattices, the quadratic Hamiltonian can be diagonalized in the Bloch band basis, which yields
\begin{align}
 \label{bloch2} \hat{\mathcal{H}} = \sum_{\alpha=1}^{N_{\rm u}} \int_{\vec{k}} \vare_{\alpha}(\vec{k}) \hat{c}^\dagger_{\vec{k} \alpha}\hat{c}_{\vec{k}\alpha}.
\end{align}
Knowledge of the energy bands $\vare_{\alpha}(\vec{k})$ encodes important information about the electronic properties of the material and typically constitutes the starting point for understanding interaction or response phenomena.

\begin{figure*}[t!]
\includegraphics[width=\linewidth]{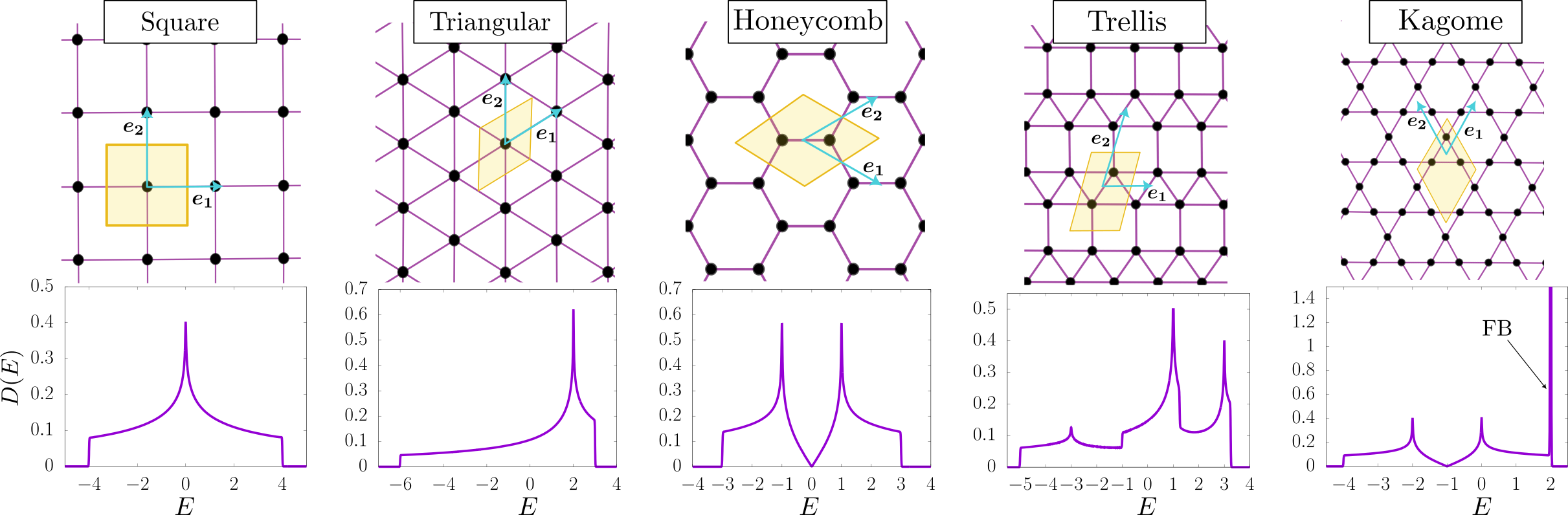}
 \caption{The density of states for the Square, Triangular, Honeycomb, Trellis, and Kagome lattices is shown alongside their respective unit cells (in golden, filled). $E$ denotes the energy in units of $t$. The lattice translation vectors $\boldsymbol{e_1}$ and $\boldsymbol{e_2}$ are illustrated in turquoise. FB represents a flat band in the Brillouin Zone.}
\label{FigPanel1}
\end{figure*}

\begin{figure}[t]
    \includegraphics[width=\linewidth]{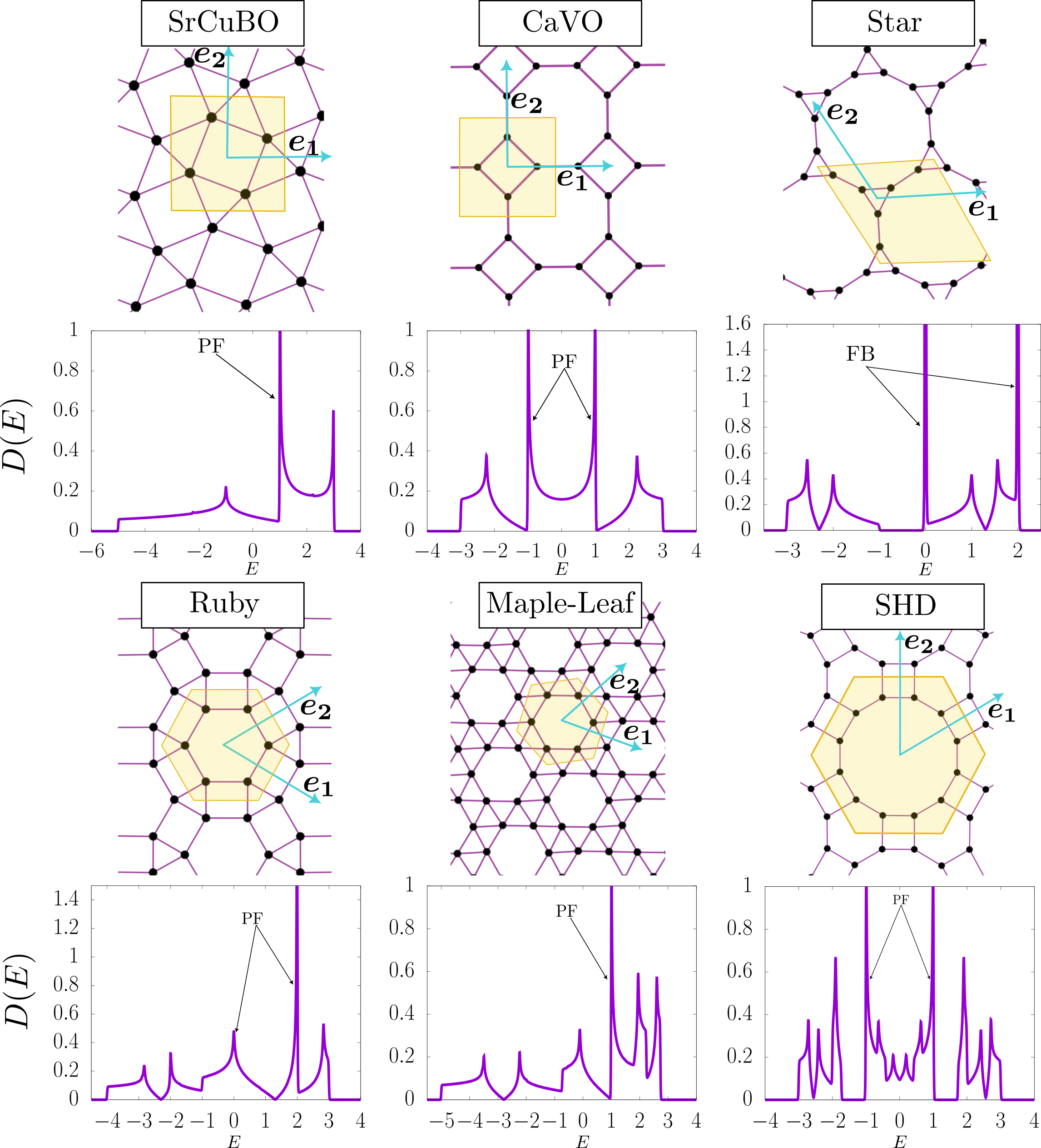}
        \caption{Same setting as in Fig. \ref{FigPanel1} but for lattices with $N_\uc\geq 4$. The density of states has new features like partial flat bands (PFs). A PF is a band that is non-dispersive along a high symmetry direction in the Brillouin zone. Note the striking resemblance of features from the Honeycomb or Triangular lattice DOSs in small neighborhoods of $E$ for these lattices.}
     \label{FigPanel2}
\end{figure}

In crystallographic terms, the Archimedean lattices split into a Bravais lattice and a unit cell. The Bravais lattice, which can be thought of as the scaffold for the tiling, is chosen among five possible choices in two dimensions, although topologically only two are relevant: the Square and Triangular lattice. This also implies that the Square and Triangular lattices are their own Bravais lattices, with a trivial unit cell of one site. More generally, the unit cell consists of $N_{\rm u}$ sites, with the concomitant number of Bloch energy bands $\vare_{\alpha}(\vec{k})$. The unit cell can be thought of as the fundamental tile of the tessellation. Topologically, it has the shape of a square or hexagon if the Bravais lattice is the Square or Triangular lattice, respectively. For this reason, the triangular lattice is also referred to as hexagonal lattice. For the Archimedean lattices, $N_{\rm u}$ takes the values $1,2,3,4,6,12$. We highlight the corresponding unit cells in golden in Figs. \ref{FigPanel1} and \ref{FigPanel2}.

\subsection{Bloch adjacency matrix}\label{SecBlochA}

In this work, we determine the $N_{\rm u}$ Bloch energy bands $\vare_{\alpha}(\vec{k})$ as the eigenvalues of the single-particle Bloch Hamiltonian 
\begin{align}
 \label{bloch3} \hat{H}(\vec{k}) = -t \hat{A}(\vec{k}).
\end{align}
Herein, $\hat{A}(\vec{k})$ is the  Bloch adjacency operator which acts on the position-space kets $\ket{i_{\rm u}}, i_{\rm u}=1,\dots,N_{\rm u}$, in the unit cell. The Bloch adjacency operator is the central object of our analysis. Since the crystal momentum $\vec{k}$ is a good quantum number, it takes on the role of a parameter in the diagonalization of $\hat{A}(\vec{k})$. More formally, $e^{\rmi k_1}, e^{\rmi k_2}\in\mathbb{S}$ label the irreducible representations of the Abelian translation group.

Define the (usual) adjacency matrix $A$ and the adjacency operator of the lattice by
\begin{align}
 \label{bloch4} \Mel{i}{\hat{A}}{j} =A_{ij} =\begin{cases} 1 & i\ \text{and } j\text{ are neighbors} \\ 0 & \text{else}\end{cases}.
\end{align}
We refer to the matrix that has components $A_{ij}$ as $A$. Analogously, we call the $N_{\rm u}\times N_{\rm u}$ matrix $A(\bk)$ the  Bloch adjacency matrix, whose components are given by $A(\bk)_{i_\uc j_\uc}=\Mel{i_\uc}{\hat{A}(\bk)}{j_{\rm u}}$.

With the adjacency matrix at hand, the Hamiltonian in Eq. (\ref{bloch1}) becomes 
\begin{align}
 \label{bloch4b} \hat{\mathcal{H}} = -t \sum_{i,j}A_{ij}\hat{c}^\dagger_i\hat{c}_j =: \sum_{i,j}H_{ij}\hat{c}^\dagger_i\hat{c}_j,
\end{align}
where we defined the single-particle Hamiltonian matrix $H=-t A$. While $A$ is formally an infinite matrix for the infinite lattice, we may consider finite subgraphs with $N$ sites, which yields finite $N\times N$ adjacency matrices. These have $N$ eigenvalues $E_i$. As the subgraph is enlarged to encompass the entire lattice as $N\to\infty$, the eigenvalues $\{E_i\}_{i=1,\dots,N}$ approach the set  $\{\vare_{\alpha}(\vec{k})\}_{\vec{k},\alpha}$. We describe the construction of suitable families of finite graphs to cover the Archimedean lattices in Sec. \ref{SecGraphs}.

To construct the Bloch adjacency matrices $A(\vec{k})$, we first identify for each lattice (i) two linearly independent directions of translation vectors $\vec{e}_1$ and $\vec{e}_2$ that form the Bravais lattice, and (ii) the $N_{\rm u}$ sites of the unit cell that tessellate the plane upon translations along these two directions. Note that $\vec{e}_1$ and $\vec{e}_2$ need not be orthogonal or coincide with the x- and y-directions. The identifications used  for the Archimedean lattices in this work are shown in Figs. \ref{FigPanel1} and \ref{FigPanel2}. Note that the unit cell sites are also referred to as sublattice sites in the literature.

After the crystallographic setting is laid out, we can clarify the role of the crystal momentum $\vec{k}=(k_1,k_2)^T$. For this, denote the unitary translation operators along the directions $\vec{e}_1$ and $\vec{e}_2$ by $\hat{T}_{\vec{e}_1}$ and $\hat{T}_{\vec{e}_2}$. Due to translation symmetry, $\hat{T}_{\vec{e}_1}$ and $\hat{T}_{\vec{e}_2}$ commute with $\hat{H}$, and all three can be diagonalized simultaneously. The eigenstates of $\hat{T}_{\vec{e}_i}$ are labelled by $\vec{k}$ as 
\begin{align}
\label{bloch5} \hat{T}_{\boldsymbol{e}_i}|\vec{k}\rangle= e^{\rmi k_i}|\vec{k}\rangle.
\end{align}
Consequently, determining the spectrum of $H=-tA$ is equivalent to finding the spectrum of  $H(\vec{k})=-tA(\vec{k})$ for all $\vec{k}$. Importantly, $A(\vec{k})$ is only an $N_{\rm u}\times N_{\rm u}$ matrix, while $A$ is infinite for the infinite lattice. Also note that we choose $\vec{e}_1$ and $\vec{e}_2$ in this work such that the Brillouin zone is always a square in momentum space given by
\begin{align}
 \label{bloch6} \vec{k}\in [0,2\pi]^2\ \Leftrightarrow\ (e^{\rmi k_1},e^{\rmi k_2})\in\mathbb{T}^2,
\end{align}
where $\mathbb{T}^2=\mathbb{S}\times\mathbb{S}$ denotes the two-dimensional torus. This choice may differ from other, equivalent Brillouin zones used in physics applications.

The matrix elements of $A(\vec{k})$ are determined as follows. The sites of the unit cell are labelled by $i_\uc=1,\dots,N_{\rm u}$. A fundamental domain of the unit cell, given by a continuous patch of the two-dimensional plane, is identified together with its boundary. In Figs. \ref{FigPanel1} and \ref{FigPanel2}, the fundamental domains are colored in golden shade and their boundaries are indicated by solid golden lines. If two sites $i_\uc$ and $j_\uc$ are connected by an edge that lies within the fundamental domain, then $A_{i_{\uc} j_{\uc}}=A_{j_{\uc}i_{\uc}}=1$. If site $i_{\uc}$ within the fundamental domain is connected to site $j_{\uc}$ in another unit cell, then $A_{i_{\uc} j_{\uc}}=(A_{j_{\uc} i_{\uc}})^*=e^{-\rmi \vec{n}\cdot\vec{k}}$, where $\vec{n}=n_1\vec{e}_1+n_2\vec{e}_2$ is the translation vector that connects the two fundamental domains. All other entries of $A(\vec{k})$ are zero. The examples for the Archimedean lattices given below should clarify the construction.

In the following, we list the Bloch adjacency matrices $A(\vec{k})$ for the eleven Archimedean lattices and, if we can compute them analytically, the Bloch energy bands $\vare_{\alpha}(\vec{k})$ as the eigenvalues of $H(\vec{k})=-tA(\vec{k})$. Plots of the energy bands for Archimedean lattices can be found in Ref. \cite{CrastodeLima2019}. Here and in the remainder of this work we set
\begin{align}
 t=1,
\end{align}
so that all energies are measured in units of $t$.

\subsubsection{Square and Triangular $(N_{\rm u}=1)$}

Since the Square ($\square$) and Triangular ($\Delta$) lattices are Bravais lattices, their Bloch adjacency matrices are real numbers. We have
\begin{align}
 \nonumber  A^{(\square)}(\boldsymbol{k}) &= e^{\rmi k_1}+e^{-\rmi k_1} + e^{\rmi k_2}+e^{-\rmi k_2}\\
 \label{bloch7}  &=2(\cos k_1+\cos k_2),\\
 \label{bloch8}  A^{(\Delta)}(\boldsymbol{k})&=2\Bigl(\cos k_1+\cos k_2+\cos(k_1-k_2)\Bigr),
 \end{align}
 with Bloch energy bands
\begin{align}
 \label{bloch9} \vare_\square(\vec{k}) &= -2(\cos k_1+\cos k_2),\\
\label{bloch10} \vare_\Delta(\vec{k}) &= -2\Bigl(\cos k_1+\cos k_2+\cos(k_1-k_2)\Bigr),
\end{align}
respectively.

\subsubsection{Honeycomb and Trellis $(N_{\rm u}=2)$}

The Bloch adjacency matrices of the Honeycomb (H) and Trellis (Tr) lattice are given by
\begin{align}
  \label{bloch11} &A^{(\rm H)}(\boldsymbol{k})  =\begin{pmatrix} 0 & 1+e^{-\rmi k_1}+e^{-\rmi k_2} \\ 1+e^{\rmi k_1}+e^{\rmi k_2} & 0 \end{pmatrix},\\
  \label{bloch12} &A^{(\rm Tr)}(\boldsymbol{k})= \begin{pmatrix} e^{-\rmi k_1}+e^{\rmi k_1} & 1+e^{-\rmi k_1}+e^{-\rmi k_2} \\  1+e^{\rmi k_1}+e^{\rmi k_2} & e^{-\rmi k_1}+e^{\rmi k_1},\end{pmatrix}
\end{align}
with Bloch energy bands
\begin{align}
 \label{bloch13} \vare_\pm^{(\rm H)}(\vec{k}) &= \pm\ \vare_{\rm H}(\vec{k}),\\
 \label{bloch14} \vare_\pm^{(\rm Tr)}(\vec{k}) &= -2\cos k_1\pm \vare_{\rm H}(\vec{k}),
\end{align}
where
\begin{align}
\label{bloch13a}
 \vare_{\rm H}(\vec{k}) = |1+e^{\rmi k_1}+e^{\rmi k_2}|.
\end{align}
Note that the Honeycomb lattice is bipartite and that the sublattices of 1- and 2-sites are triangular lattices. Thus we have
\begin{align}
 \label{bloch15} \vare_{\rm H}(\vec{k}) = \sqrt{3-\vare_\Delta(\vec{k})}.
\end{align}

\subsubsection{Kagome $(N_{\rm u}=3)$}

The Bloch adjacency matrix of the Kagome (K) lattice is given by
\begin{equation}
   \label{bloch16}  A^{(\rm K)}(\boldsymbol{k})=\begin{pmatrix} 0 & 1+e^{-\rmi k_2} & 1+e^{\rmi (k_1-k_2)} \\ 1+e^{\rmi k_2} & 0 & 1+e^{\rmi k_1} \\  1+e^{\rmi (k_2-k_1)} & 1+e^{-\rmi k_1} & 0 \end{pmatrix},
\end{equation}
with Bloch energy bands
\begin{align}
 \label{bloch17} \vare^{(\rm K)}_\pm(\vec{k}) &= -1\pm \vare_{\rm H}(\vec{k}),\\
 \label{bloch18} \vare^{(\rm K)}_{\rm FB}(\vec{k}) &= 2.
\end{align}
Note that the Kagome lattice is the line-graph of the honeycomb lattice and thus features a flatband (FB)---defined as a non-dispersive or constant band $\vare_\alpha(\vec{k})$---and the dispersive bands are given by $\vare_\pm^{(\rm H)}(\vec{k})$ up to a constant shift.

\subsubsection{CaVO and SrCuBO $(N_{\rm u}=4)$}

The Bloch adjacency matrices of the CaVO (Ca) and SrCuBO (Sr) lattices are given by
\begin{align}
 \label{bloch21}   &A^{(\rm Ca)}(\boldsymbol{k})=\begin{pmatrix}  0 & 1 & e^{-\rmi k_1} & 1 \\  1 & 0 & 1 & e^{-\rmi k_2} \\  e^{\rmi k_1} & 1 & 0 & 1 \\  1 & e^{\rmi k_2} & 1 & 0 \\
\end{pmatrix},\\
  \label{bloch20}   &A^{(\rm Sr)}(\boldsymbol{k})=\begin{pmatrix}  0 & 1+e^{\rmi k_1} & e^{\rmi k_2} & 1+e^{\rmi k_2} \\  1+e^{-\rmi k_1} & 0 & 1+e^{\rmi k_2} & e^{-\rmi k_1} \\  e^{-\rmi k_2} & 1+e^{-\rmi k_2} & 0 & 1+e^{-\rmi k_1} \\  1+e^{-\rmi k_2} & e^{\rmi k_1} & 1+e^{\rmi k_1} & 0 \\ \end{pmatrix}.
\end{align}
There is no closed formula for the corresponding Bloch energy bands $\vare_{\alpha}(\vec{k})$, but they can be determined numerically for fixed $\vec{k}$.

\subsubsection{Star, Ruby, Maple-Leaf  $(N_{\rm u}=6)$, and SHD $(N_{\rm u}=12)$}

The Bloch adjacency matrix for the Star (St) lattice is given by
\begin{align} \label{bloch22} 
 A^{(\rm St)}(\boldsymbol{k})&=\begin{pmatrix}
 0 & 1 & 0 & e^{-\rmi k_1} & 0 & e^{-\rmi (k_1+k_2)} \\
 1 & 0 & 1 & 0 & 1 & 0 \\
 0 & 1 & 0 & 1 & 1 & 0 \\
 e^{\rmi k_1} & 0 & 1 & 0 & 0 & e^{-\rmi k_2} \\
 0 & 1 & 1 & 0 & 0 & 1 \\
 e^{\rmi (k_1+k_2)} & 0 & 0 & e^{\rmi k_2} & 1 & 0 \\
\end{pmatrix}.
\end{align}
The corresponding Bloch energy bands are given by
\begin{align}
 \label{bloch22b} \vare^{(\rm St)}_1(\vec{k}) &=0,\ \vare^{(\rm St)}_2(\vec{k}) =2,\\
 \label{bloch22d} \vare^{(\rm St)}_{3,4}(\vec{k}) &=-\frac{1}{2}\Bigl(1\pm\sqrt{13+4|1+e^{\rmi k_1}+e^{-\rmi k_2}|}\Bigr),\\
 \label{bloch22e} \vare^{(\rm St)}_{5,6}(\vec{k}) &=-\frac{1}{2}\Bigl(1\pm\sqrt{13-4|1+e^{\rmi k_1}+e^{-\rmi k_2}|}\Bigr).
\end{align}
In particular, there are two flatbands at energies 0 and 2. For the Ruby (R), Maple-Leaf (ML), and SHD lattices we have
\begin{widetext}
\begin{align}
 \label{bloch23} A^{(\rm R)}(\boldsymbol{k}) &= \begin{pmatrix}
 0 & 1 & e^{\rmi k_2} & 0 & e^{\rmi k_1} & 1 \\
 1 & 0 & 1 & e^{\rmi k_1} & 0 & e^{\rmi (k_1-k_2)} \\
 e^{-\rmi k_2} & 1 & 0 & 1 & e^{\rmi (k_1-k_2)} & 0 \\
 0 & e^{-\rmi k_1} & 1 & 0 & 1 & e^{-\rmi k_2} \\
 e^{-\rmi k_1} & 0 & e^{\rmi (k_2-k_1)} & 1 & 0 & 1 \\
 1 & e^{\rmi (k_2-k_1)} & 0 & e^{\rmi k_2} & 1 & 0 \\
 \end{pmatrix},\\
 \label{bloch24} A^{(\rm ML)}(\boldsymbol{k}) &=\begin{pmatrix}
 0 & 1 & e^{\rmi (k_1-k_2)} & e^{\rmi k_1} & e^{\rmi k_1} & 1 \\
 1 & 0 & 1 & e^{\rmi k_1} & e^{\rmi k_2} & e^{\rmi k_2} \\
 e^{\rmi (k_2-k_1)} & 1 & 0 & 1 & e^{\rmi k_2} & e^{\rmi (k_2-k_1)} \\
 e^{-\rmi k_1} & e^{-\rmi k_1} & 1 & 0 & 1 & e^{\rmi (k_2-k_1)} \\
 e^{-\rmi k_1} & e^{-\rmi k_2} & e^{-\rmi k_2} & 1 & 0 & 1 \\
 1 & e^{-\rmi k_2} & e^{\rmi (k_1-k_2)} & e^{\rmi (k_1-k_2)} & 1 & 0 \\
\end{pmatrix},\\
 \label{bloch25}  A^{(\rm SHD)}(\boldsymbol{k}) &=\left(
\begin{array}{cccccccccccc}
 0 & 1 & 0 & 0 & 0 & e^{\rmi k_2} & 0 & 0 & 0 & 0 & 0 & 1 \\
 1 & 0 & 1 & 0 & 0 & 0 & 0 & 0 & e^{\rmi k_1} & 0 & 0 & 0 \\
 0 & 1 & 0 & 1 & 0 & 0 & 0 & e^{\rmi k_1} & 0 & 0 & 0 & 0 \\
 0 & 0 & 1 & 0 & 1 & 0 & 0 & 0 & 0 & 0 & e^{\rmi (k_1-k_2)} & 0 \\
 0 & 0 & 0 & 1 & 0 & 1 & 0 & 0 & 0 & e^{\rmi (k_1-k_2)} & 0 & 0 \\
 e^{-\rmi k_2} & 0 & 0 & 0 & 1 & 0 & 1 & 0 & 0 & 0 & 0 & 0 \\
 0 & 0 & 0 & 0 & 0 & 1 & 0 & 1 & 0 & 0 & 0 & e^{-\rmi k_2} \\
 0 & 0 & e^{-\rmi k_1} & 0 & 0 & 0 & 1 & 0 & 1 & 0 & 0 & 0 \\
 0 & e^{-\rmi k_1} & 0 & 0 & 0 & 0 & 0 & 1 & 0 & 1 & 0 & 0 \\
 0 & 0 & 0 & 0 & e^{\rmi (k_2-k_1)} & 0 & 0 & 0 & 1 & 0 & 1 & 0 \\
 0 & 0 & 0 & e^{\rmi (k_2-k_1)} & 0 & 0 & 0 & 0 & 0 & 1 & 0 & 1 \\
 1 & 0 & 0 & 0 & 0 & 0 & e^{\rmi k_2} & 0 & 0 & 0 & 1 & 0 \\
\end{array}
\right).
\end{align}
\end{widetext}
Again, the Bloch energy bands $\vare_{\alpha}(\vec{k})$ for these three lattices cannot be determined in closed form, but can be found numerically for fixed $\vec{k}$.

\subsection{Density of states}\label{SecDOS}

The Bloch energy bands determine the DOS of single-particle excitations according to
\begin{align}
 \label{bloch27} D(E) = \frac{1}{N_{\rm u}}\int_{\vec{k}} \sum_{\alpha=1}^{N_{\rm u}}\delta(E-\vare_{\alpha}(\vec{k})).
\end{align}
We show the resulting functions for the Archimedean lattices in Figs. \ref{FigPanel1} and \ref{FigPanel2}. For the plots, we determine the eigenvalues $\vare_{\alpha}(\vec{k})$ analytically or numerically by diagonalizing the matrix $H(\vec{k})=-A(\vec{k})$ and sample the $\vec{k}$-integral uniformly in 2500 points across the Brillouin zone, using the representation $\delta(x)=\frac{\nu}{2}\sech^2(\frac{\nu x}{2})$ with $\nu = 100$. We normalize the DOS such that
\begin{align}
 \int_{-\infty}^\infty \mbox{d}E\ D(E) =1.
\end{align}
Note that for a regular lattice with coordination number $q$, the function $D(E)$ is supported on a region contained in the real interval $[-q,q]\subset \mathbb{R}$.

In the following, we collect explicit expressions for the DOS of seven of the Archimedean lattices. For the Square lattice, Eq. (\ref{intro2}) applies, namely
\begin{align}
 \label{bloch28} D_\square(E) = \frac{1}{2\pi^2}K\Bigl(1-\frac{E^2}{16}\Bigr)
\end{align} 
for $|E|\leq4$, with complete elliptic integral of the first kind given by
\begin{align}
 \label{bloch29} K(k^2) =  \int_0^1\frac{\mbox{d}t}{\sqrt{(1-t^2)(1-k^2t^2)}}.
\end{align}
We derive this formula from the returning walk numbers in Sec. \ref{SecReturnDOS}. For the Triangular and Honeycomb lattices, we adopt the notation from Ref. \cite{Kogan2021-fy} and write
\begin{align}
 \label{bloch30}  Z_0(E) &= \begin{cases}(3-|E|)(|E|+1)^3 & |E|<1\\ 16|E| & 1\leq |E|\leq 3\end{cases},\\
 \label{bloch31}Z_1(E) &= \begin{cases} 16|E| & |E|<1 \\ (3-|E|)(|E|+1)^3 & 1\leq |E|\leq 3\end{cases},
\end{align}
so that
\begin{align}
  \label{bloch32} D_\triangle(E) &=\frac{2}{\pi^2\sqrt{Z_0(\sqrt{3-E})}} K\Biggl(\frac{Z_1(\sqrt{3-E})}{Z_0(\sqrt{3-E})}\Biggr)
\end{align}
for $-6\leq E<3$, and
\begin{align}
 \label{bloch33} D_{\rm H}(E) &=\frac{2|E|}{\pi^2\sqrt{Z_0(E)}} K\Biggl(\frac{Z_1(E)}{Z_0(E)}\Biggr)
\end{align}
for $|E|\leq3$, respectively. Note that
\begin{align}
 \label{bloch34} D_{\rm H}(E) = |E|D_{\triangle}(3-E^2).
\end{align}
For the Kagome lattice we have
\begin{align}
 \label{bloch35} D_{\rm K}(E) =\frac{1}{3}\Bigl[\delta(E-2)+ 2|E+1| D_{\Delta}\Bigl(2-E(E+2)\Bigr)\Bigr]
\end{align}
for $-4\leq E\leq 2$. For the Star lattice we have
\begin{equation}\label{bloch36} 
\begin{aligned}
 D_{\rm St}(E) ={}&\frac{1}{3}\Biggl[\frac{\delta(E-2)}{2}+\frac{\delta(E)}{2}\\
  &+|(1+2E)(E^2+E-3)|\\& \times D_{\triangle}\Bigl((3-E^2)(E(E+2)-2)\Bigr)\Biggr]
\end{aligned}
\end{equation} for $-3\leq E<-1$ and $0<E\leq 2$, while $D_{\rm St}(E)=0$ for $-1\leq E\leq 0$. For the CaVO lattice, the DOS is given by
\begin{align}
   \nonumber D_{\rm Ca}(E) &=  \int_{-1}^1\frac{\mbox{d}u}{\pi^2} \frac{|2u+E(E^2-3)|}{\sqrt{(1-u^2)\mathcal{Q}(E,u)}} \Theta(\mathcal{Q}(E,u)),\\
    \mathcal{Q}(E,u) &= 16(E-u)^2-(1-6E^2+E^4+4Eu)^2
\end{align}
for $|E|\leq 3$, where $\Theta(x)$ is the Heaviside step function. The DOS $D_{\rm Sr}(E)$ of the SrCuBO lattice in terms of a one-dimensional integral over a variable $u$ is given  in Eqs. (\ref{EDOS3}) and (\ref{EDOS3b}).

\section{Constructing large flakes and clusters of periodic tilings}\label{SecGraphs}

One way to compute the returning walk numbers $S_n$ on periodic tilings of the plane is to compute the $n$-th moments of the adjacency matrix $A$ for large, but finite lattices (see Sec. \ref{SecWalks}). This can then be tested for convergence as the size of the graph is increased. Since the construction of such finite graphs is of practical relevance for many applications in condensed matter and statistical physics, we describe here a general method how to construct adjacency matrices for finite tilings of arbitrary size, with both open and periodic boundary conditions. It applies to all periodic tilings of the Euclidean plane, as long as the Bravais lattice and unit cell have been identified, not just Archimedean lattices. Moreover, the methodology can be generalized to periodic tilings in $\mathbb{R}^d$ for $d\geq 3$.

In the following, we consider two-dimensional tilings that are built from periodically repeating a unit cell consisting of $N_{\rm u}$ sites. For positive integers $p$ and $q$, we denote by $\Lambda_{p,q}$ the graph obtained by stitching together $p$ ($q$) repetitions of the unit cell in the $\boldsymbol{e}_1$ ($\boldsymbol{e}_2$) direction, and denote its $N\times N$  adjacency matrix by $A_{p,q}$. The associated graph has $N=pqN_{\rm u}$ sites with either open boundary conditions (OBC) or periodic boundary conditions (PBC). In the case of OBC, the resulting graph is planar and will be called a ``flake''. In the case of PBC, the resulting graph wraps around a torus of genus 1 and will be called a ``cluster''. The central idea behind this construction is to view the graph as a placement of the unit cell onto the vertices of a Bravais lattice.

Before we proceed, we want to highlight two perspectives that naturally emerge in the pursuit of constructing $A_{p,q}$. The first is along the lines of constructing periodic tessellations given an arbitrary  unit cell $\Lambda_\uc$ described by its adjacency matrix $A_\uc$. We may envision a lattice made up of periodic repetitions of $A_\uc$ with some tailored connections (edges) between different unit cells. This method is outlined below. On the other hand, we may start with an existing lattice (such an Archimedean lattice) and identify the unit cell and its corresponding Bloch adjacency matrix $A(\bk)$. We could then use $A(\bk)$ as a reference to construct the finite graph. This is because $A(\bk)$ efficiently encodes the information pertaining to both the inter- and 
intra-unit cell connections (see App. \ref{AppApq}).

Some definitions are necessary to state the master formula for $A_{p,q}$ in Eq. (\ref{mastereq1}) below. First, define the $N_{\rm u}\times N_{\rm u}$ adjacency matrix of the unit cell, $A_{\rm u}$, which corresponds to the graph that one wishes to use to tessellate the plane. Its matrix elements specify the connections within the unit cell. Alternatively, if one possesses $A(\bk)$, then one can obtain $A_\uc$ by erasing all entries from $A(\vec{k})$ that are not 1. To account for edges that leave the unit cell, define for $i,j\in\{1,\dots, N_{\rm u}\}$ the $N_{\rm u}\times N_{\rm u}$ connection matrices
 \begin{equation}
   \label{flake2} (\Gamma_{(i,j)})_{kl}=\begin{cases}
    1 \quad k=i \text{ and } l=j \\
    0 \quad \text{else}
    \end{cases},
 \end{equation}
and the $m\times m$ right- and left-shift matrices $R_m$ and $L_m$, respectively, for OBC and PBC via
\begin{align}
  \label{flake3} \text{OBC}:\ (R_m)_{ij}&=\begin{cases}
    1 \quad j=i+1 \\
    0 \quad \text{else}
    \end{cases},\\
 \label{flake4} \text{PBC}:\ (R_m)_{ij}&=\begin{cases}
    1 \quad j=i+1 \text{ or } (i,j)=(m,1)\\
    0 \quad \text{else}
    \end{cases},
\end{align}
and
\begin{align}
 \label{flake5} L_m = (R_m)^T.
\end{align}
The shift matrices are the adjacency matrices of directed linear graphs (i.e  directed chains) with $m$ vertices, with either OBC or PBC. They are used to stitch together neighboring unit cells.

At last, let $\mathcal{I}(\boldsymbol{d})$ denote the set of directed edges leaving the unit cell in the $\boldsymbol{d}$-direction. Each entry of $\mathcal{I}(\vec{d})$ for a given $\vec{d}$ is of the form $(i,j)$ with some $i,j\in\{1,\dots,N_{\rm u}\}$. For tessellating $\mathbb{R}^2$, we only need the cases where $\boldsymbol{d}\in\{\boldsymbol{e}_1,\boldsymbol{e}_2,\boldsymbol{e}_1+\boldsymbol{e}_2,\boldsymbol{e}_1-\boldsymbol{e}_2\}$. Note that we will account for $-\bd{d}$ separately below. The set $\mathcal{I}(\bd{d})$ cata\-logs all directed edges that leave a single, fixed unit cell and end in a neighboring unit cell. Examples for how to construct the $\mathcal{I}(\vec{d})$ are given below and in App. \ref{AppApq}.

With these definitions at hand, the graph adjacency matrix $A_{p,q}$ for the $N$-site flake or cluster is given by 
\begin{equation}
\begin{aligned}\label{mastereq1}
A_{p,q}&=(\mathbb{1}_p\otimes \mathbb{1}_q)\otimes A_{\rm u} \\
&+ \sum_{(i,j)\in\mathcal{I}(\boldsymbol{e}_1)}\left((\mathbb{1}_p\otimes R_q)\otimes \Gamma_{(i,j)}+ \text{h.c.} \right)\\
&+ \sum_{(i,j)\in\mathcal{I}(\boldsymbol{e}_2)}\left((R_p\otimes \mathbb{1}_q)\otimes \Gamma_{(i,j)}+\text{h.c.}\right)\\
& + \sum_{(i,j)\in\mathcal{I}(\boldsymbol{e}_1+\boldsymbol{e}_2)}\left((R_p\otimes R_q)\otimes \Gamma_{(i,j)}+\text{h.c.}\right)\\
&+ \sum_{(i,j)\in\mathcal{I}(\boldsymbol{e}_1-\boldsymbol{e}_2)}\left((L_p\otimes R_q)\otimes \Gamma_{(i,j)}+\text{h.c.}\right).
\end{aligned}
\end{equation}
Herein $\mathbb{1}_n$ is the $n\times n$ unit matrix, and the right- and left-shift matrices have to be chosen according to the desired boundary conditions.  We emphasize the inclusion of the Hermitian conjugate, which accounts for the edges leaving in the $-\boldsymbol{d}$ direction. If $\mathcal{I}(\boldsymbol{d})=\emptyset$ for some $\boldsymbol{d}$, we exclude the corresponding term. Finally, we note that if $\mathcal{I}(\bd{e}_1+\bd{e}_2)\neq\emptyset$ then $\mathcal{I}(\bd{e}_1-\bd{e}_2)=\emptyset$, and vice versa, in order to produce a periodic tiling that corresponds to a planar lattice. If both are non-empty simultaneously, then the graph would no longer be planar due to the simple fact that in $\mathbb{R}^2$ there are only two topologically distinct Bravais lattices. They are the Square and Triangular lattices, which we will call Type S and Type T in the following. We show below that Eq. (\ref{mastereq1}) can be interpreted as a Type S or Type T graph wherein each vertex is replaced with a unit cell, producing the graph $\Lambda_{p,q}$. 

We call the graph that arises from identifying each unit cell with a vertex (called \textit{quotient vertex}) the \textit{quotient graph} of the lattice $\Lambda$ \cite{GARDINER1974255,Bader2013}.  In our case, the quotient graph consists of quotient vertices $\mathcal{V}$ and \textit{quotient edges} $\mathcal{E}$, wherein a quotient vertex is identified with a unit cell on the underlying lattice, and a quotient edge exists between quotient vertices $I,J\in\mathcal{V}$ if there is an edge connecting the unit cells $I$ and $J$. More formally, we quotient the original graph $\Lambda$ with the equivalence relation $R$ where two vertices $i$ and $j$ of the original lattice are equivalent $i \sim j$ (are identified with the same quotient vertex) if they belong to the same unit cell. Thus, the quotient graph is given by $\Lambda/R$. For the Archimedean lattices, the quotient graph is always a Bravais lattice and therefore can only be either Type S or Type T. 

We illustrate the method with two examples from the eleven Archimedean lattices. These two examples also illustrate the difference between Type S and Type T quotient graphs. The remaining nine lattices are covered in App. \ref{AppApq}.

\begin{figure}[t!]
\includegraphics[width=1\linewidth]{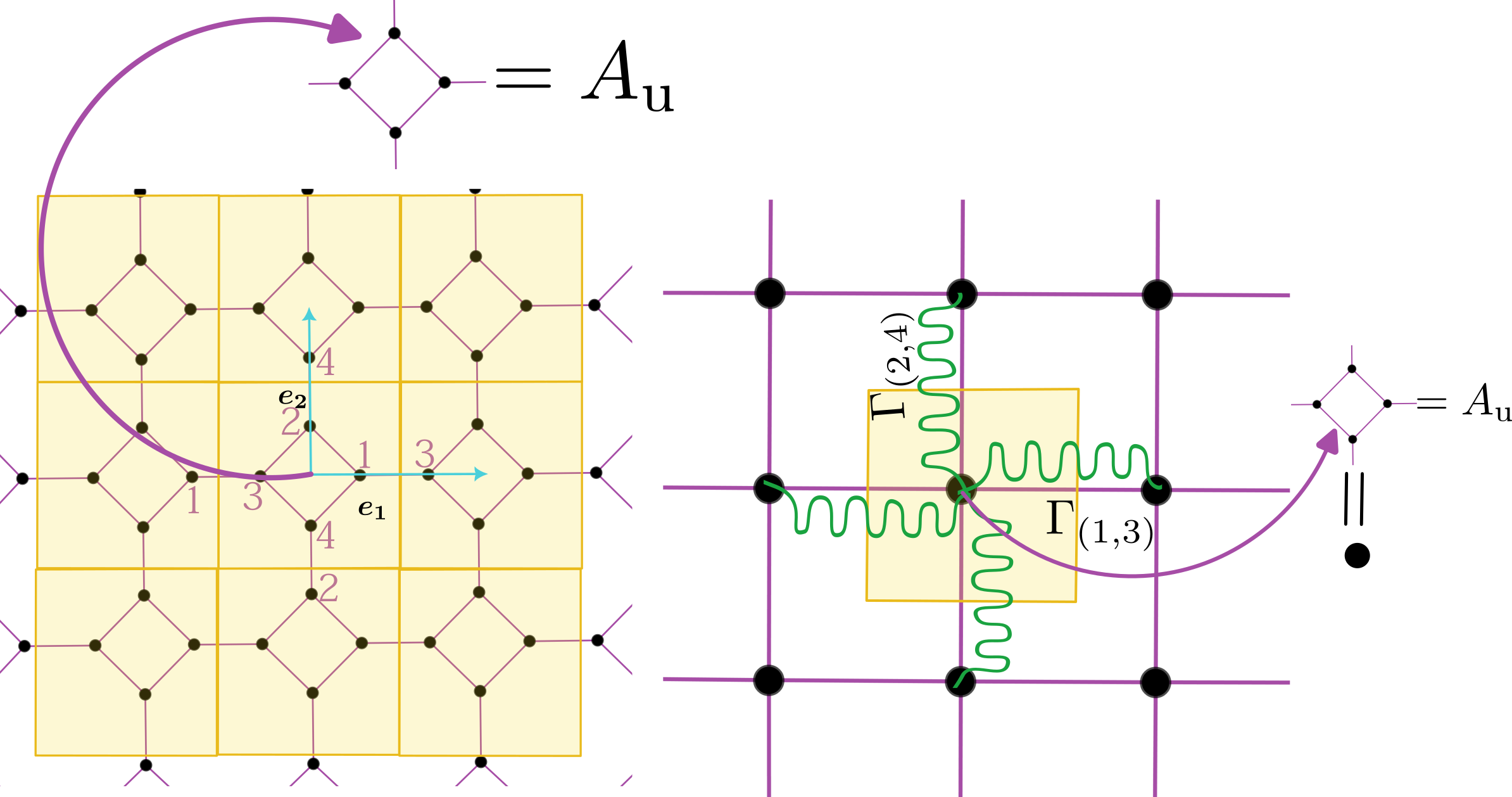}
\caption{\emph{Left.} CaVO lattice shown together with its unit cell $A_{\rm u}$. Note the numbering scheme within the unit cell, which is applied consistently among the repeated unit cells. \emph{Right.} The quotient graph underlying the CaVO lattice, consisting of black dots and purple lines, is of Type S. Each quotient vertex (black dot) corresponds to a unit cell $A_{\rm u}$ and each squiggly line (repeated for the entire quotient graph) corresponds to connections $\Gamma_{(i,j)}$ between sites from different unit cells.}
\label{cavotess}
 \end{figure}

First consider the CaVO lattice, which tessellates the plane by a $4$-site unit cell as illustrated in Fig. \ref{cavotess}a. The adjacency matrix of the unit cell is 

\begin{equation}
    A_\uc^{(\rm Ca)}=\begin{pmatrix}    0 & 1 & 0 & 1 \\ 1 & 0 & 1 & 0 \\ 0 & 1 & 0 & 1 \\ 1 & 0 & 1  & 0
\end{pmatrix}.
\end{equation}
The quotient graph has the form shown in Fig. \ref{cavotess}b. The edges that leave the unit cell in the $\bd{e}_1$- and $\bd{e}_2$- directions are given by $\mathcal{I}(\bd{e}_1)=\{(1,3)\}$ and $\mathcal{I}(\bd{e}_2)=\{(2,4)\},$ respectively. The lattice is of S-type and thus $\mathcal{I}(\bd{e}_1\pm\bd{e}_2)=\emptyset$. We arrive at
\begin{align}
 \nonumber A_{p,q}^{(\rm Ca)} ={}& (\mathbb{1}_p\otimes \mathbb{1}_q)\otimes A_{\rm u}^{(\rm Ca)} \\
 \label{flake11} &+ [(\mathbb{1}_p\otimes R_q)\otimes \Gamma_{(1,3)}+ \text{h.c.}]\\
 \nonumber &+ [(R_p\otimes \mathbb{1}_q)\otimes \Gamma_{(2,4)}+\text{h.c.}].
\end{align}
The corresponding flake or cluster has $N=4pq$ sites.

\begin{figure}[t!]
 \includegraphics[width=0.4\linewidth]{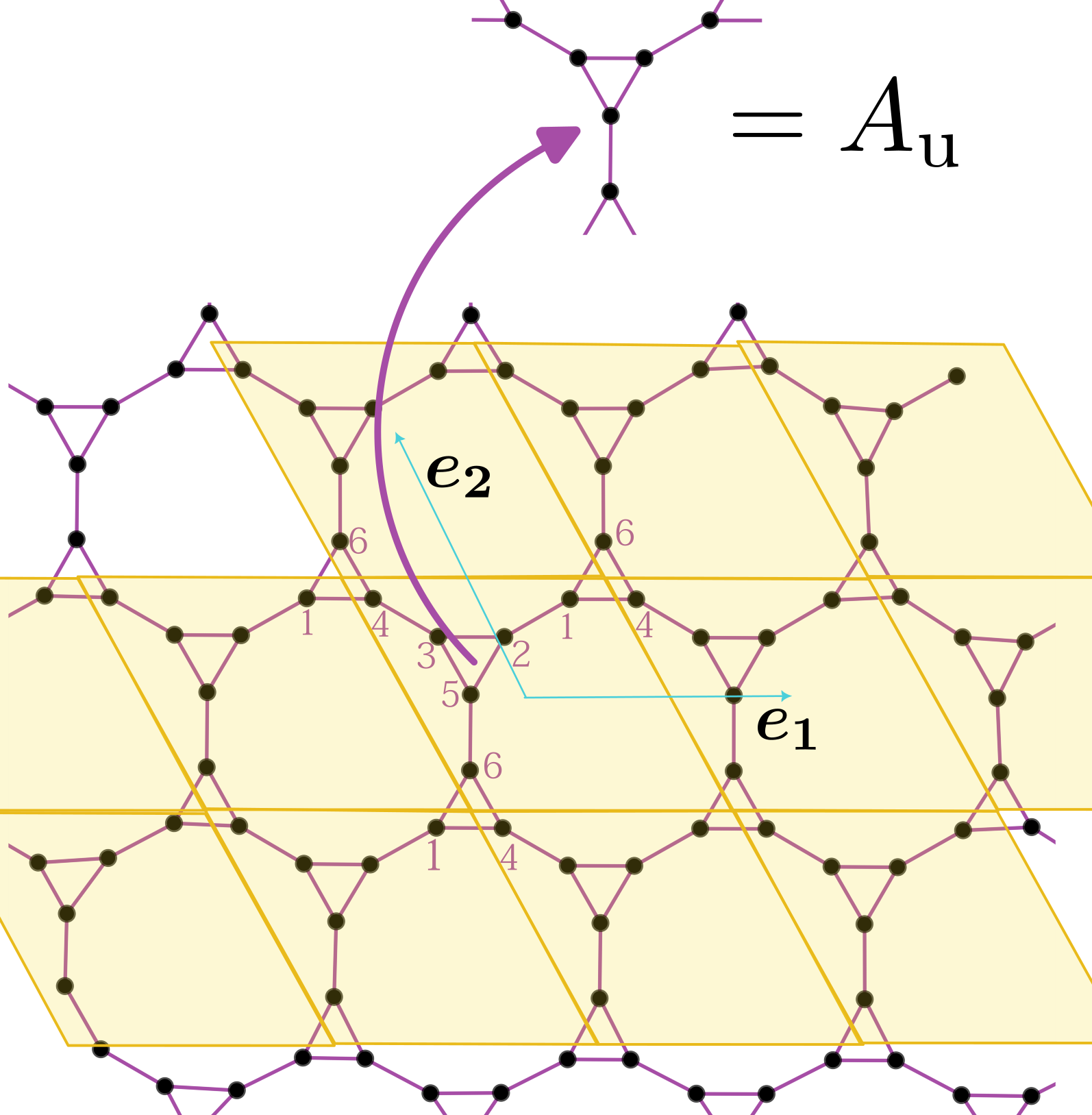}
 \includegraphics[width=0.57\linewidth]{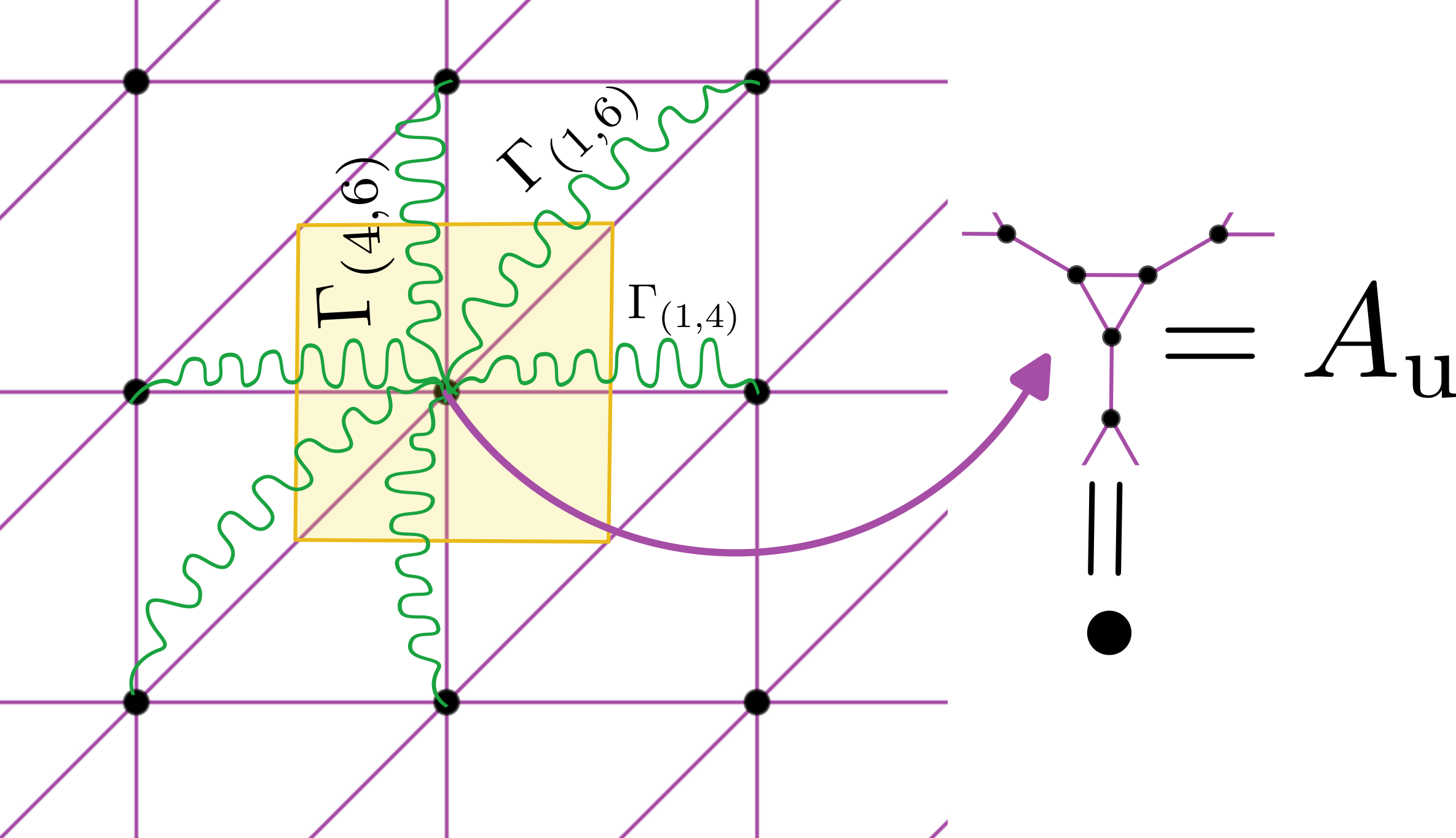}
\caption{Same as in Fig. \ref{cavotess}, but for the Star lattice, where the unit cell now consists of $N_\uc=6$ sites. The underlying quotient graph, indicated by the black dots and purple lines, is of Type T. This can be seen from the fact that there exists a diagonal link between the unit cells.}
\label{startess}
\end{figure}

Next consider the Star lattice, which tessellates the plane with the 6-site unit cell as shown in Fig. \ref{startess}a. The adjacency matrix of the unit cell is
\begin{align}
 \label{flake12} A_{\rm u }^{(\rm St)}=\begin{pmatrix}
 0 & 1 & 0 & 0 & 0 & 0 \\
 1 & 0 & 1 & 0 & 1 & 0 \\
 0 & 1 & 0 & 1 & 1 & 0 \\
 0 & 0 & 1 & 0 & 0 & 0 \\
 0 & 1 & 1 & 0 & 0 & 1 \\
 0 & 0 & 0 & 0 & 1 & 0 \\
\end{pmatrix}.
\end{align}
The associated quotient graph is shown in Fig. \ref{startess}b. We have $\mathcal{I}(\vec{e}_1) = \{(1,4)\}$, $\mathcal{I}(\vec{e}_2) = \{(4,6)\}$, $\mathcal{I}(\vec{e}_1+\vec{e}_2) = \{(1,6)\}$, and $\mathcal{I}(\vec{e}_1-\vec{e}_2) = \emptyset$. We arrive at
\begin{align}
 \nonumber A_{p,q}^{(\rm St)} ={}& (\mathbb{1}_p\otimes \mathbb{1}_q)\otimes A_{\rm u}^{(\rm St)} \\
 \nonumber &+ [(\mathbb{1}_p\otimes R_q)\otimes \Gamma_{(1,4)}+ \text{h.c.}]\\
 \nonumber &+ [(R_p\otimes \mathbb{1}_q)\otimes \Gamma_{(4,6)}+\text{h.c.}]\\
  \label{flake17} &+ [(R_p\otimes R_q)\otimes \Gamma_{(1,6)}+\text{h.c.}].
\end{align}
The corresponding flake or cluster has $N=6pq$ sites.

 To summarize, the adjacency matrix $A_{p,q}$ for a graph consisting of $p\times q$ unit cells can be constructed by first constructing a quotient graph whose vertices are replicas of the unit cell. For this, we introduce a grid with $p\times q$ disconnected vertices, to each of which we attach the adjacency matrix of the unit cell $A_\uc$ to obtain the matrix $(\mathbb{1}_p\otimes \mathbb{1}_q)\otimes A_\uc$.  Then we connect the vertices of the quotient graph through the edges between unit cells in $\pm\bd{e}_1$-directions by adding the terms $\sum_{(i,j)\in\mathcal{I}(\boldsymbol{e}_1)}\left( (\mathbb{1}_p\otimes R_q)\otimes \Gamma_{(i,j)}+ \text{h.c.} \right)$, and repeat the procedure for $\vec{e}_2$, $\vec{e}_1+\vec{e}_2$, and $\vec{e}_1-\vec{e}_2$. Connections in the $(\vec{e}_1\pm\vec{e}_2)$-directions only exists for Type T lattices, while they are absent for Type S.

\section{Returning walks on Archimedean lattices}\label{SecWalks}

In this section, we illustrate how to compute the returning walks numbers $S_n$ on a two-dimensional Euclidean lattice $\Lambda$ on the example of the Archimedean lattices. We introduce the generating function $\mathcal{G}(z)=\sum_{n\geq 0}S_n z^n$ for the integers $S_n$ and show how it can be computed efficiently from the Bloch Hamiltonian matrix. We establish the connection to the density of states $D_\Lambda(E)$.

\subsection{Definition and generating function}\label{SecWalksDef}

Given a graph $\Lambda$ and its adjacency matrix $A$, we define 
\begin{equation}
 \label{def1} S^{(i)}_n=(A^{n})_{ii} = \langle i|\hat{A}^n|i\rangle,
\end{equation} 
which counts the number of walks that start at site $i$ and return to site $i$ in $n$ steps. We 
call such walks \textit{returning walks} and the associated numbers $S_n^{(i)}$ returning walk numbers at site $i$. In Eq. (\ref{def1}), one can identify $|i\rangle \longleftrightarrow (0,\dots,0,1,0,\dots,0)^T$ with the $N$-component unit vector that is nonzero at $i$. For a graph with $N$ sites, we define the average number of returning walks in $n$ steps by
\begin{equation}
 \label{avgTr} S_n= \frac{1}{N} \mbox{Tr}[\hat{A}^n]=\frac{1}{N}\sum_{i=0}^N S^{(i)}_n.
\end{equation}
For Archimedean lattices, vertex transitivity implies
\begin{align}
 \label{def3} S_n=S_n^{(i)}
\end{align}
for all $i$, i.e. $S_n^{(i)}$ is independent of the site $i$ chosen.

The returning walk numbers $S_n$ of Archimedean lattices for small $n$ can be computed from large enough clusters as constructed in Sec. \ref{SecGraphs}. For this we compute the adjacency matrices $A_{p,q}$ for $p\times q$ unit cells, increase $p,q$, and check for convergence of $S_n$ for each fixed $n$. The resulting walk numbers $S_n$ are given in Table \ref{TabWalks}. Note that for bipartite lattices  $S_n=0$ for $n$ odd. The reason is that a bipartite lattice can be divided into two sublattices of, say, red and blue sites, and edges are only between blue and red sites. Starting at a blue site, it is impossible to return to this (or any other) blue site in an odd number of steps.

Following P\'olya \footnote{{\it``A generating function is a device somewhat similar to a bag. Instead of carrying many little objects detachedly, which could be embarrassing, we put them all in a bag, and then we have only one object to carry, the bag.''}, G. P\'olya, Mathematics and plausible reasoning (1954)}, 
we conveniently collect the returning walk numbers in the generating functions
\begin{align}
 \label{def3b} \mathcal{G}^{(i)}(z) &= \sum_{n\geq 0}S_n^{(i)}z^n,\\
 \label{def3c} \mathcal{G}(z) &= \sum_{n\geq 0}S_nz^n.
\end{align}
These are a formal power series in the variable $z$ and as such not required to converge. The integer $S_n$ appears as the coefficient of $z^n$, which we write as
\begin{align}
 \label{gf2} S_n = [z^n] \mathcal{G}(z),
\end{align}
and analogous for the other generating function.

\subsection{Bloch generating function}

In this section, we show that the returning walk numbers for a periodic tessellation consisting of a unit cell with sites $i_\uc=1,\dots N_\uc$ can be computed from the $N_{\rm u}\times N_{\rm u}$ Bloch adjacency matrix $A(\vec{k})$ through
\begin{equation}
    \label{gf5b}
    S_{n}^{(i_\uc)}=\int_{\bk}[A(\bk)^n]_{i_\uc i_\uc},
\end{equation}
with the integrand being the $(i_{\rm u},i_{\rm u})$-matrix element of $A(\vec{k})$. The averaged returning walk number $S_n$ from Eq. (\ref{avgTr}) is given by
\begin{align}
 \label{gf5} S_n = \frac{1}{N_{\rm u}} \int_{\vec{k}} \mbox{Tr}[A(\vec{k})^n].
\end{align}
These equations are the central formulas of our work. It connects the graph-theoretic definition of the returning walk numbers to the Bloch energy bands. Importantly, in contrast to the trace over all $N$ sites of the lattice in the original definition of $S_n$ in Eq. (\ref{avgTr}), with the limit $N\to \infty$ understood for infinite lattices, the trace here only comprises the manifestly finite number of $N_{\rm u}$ sites of the unit cell. An infinite lattice is incorporated by the Bloch momenta $\vec{k}$ that densely fill the Brillouin zone. 
In the following, we give a self-contained derivation in the light of Archimedean lattices that generalizes to other two- or higher-dimensional tessellations.

To establish Eq. (\ref{gf5}), some definitions are necessary. For fixed $\vec{k}$, we denote
\begin{align}
 \label{gf0a} S_n^{(i_\uc)}(\vec{k}) &= [A(\vec{k})^n]_{i_\uc i_\uc},\\
 \label{gf0b} S_n(\vec{k}) &= \frac{1}{N_{\rm u}} \sum_{i_{\rm u}=1}^{N_{\rm u}} [A(\vec{k})^n]_{i_\uc i_\uc} = \frac{1}{N_{\rm u}}\mbox{Tr}[A(\vec{k})^n],
\end{align}
where $i_\uc$ corresponds to a site in the unit cell. We further introduce the Bloch generating functions
\begin{align}
 \mathcal{G}^{(i_\uc)}(z,\vec{k}) &= \sum_{n\geq 0}S_n^{(i_\uc)}(\vec{k})z^n,\\
 \label{gf3more} \mathcal{G}(z,\vec{k}) &= \sum_{n\geq 0}S_n(\vec{k})z^n
\end{align}
in analogy to Eqs. (\ref{def3b}) and (\ref{def3c}). These functions satisfy
\begin{align}
 \label{gf3a} &\mathcal{G}^{(i_\uc)}(z,\vec{k}) = \Bigl(\frac{1}{\mathbb{1}-zA(\vec{k})}\Bigr)_{i_\uc i_\uc},\\
 \label{gf3} &\mathcal{G}(z,\vec{k}) = \frac{1}{N_{\rm u}}\mbox{Tr}\Bigl(\frac{1}{\mathbb{1}-zA(\vec{k})}\Bigr)
\end{align}
and
\begin{align}
 \label{gf4c} \mathcal{G}(z,\vec{k}) = \frac{1}{N_{\rm u}}\sum_{i_\uc=1}^{N_{\rm u}}\mathcal{G}^{(i_\uc)}(z,\vec{k}).
\end{align}
Equations (\ref{gf3a}) and (\ref{gf3}) can be obtained by expanding the matrix inverse in a geometric series. We list the generating functions $\mathcal{G}(z,\vec{k})$ for the Archimedean lattices in Tab. \ref{TabGF}.

\renewcommand{\arraystretch}{2.5}
\begin{table*}[t!] 
\begin{tabular}{|c|c|}
\hline
\centering
Lattice & $\mathcal{G}(z,\vec{k})$\\ 
\hline\hline
 Square &  \ {\large $\frac{1}{1+z\varepsilon_\square(\vec{k})}$} \\ 
\hline
 Triangular \ &  \	{\large $\frac{1}{1+z\varepsilon_\Delta(\vec{k})}$}\\
\hline
 \ Honeycomb	\ &  \	{\large $\frac{1}{1-z^2(3-\varepsilon_\Delta(\vec{k}))}$}\\
\hline
  Kagome  &  	{\large $\frac{1}{3}\Bigl(\frac{1}{1+2 z}+\frac{2 (1-z)}{1-2 z (1+z)+z^2 \varepsilon_{\Delta }(\vec{k})}\Bigr)$}\\
 \hline
 Star & {\large $\frac{1}{3}\Bigl(\frac{1+z}{1+2 z}+\frac{(2-z)(1-z-3 z^2)}{(1-3 z^2) [1-2 z (1+z)]+z^4 \varepsilon _{\Delta }(\vec{k})}\Bigr)$}\\ 
 \hline
 Trellis & {\large $\frac{1-2zc_1}{1-3z^2-4zc_1+4z^2c_1^2 +z^2\vare_\Delta(\vec{k})}$}\\
 \hline
  CaVO & {\large $\frac{1-3z^2-z^3(c_1+c_2)}{1-6z^2+z^4-4z^3(c_1+c_2)-4z^4c_1c_2}$ }\\
   \hline
 SrCuBO &  {\large $\frac{1-5z^2-2z^3-2z^2(1+z)(c_1+c_2)-2z^3c_1c_2}{1-10z^2-8z^3+z^4-4z^2(1+z)^2(c_1+c_2)+ 4z^4(c_1^2+c_2^2)-8z^3(1+2z)c_1c_2}$}\\
   \hline
 Ruby & {\large $\frac{1-8z^2+z^3+12z^4-\frac{1}{3}(3+4z)z^3\vare_{\rm H}^2}{1 -12z^2 +2z^3+36z^4 +9z^6-2z^3(1+2z+3z^3)\vare_{\rm H}^2-4z^6(\vare_{\rm H}')^2+z^6\vare_{\rm H}^4}$}\\
   \hline 
 Maple-Leaf & \ {\large $\frac{1-10z^2-2z^3+21z^4+8z^5-z^3(2+4z+z^2)\vare_{\rm H}^2-\frac{1}{3}z^5(\vare_{\rm H}')^2}{1-15z^2-4z^3+63z^4+48z^5+4z^6-2z^3(2+6z+3z^2+2z^3)\vare_{\rm H}^2-2z^5(1+3z)(\vare_{\rm H}')^2+z^6\vare_{\rm H}^4}$} \ \\
   \hline
 SHD & {\large $\frac{1-15z^2+74z^4-131z^6+59z^8-z^6(3-10z^2+5z^4)\vare_{\rm H}^2-\frac{2}{3}z^{10}(\vare_{\rm H}')^2}{1 - 18z^2+111z^4 -262z^6+177z^8 -6z^6(1-5z^2+5z^4)\vare_{\rm H}^2-4z^{10}(\vare_{\rm H}')^2 + z^{12}\vare_{\rm H}^4}$}\\
   \hline
\end{tabular}
\caption{Bloch generating functions $\mathcal{G}(z,\bk)=\sum_{n\geq 0} S_n(\bk) z^n$ for the Archimedean lattices. Performing the $\vec{k}$-integrals $\mathcal{G}(z)=\int_{\vec{k}}\mathcal{G}(z,\vec{k})$ and $S_n=\int_{\vec{k}}[z^n]\mathcal{G}(z,\vec{k})$ yields the generating functions and returning walks numbers from Tab. \ref{TabWalks}. We abbreviate $c_i=\cos(k_i)$, $\vare_{\rm H}=\vare_{\rm H}(\bk)$ as in Eq. (\ref{bloch13a}), and $(\vare^{\prime}_{\rm H})^2=3+2\cos(2k_1-k_2)+2\cos(k_1-2k_2)+2\cos(k_1+k_2)$ as in Eq. (\ref{uv52}), while $\vare_\square(\vec{k})$ and $\vare_\Delta(\vec{k})$ are the dispersion relations of the Square and Triangular lattice in Eqs. (\ref{bloch9}) and (\ref{bloch10}), respectively. The expressions listed here have been obtained from computing the trace of the resolvent according to Eq. (\ref{gf3}). For the Star lattice, we have shifted $k_2\to-k_2$ with respect to $A^{(\rm St)}(\vec{k})$ in Eq. (\ref{bloch22}) to simplify the expression for $\mathcal{G}_{\rm St}(z,\vec{k})$.}
\label{TabGF}
\end{table*}
\renewcommand{\arraystretch}{1}

We now show the crucial property
\begin{align}
 \label{gf0c} S_n^{(i_u)} &=\int_{\vec{k}} S_n^{(i_u)}(\vec{k}),\\
 \label{gf0d} S_n &=\int_{\vec{k}} S_n(\vec{k}).
\end{align}
Note that Eq. (\ref{gf0c}) implies Eq. (\ref{gf0d}) when summed over all unit cell sites $i_{\rm u}$, and that Eq. (\ref{gf0d}) is equivalent to Eq. (\ref{gf5}). When applied to the whole generating functions, we have
\begin{align}
 \label{gf4b} &\mathcal{G}^{(i_\uc)}(z) = \int_{\vec{k}} \mathcal{G}^{(i_\uc)}(z,\vec{k}),\\
 \label{gf4} &\mathcal{G}(z) =\int_{\vec{k}} \mathcal{G}(z,\vec{k}).
 \end{align}
The corresponding $\vec{k}$-integrals can be evaluated with an algebraic technique described in Sec. \ref{uvMethod} or, of course, numerically. While for Archimedean lattices $S_n^{(i_\uc)}(\vec{k})$ depends on the choice of $i_\uc$, the integral $\int_{\vec{k}} S_n^{(i_\uc)}(\vec{k})$ does not, and coincides with $S_n$. We confirm that the correct walk numbers $S_n$ listed in Tab. \ref{TabWalks} are reproduced from $S_n=\int_{\vec{k}}  [z^n]\mathcal{G}(z,\vec{k})$ for the Archimedean lattices.

To prove Eq. (\ref{gf0c}), which in turn implies Eqs. (\ref{gf0d}) and (\ref{gf5}), label the sites on the whole lattice $i=1,2,\dots,N$ and the sites of the first unit cell, which comprises the fundamental domain of the tessellation, by $i_{\rm u}=1,\dots,N_{\rm u}$. Any site $|i\rangle$ can be uniquely identified with a pair $|i_{\rm u},\vec{v}\rangle$, where $i_{\rm u}$ is a site in the unit cell and $\vec{v}=v_1\vec{e}_1+v_2\vec{e}_2$ is a Bravais lattice vector with $(v_1,v_2)\in\mathbb{Z}^2$. For the sites in the first unit cell we have $|i_{\rm u}\rangle=|i_{\rm u},\vec{0}\rangle$. In the following, we identify the Bravais lattice vertices for simplicity with $\mathbb{Z}^2$ and write $\vec{v}=(v_1,v_2)^T\in\mathbb{Z}^2$. We define the translation operator $\hat{T}_{\vec{v}}$ on the Bravais lattice through its action
\begin{align}
 \label{gf6} \hat{T}_{\vec{v}} |i_{\rm u},\vec{w}\rangle = |i_{\rm u},\vec{v}+\vec{w}\rangle.
\end{align}
Note that $\hat{T}_{\vec{v}}$ is unitary and $\hat{T}_{\vec{v}}^\dagger=\hat{T}_{-\vec{v}}$. We have $|i_{\rm u},\vec{v}\rangle=\hat{T}_{\vec{v}}|i_{\rm u},\vec{0}\rangle$. The Bloch momentum $\vec{k}$ labels the states that are dual to the Bravais lattice sites $\vec{v}$.

The matrix elements of the Bloch adjacency matrix are given by
\begin{align}
 \label{gf11}  A(\vec{k})_{i_{\rm u}j_{\rm u}} &= \langle i_{\rm u}|\hat{A}(\vec{k})|j_{\rm u}\rangle=\sum_{\vec{v}\in\mathbb{Z}^2} \langle i_{\rm u}|\hat{A}\hat{T}_{\vec{v}}|j_{\rm u}\rangle e^{-\rmi \vec{v}\cdot\vec{k}},
\end{align} 
as derived in App. \ref{AppBAdj}. This equation may be taken as the definition of $A(\vec{k})$. The spectrum of $\hat{A}(\bk)$ for $\bk \in [0,2\pi)\times [0,2\pi)$ is equivalent to the spectrum of the lattice described by $\hat{A}$. Through induction in $n$ we find
\begin{align}
 \label{gf12} [A(\vec{k})^n]_{i_{\rm u}i_{\rm u}}=\sum_{\vec{v}\in\mathbb{Z}^2} \langle i_{\rm u}|\hat{A}^n\hat{T}_{\vec{v}}|i_{\rm u}\rangle e^{-\rmi \vec{v}\cdot\vec{k}}
\end{align}
for $n\geq 1$, see App. \ref{AppBAdj}. Then, using $\int_{\vec{k}} e^{-\rmi \vec{v}\cdot\vec{k}}=\delta_{\vec{0},\vec{v}}$ and $\hat{T}_{\vec{0}}=\hat{\mathbb{1}}$, we arrive at
\begin{align}
 \nonumber \int_{\vec{k}}[A(\vec{k})^n]_{i_{\rm u}i_{\rm u}} &= \sum_{\vec{v}\in\mathbb{Z}^2} \langle i_{\rm u}|\hat{A}^n\hat{T}_{\vec{v}}|i_{\rm u}\rangle \delta_{\vec{0},\vec{v}} \\
 \label{gf13} &= (A^n)_{i_{\rm u}i_{\rm u}} =S_n^{(i_{\rm u})},
\end{align}
which shows Eq. (\ref{gf0c}), and thus Eq. (\ref{gf5}). Remarkably, Eq. (\ref{gf5}) holds generally for any periodic tiling of the plane. For a more detailed derivation involving the thermodynamic limit and Riemann sums, see App \ref{Trid}. 

Equation (\ref{gf5}) implies that we have
\begin{align}
 \label{gf14} S_n = \frac{(-1)^n}{N_{\rm u}} \sum_{\alpha=1}^{N_{\rm u}} \int_{\vec{k}} \vare_\alpha(\vec{k})^n,
\end{align}
where $\vare_\alpha(\vec{k})$ are the eigenvalues of the $N_{\rm u}\times N_{\rm u}$ Hermitian matrix $H(\vec{k})=-A(\vec{k})$. This formula, albeit remarkable, is of limited practical use since the eigenvalues $\vare_\alpha(\vec{k})$ cannot always be determined in closed analytic form. However, we can always analytically compute the matrix inverse $[\mathbb{1}-z A(\vec{k})]^{-1}$ in practice, for instance by using \texttt{Mathematica} or similar computer algebra software, or from the first few  moments of $A(\bk)$ through application of the Cayley--Hamilton theorem. The diagonal elements then give $\mathcal{G}^{(i_{\rm u})}(z,\vec{k})$ through Eq. (\ref{gf3a}). With this, we confirm that the correct walk numbers $S_n$ listed in Tab. \ref{TabWalks} are reproduced by $S_n = \int_{\vec{k}}[z^n]\mathcal{G}(z,\vec{k})$ using the Bloch generating functions from Tab. \ref{TabGF}.

\subsection{Evaluation of momentum integral and explicit formulas for $S_n$ and $\mathcal{G}(z)$}\label{uvMethod}

We now present a purely algebraic technique to evaluate the momentum integral in
\begin{align}
 \label{uv1} S_n = \int_{\vec{k}} S_n(\vec{k})
\end{align}
and other similar expressions. Since $S_n(\vec{k})$ is obtained from the Bloch adjacency matrix $A(\vec{k})$, it is a meromorphic function in the two variables
\begin{align}
 \label{uv2} u = e^{\rmi k_1},\ v=e^{\rmi k_2}.
\end{align} 
Indeed, all matrix elements of $A(\vec{k})$ and their associated traces and eigenvalues are functions of $u,v,$ and their inverses
\begin{align}
 \label{uv3} \frac{1}{u} = e^{-\rmi k_1},\ \frac{1}{v}=e^{-\rmi k_2}.
\end{align}
Integrals like $S_n=\int_{\vec{k}} [z^n]\mathcal{G}(z,\vec{k})$ can then be computed using the formula
\begin{align}
 \label{uv4} \int_{\vec{k}} u^m v^n = \delta_{m,0}\delta_{n,0}.
\end{align}
In particular, for a meromorphic function $F(u,v)$ of $u,v$, we define the constant part of $F(u,v)$ as
\begin{align}
 \label{uv5} \mathscr{C}[F(u,v)] = \text{constant part of }F(u,v) = F_{00},
\end{align}
where $F(u,v)=\sum_{m,n\in\mathbb{Z}}F_{mn}u^mv^n$ is the Laurent series of $F$.  Note that $F_{00}\neq F(0,0)$. Further observe that taking the constant part is linear in its arguments,
\begin{align}
 \nonumber &\mathscr{C}[c_1 F_1(u,v)+c_2 F_2(u,v)] \\
 &= c_1 \mathscr{C}[F_1(u,v)] +c_2 \mathscr{C}[F_2(u,v)],
\end{align}
and the constant part of the product of two functions $\phi(u)$ and $\sigma(v)$ is given by
\begin{equation}
    \mathscr{C}[\phi(u) \sigma(v)]= \mathscr{C}[\phi(u)]\mathscr{C}[\sigma(v)].
\end{equation}
In the following, we derive explicit formulas for $S_n$ for some Archimedean lattices, which will illuminate how this powerful technique works in practice. In cases where $S_n$ cannot be explicitly determined, we compute the generating function $\mathcal{G}(z)$. The results of this analysis are summarized in Tab. \ref{TabWalks}.

\subsubsection{Square lattice}

For the Square lattice we have
\begin{align}
 \label{uv6} S_n^{(\square)} =  \int_{\vec{k}} [z^n] \frac{1}{1+z\vare_\square(\vec{k})}= (-1)^n\int_{\vec{k}} \vare_\square(\vec{k})^{n},
\end{align}  
where
\begin{align}
 \nonumber \vare_\square(\vec{k}) &= -2[\cos k_1+\cos k_2] \\
 \label{uv7} &= -\Bigl(u+\frac{1}{u}\Bigr)-\Bigl(v+\frac{1}{v}\Bigr).
\end{align}
Since the integral over an odd number of cosines is zero, we have $S_n=0$ for odd $n$ and only need to consider $S_{2n}$. This is, of course, because the Square lattice is bipartite. Thus
\begin{align}
 \nonumber S_{2n}^{(\square)} &= \mathscr{C}\Biggl[\left(\Bigl(u+\frac{1}{u}\Bigr)+\Bigl(v+\frac{1}{v}\Bigr)\right)^{2n} \Biggr]\\
  \label{uv8} &=\mathscr{C}\Biggl[\sum_{p=0}^{2n} \binom{2n}{p}\Bigl(u+\frac{1}{u}\Bigr)^p \Bigl(v+\frac{1}{v}\Bigr)^{2n-p}\Biggr].
\end{align}
In this sum of products, the constant parts of both the $u$- and $v$-dependent terms need to be nonzero independently. This can only occur if the exponents $p$ and $2n-p$ are even, i.e. if $p=2\ell$ is even, in which case we can use
\begin{align}
 \label{uv9} \mathscr{C}\Biggl[\Bigl(u+\frac{1}{u}\Bigr)^{2\ell}\Biggr] = \binom{2\ell}{\ell},
\end{align}
where the right-hand side is the central binomial coefficient. We then arrive at
\begin{equation}
\label{uv10} 
\begin{aligned}
 S_{2n}^{(\square)} &=\mathscr{C}\Biggl[\sum_{\ell=0}^{n} \binom{2n}{2\ell}\Bigl(u+\frac{1}{u}\Bigr)^{2\ell} \Bigl(v+\frac{1}{v}\Bigr)^{2n-2\ell}\Biggr]\\
 &= \sum_{\ell=0}^{n}  \binom{2n}{2\ell} \binom{2\ell}{\ell}\binom{2n-2\ell}{n-\ell}\\
 &=\sum_{\ell=0}^{n}\frac{(2n)!(2\ell)!(2n-2\ell)!}{(2\ell)!(2n-2\ell)!\ell!^2(n-\ell)!^2}\\
 &= \sum_{\ell=0}^{n}\frac{(2n)!}{\ell!^2(n-\ell)!^2}= \sum_{\ell=0}^{n}\frac{(2n)!}{n!^2}\frac{n!^2}{\ell!^2(n-\ell)!^2}\\
&= \binom{2n}{n} \sum_{\ell=0}^n\binom{n}{\ell}^2 = \binom{2n}{n}^2.
\end{aligned}
\end{equation}
Here we used the identity
\begin{align}
\label{uv11} \sum_{\ell=0}^n \binom{n}{\ell}^2 = \binom{2n}{n}.
\end{align}

\subsubsection{Honeycomb lattice}
For the Honeycomb lattice we have
\begin{align}
 \label{uv12} S_{2n}^{(\rm H)} &= \frac{1}{2}\int_{\vec{k}}\Bigl(\vare_+(\vec{k})^{2n}+\vare_-(\vec{k})^{2n}\Bigr) \\
 \label{uv14} &= \int_{\vec{k}}\vare_{\rm H}(\vec{k})^{2n}= \mathscr{C}[\vare_{\rm H}(u,v)^{2n}],
\end{align}
where we have ignored $S_n=0$ for odd $n$ since the lattice is bipartite. We have
\begin{align}
 \vare_{\rm H}(\vec{k}) &= |1+e^{\rmi k_1}+e^{\rmi k_2}|
\end{align}
and, consequently,
\begin{align}
 \label{uv20} \vare_{\rm H}(u,v)^2 = (1+u+v)\Bigl(1+\frac{1}{u}+\frac{1}{v}\Bigr).
\end{align}
We have
\begin{align}
 \nonumber \vare_{\rm H}(u,v)^{2n} ={}& \sum_{\ell_1=0}^n\sum_{\ell_2=0}^n \binom{n}{\ell_1}\binom{n}{\ell_2} (1+u)^{\ell_1} \\
 \label{uv15}&\times \Bigl(1+\frac{1}{u}\Bigr)^{\ell_2}v^{n-\ell_1}\Bigl(\frac{1}{v}\Bigr)^{n-\ell_2}.
\end{align}
In the following, we write ``$+\text{nc}$" for non-constant terms that drop out when taking the constant part. The $v$-terms can only contribute for $\ell_1=\ell_2=\ell$, hence
\begin{equation}
\label{uv16} 
\begin{aligned}
\vare_{\rm H}(u,v)^{2n}  &= \sum_{\ell=0}^n\binom{n}{\ell}^2 (1+u)^{\ell}\Bigl(1+\frac{1}{u}\Bigr)^{\ell} +\text{nc}.\\
 &= \sum_{\ell=0}^n\binom{n}{\ell}^2 \frac{1}{u^\ell} (1+u)^{2\ell} +\text{nc}\\
&= \sum_{\ell=0}^n\binom{n}{\ell}^2 \sum_{k=0}^{2\ell} \binom{2\ell}{k} \frac{1}{u^\ell}u^k +\text{nc}\\
&\stackrel{k=\ell}{=} \sum_{\ell=0}^n\binom{n}{\ell}^2 \binom{2\ell}{\ell}+\text{nc}.
\end{aligned}
\end{equation}
We used $(1+\frac{1}{u})=\frac{1}{u}(1+u)$. Thus, taking the constant part, we arrive at
\begin{align}
 \label{uv17}  S_{2n}^{(\rm H)} =  \sum_{\ell=0}^n \binom{2\ell}{\ell}\binom{n}{\ell}^2.
\end{align}

\subsubsection{Kagome lattice}

For the Kagome lattice we have
\begin{align}
 \label{uv18} \vare_{1,2}^{(\rm K)}(\vec{k}) &= 1 \pm \vare_{\rm H}(\vec{k}),\\
 \label{uv19} \vare_3^{(\rm K)}(\vec{k}) &= -2.
\end{align}
This yields
\begin{align}
 \nonumber S_n &=\frac{1}{3}\int_{\vec{k}}\Bigl( (1+\vare_{\rm H})^n + (1-\vare_{\rm H})^n +(-2)^n\Bigr)\\
 \nonumber &= \frac{1}{3}\mathscr{C}\Biggl[(1+\vare_{\rm H})^n+(1-\vare_{\rm H})^n+(-2)^n\Biggr]\\
 \nonumber &=\frac{1}{3}\sum_{\ell=0}^n \binom{n}{\ell}(1+(-1)^\ell) \mathscr{C}[\vare_{\rm H}^\ell]+\frac{1}{3}(-2)^n\\
  \nonumber &=\frac{1}{3}2\sum_{\ell=0}^{[n/2]} \binom{n}{2\ell}\mathscr{C}[\vare_{\rm H}^{2\ell}]+\frac{1}{3}(-2)^n\\
 \label{uv21}  &= \frac{2}{3}\sum_{\ell=0}^{[n/2]} \binom{n}{2\ell}S^{(\rm H)}_{2\ell}+\frac{1}{3}(-2)^n,
\end{align}
where we used Eq. (\ref{uv14}) in the last line.

\subsubsection{Triangular lattice}

For the Triangular lattice we have
\begin{align}
 \label{uv23}\vare_\Delta(\vec{k})&=-\Bigl(\vare_{\rm H}(\vec{k})^{2}-3\Bigr),
 \end{align}
thus
\begin{equation}
\label{uv24} 
\begin{aligned}
 S_n^{(\Delta)} &= (-1)^n\int_{\vec{k}}\vare_\Delta(\vec{k})^n\\
  &= \mathscr{C}\Biggl[\Bigl([\vare_{\rm H}(u,v)]^{2}-3\Bigr)^n\Biggr]\\
  &=\sum_{\ell=0}^n \binom{n}{\ell} (-3)^{n-\ell} \mathscr{C}[\vare_{\rm H}(u,v)^{2\ell}]\\
  &=\sum_{\ell=0}^n \binom{n}{\ell} (-3)^{n-\ell} S_{2\ell}^{(\rm H)}.
 \end{aligned}
 \end{equation}
We used again Eq. (\ref{uv14}). An alternative formula can be obtained from writing
\begin{align}
 \label{uv25} \vare_\Delta(\vec{k})=-\Bigl[-2 + 8\cos\Bigl(\frac{k_1}{2}\Bigr)\cos\Bigl(\frac{k_2}{2}\Bigr)\cos\Bigl(\frac{k_1-k_2}{2}\Bigr)\Bigr],
\end{align}
which yields
\begin{align}
 \nonumber S_n^{(\Delta)} ={}& \int_{\vec{k}}  \Bigl[-2 + 8\cos\Bigl(\frac{k_1}{2}\Bigr)\cos\Bigl(\frac{k_2}{2}\Bigr)\cos\Bigl(\frac{k_1-k_2}{2}\Bigr)\Bigr]^n\\
 \nonumber ={}&\sum_{\ell=0}^n \binom{n}{\ell} (-2)^{n-\ell} 8^{\ell}\\
 \nonumber &\times\int_0^{\pi}\frac{\mbox{d}s}{\pi} \int_0^{\pi} \frac{\mbox{d}t}{\pi}(\cos s)^\ell(\cos t)^\ell[\cos(s-t)]^\ell\\
\label{uv26}={}&\sum_{\ell=0}^n \binom{n}{\ell} (-2)^{n-\ell}\sum_{j=0}^\ell \binom{\ell}{j}^3,
\end{align}
where we used an identity for trigonometric integrals.

\subsubsection{Trellis lattice}

For the Trellis lattice we have
\begin{align}
 \label{uv27} \vare_{1,2}^{(\rm Tr)}(\vec{k}) = -2\cos k_1 \pm \vare_{\rm H}(\vec{k}),
\end{align}
thus
\begin{align}
 \nonumber S_n^{(\rm Tr)} &= \frac{(-1)^n}{2}\int_{\vec{k}}\Bigl( \vare_1^n + \vare_2^n\Bigr)\\
 \nonumber &= \frac{1}{2}\int_{\vec{k}}\Biggl[ \Bigl(u+\frac{1}{u}+\vare_{\rm H}\Bigr)^n+\Bigl(u+\frac{1}{u}-\vare_{\rm H}\Bigr)^n\Biggr]\\
 \label{uv28} &= \sum_{\ell=0}^{[n/2]}\binom{n}{2\ell} \mathscr{C}\Biggl[\Bigl(u+\frac{1}{u}\Bigr)^{n-2\ell} \vare_{\rm H}(u,v)^{2\ell}\Biggr].
\end{align}
Here $ \vare_{\rm H}^{2\ell}$ is given by Eq. (\ref{uv15}) with $\ell_1=\ell_2=m$ due to taking the constant part with respect to $v$, i.e.
\begin{align}
 \label{uv29} \vare_{\rm H}^{2\ell} &= \sum_{m=0}^\ell  \binom{\ell}{m}^2 (1+u)^{m}\Bigl(1+\frac{1}{u}\Bigr)^{m}+\text{nc. in }v.
\end{align}
We then arrive at
\begin{align}
 \nonumber S_n^{(\rm Tr)} ={}& \sum_{\ell=0}^{[n/2]} \sum_{m=0}^\ell\binom{n}{2\ell}   \binom{\ell}{m}^2 \\
 &\times \mathscr{C}\Biggl[\Bigl(u+\frac{1}{u}\Bigr)^{n-2\ell} (1+u)^{m}\Bigl(1+\frac{1}{u}\Bigr)^{m}\Biggr]\\
 \nonumber ={}& \sum_{\ell=0}^{[n/2]} \sum_{m=0}^\ell\binom{n}{2\ell}  \sum_{k=0}^{n-2\ell} \sum_{s=0}^{2m}\binom{\ell}{m}^2\ \binom{n-2\ell}{k}\\
 &\times \binom{2m}{s} \mathscr{C}\Bigl[u^{n-2\ell-2k-m+s}\Bigr].
\end{align}
To take the constant part, we set  $s=2\ell+2k+m-n$ and obtain the closed expression
\begin{align}
  \nonumber S_n^{(\rm Tr)} ={}& \sum_{\ell=0}^{[n/2]}  \sum_{m=0}^\ell  \sum_{k=0}^{n-2\ell} \binom{n}{2\ell} \binom{\ell}{m}^2\\
   \label{uv32}&\times  \binom{n-2\ell}{k}\binom{2m}{2\ell+2k+m-n}
\end{align}
for the Trellis lattice.

\subsubsection{Star lattice}

Consider the Star lattice with Bloch energies $\vare_1(\vec{k}),\dots,\vare_6(\vec{k})$ given by Eq. (\ref{bloch22b})-(\ref{bloch22e}). The flat band at zero energy only contributes for $n=0$ and we have
\begin{align}
 \label{uv33} S_n^{(\rm St)} = \frac{(-1)^n}{6}\Bigl[ \delta_{n0} + 2^n+\int_{\vec{k}}(\vare_3^n+\vare_4^n+\vare_5^n+\vare_6^n)\Bigr].
\end{align}
We shift $k_2\to -k_2$ so that 
\begin{align}
 \vare_{3,\dots,6}(\vec{k})=-\frac{1}{2}(1\pm\sqrt{13\pm 4\vare_{\rm H}(\vec{k})}).
\end{align}
Then, applying the binomial theorem twice, we find
\begin{align}
 \nonumber S_n^{(\rm St)} ={}&\frac{(-1)^n}{6}\Biggl[ \delta_{n0}+2^n +4\Bigl(-\frac{1}{2}\Bigr)^n\sum_{\ell=0}^{[n/2]} \sum_{m=0}^{[\ell/2]} \binom{n}{2\ell}\binom{\ell}{2m}\\
 \label{uv34} &\times 13^{\ell-2m}16^m\int_{\vec{k}}\vare_{\rm H}(\vec{k})^{2m}\Biggr].
\end{align}
Employing Eq. (\ref{uv20}) and slightly rearranging the coefficients we arrive at the closed expression
\begin{align}
 \nonumber S_n^{(\rm St)} ={}&\frac{1}{6}\Biggl[ \delta_{n0} + (-2)^{n}\\
 \label{uv35} &+\sum_{\ell=0}^{[n/2]} \sum_{m=0}^{[\ell/2]} \binom{n}{2\ell}\binom{\ell}{2m}13^{\ell-2m}2^{4m-n+2}S_{2m}^{(\rm H)}\Biggr]
\end{align}
for the Star lattice.

\subsubsection{CaVO lattice}

For the CaVO lattice, the Bloch energy bands cannot be determined in closed form. The generating function $\mathcal{G}^{(2)}_{\rm Ca}(z)$, however, can be given in closed form. For this we start from
\begin{align}
 \nonumber &\mathcal{G}_{\rm Ca}^{(2)}(z,\vec{k}) = \Bigl(\frac{1}{\mathbb{1}-zA^{(\rm Ca)}(\vec{k})}\Bigr)_{22} \\
 \label{uv37} &= \frac{1-3z^2-2z^3\cos k_1}{1-6z^2+z^4-4z^3(\cos k_1+\cos k_2)-4z^4\cos k_1\cos k_2}
\end{align}
to compute $\mathcal{G}_{\rm Ca}(z)=\int_{\vec{k}}\mathcal{G}_{\rm Ca}^{(2)}(z,\vec{k})$. Choosing the $22$-component allows us to perform the $k_2$ integration by utilizing
\begin{align}
 \label{uv39} \int_{0}^{2\pi} \frac{\mbox{d}k_2}{2\pi}\frac{C}{A-B\cos k_2} = \frac{C}{\sqrt{A^2-B^2}}
\end{align}
for $A^2>B^2$. In our case, we have
\begin{align}
 A&= 1-6z^2+z^4-4z^3\cos k_1,\\
 B&= 4z^3(1+z\cos k_1),\\
 C&=1-3z^2-2z^3\cos k_1,
\end{align}
which satisfy $A>B>0$ for sufficiently small $z>0$. We arrive at
\begin{align}
\label{uv40} \mathcal{G}_{\rm Ca}(z) &= \int_{0}^{2\pi} \frac{\mbox{d}k_1}{2\pi} \frac{1-3z^2-2z^3\cos k_1}{\sqrt{A^2-B^2}}.
\end{align}
This formula works very well in practice and can be used to compute the numbers $S_{2n}^{(\rm Ca)}$ numerically. To give a closed analytic expression, introduce 
\begin{align}
 \label{uv41} C_\pm = \frac{1-5z^2\pm z(1-z^2)}{4z^3}
\end{align}
so that
\begin{widetext}
\begin{align}
 \nonumber &\mathcal{G}_{\rm Ca}(z)= \int_{0}^{2\pi} \frac{\mbox{d}k_1}{2\pi} \frac{1-3z^2-2z^3\cos k_1}{\sqrt{16z^6(1-z^2)(C_+-\cos k_1)(C_--\cos k_1)}}\\
 \nonumber &=\int_{0}^{2\pi} \frac{\mbox{d}k_1}{2\pi} \frac{1-3z^2-2z^3\cos k_1}{\sqrt{16z^6(1-z^2)C_+C_-}}\sum_{n_1\geq0}\sum_{n_2\geq 0} \binom{-1/2}{n_1}\binom{-1/2}{n_2}\frac{(-\cos k_1)^{n_1+n_2}}{C_+^{n_1}C_-^{n_2}}\\
 \nonumber &=\frac{1}{\sqrt{16z^6(1-z^2)C_+C_-}}\sum_{\ell\geq 0} \sum_{n_1\geq0} \binom{2\ell}{\ell}\binom{-1/2}{n_1}\frac{1}{4^\ell}\binom{2\ell}{\ell}\Biggl(\frac{1-3z^2}{C_+^{n_1}C_-^{2\ell-n_1}}\binom{-1/2}{2\ell-n_1}+\frac{2z^3}{C_+^{n_1}C_-^{2\ell-1-n_1}}\binom{-1/2}{2\ell-n_1-1}\Biggr)\\
 \label{uv42} &=\frac{1}{\sqrt{16z^6(1-z^2)C_+C_-}}\sum_{\ell\geq 0} \sum_{n_1\geq0} \binom{2\ell}{\ell}\binom{2n_1}{n_1}\frac{1}{(8C_-)^{2\ell}}\Bigl(\frac{C_-}{C_+}\Bigr)^{n_1}\Biggl[(1-3z^2)\binom{2(2\ell-n_1)}{2\ell-n_1}-8z^3C_-\binom{2(2\ell-1-n_1)}{2\ell-1-n_1}\Biggr].
\end{align}
\end{widetext}
Here we used the identities
\begin{align}
 \label{uv43} &\frac{1}{\sqrt{1-x}} = \sum_{n\geq 0} \binom{-1/2}{n} (-x)^n,\\
 \label{uv44} &\binom{-1/2}{k} = \binom{2k}{k}\frac{(-1)^{k}}{2^{2k}},\\
 \label{uv45} &\int_0^{2\pi} \frac{\mbox{d}k_1}{2\pi} (\cos k_1)^n = \sum_{\ell\geq 0}  \frac{1}{4^\ell} \binom{2\ell}{\ell}\delta_{n,2\ell}.
\end{align}
Expanding the expression in Eq. (\ref{uv42}) we find, indeed,
\begin{align}
 \nonumber \mathcal{G}_{\rm Ca}(z)={}& 1+3z^2+ 17z^4+ 111z^6+ 773z^8\\
 \label{uv46} &+ 5623z^{10}+ 42269z^{12}+\dots
\end{align}

\subsubsection{SrCuBO Lattice}

Explicit expressions for the generating functions $\mathcal{G}(z)$ can be obtained from the momentum-dependent expressions $\mathcal{G}(z,\vec{k})$ by expanding the integrand to obtain a polynomial in $c_1=\cos k_1$ and $c_2=\cos k_2$ and then applying Eq. (\ref{uv45}). The resulting formulas are lengthy, but can be used in practice to compute the $S_n$. We demonstrate the procedure here for the SrCuBO lattice.

We compute the generating function $\mathcal{G}_{\rm Sr}(z)$. For this we start from
\begin{widetext}
\begin{align}
 \label{uv48} \mathcal{G}_{\rm Sr}(z,\vec{k}) = \frac{1-5z^2-2z^3-4z^2(1+z)c_1-2z^3c_1c_2}{Q_{\rm Sr}(z)-4z^2(1+z)^2(c_1+c_2)+ 4z^4(c_1^2+c_2^2)-8z^3(1+2z)c_1c_2}
\end{align}
with $Q_{\rm Sr}(z) = 1-10z^2-8z^3+z^4$. Compared to the expression in Tab. \ref{TabGF}, we utilized the fact that the denominator is symmetric with respect to exchanging $k_1\leftrightarrow k_2$ and replaced $c_1+c_2\to 2c_1$ in the numerator. Using the geometric series and binomial theorem, we expand 
\begin{align}
 \nonumber \frac{1}{Q_{\rm Sr}(z)-4z^2(1+z)^2(c_1+c_2)+\dots} ={}&\sum_{n\geq 0}\frac{\Bigl(4z^2(1+z)^2(c_1+c_2)-4z^4(c_1^2+c_2^2)+8z^3(1+2z)c_1c_2\Bigr)^{n}}{Q_{\rm Sr}(z)^{n+1}}\\
 \nonumber  ={}&\sum_{n\geq 0} \sum_{m=0}^{n}\sum_{s_1=0}^m\sum_{s_2=0}^{n-m}\sum_{s_3=0}^{s_2}\binom{n}{m}\binom{m}{s_1}\binom{n-m}{s_2}\binom{s_2}{s_3}\frac{1}{Q_{\rm Sr}(z)^{n+1}}(-1)^{n-s_1-s_2}2^{2n+s_3}\\
 \label{uv49} &\ \times z^{4n-2s_1-2s_2+s_3}(1+z)^{2(s_1+s_2-s_3)}(1+2z)^{s_3} c_1^{2m-s_1+s_3} c_2^{2n-2m-s_2}.
\end{align}
Applying Eq. (\ref{uv45}) for both the $k_1$- and $k_2$-integrations, we arrive at
\begin{align}
 \nonumber \mathcal{G}_{\rm Sr}(z) &= \sum_{\ell_1\geq 0}\sum_{\ell_2\geq0}\sum_{n\geq 0} \sum_{m=0}^{n}\sum_{s_1=0}^m\binom{2\ell_1}{\ell_1}\binom{2\ell_2}{\ell_2}\binom{n}{m}\binom{m}{s_1} \frac{1}{Q_{\rm Sr}(z)^{n+1}}(-1)^{n-s_1}2^{2n-2m+s_1-2\ell_2}z^{2m+2\ell_1+4\ell_2-s_1} \\
 \nonumber &\times (1+z)^{4(n-\ell_1-\ell_2)}(1+2z)^{2\ell_1-2m+s_1} \Biggl[[1-5z^2-2z^3]\binom{n-m}{2n-2m-2\ell_2}\binom{2n-2m-2\ell_2}{2\ell_1-2m+s_1}\\
 \label{uv50}  &-\frac{2z(1+z)^3}{1+2z}\binom{n-m}{2n-2m-2\ell_2}\binom{2n-2m-2\ell_2}{2\ell_1-1-2m+s_1}+\frac{(1+z)^4}{1+2z}\binom{n-m}{2n-2m-2\ell_2+1}\binom{2n-2m-2\ell_2+1}{2\ell_1-1-2m+s_1}\Biggr].
\end{align}
\end{widetext}
We confirm that we have
\begin{align}
 \nonumber \mathcal{G}_{\rm Sr}(z) ={}& 1 + 5 z^2 + 6 z^3 + 53 z^4 + 140 z^5 + 797 z^6 \\
 \label{uv51} &+ 2898 z^7 + 14317 z^8 + 59784 z^9  +\dots,
\end{align}
which reproduces the right returning walks numbers from Tab. \ref{TabWalks}.

\subsubsection{Ruby, Maple-Leaf, and SHD lattice}

We now consider the Ruby (R), Maple-Leaf (ML), and SHD lattices, whose Bravais lattice is the Triangular lattice. We define $\vec{k}'=(2k_1-k_2,k_1-2k_2)$ and
\begin{align}
 \nonumber (\vare_{\rm H}')^2={}& \vare_{\rm H}(\vec{k}')^2 \\
 \nonumber ={}& 3+2\cos(2k_1-k_2)\\
 \label{uv52} &+2\cos(k_1-2k_2)+2\cos(k_1+k_2).
\end{align}
Using the same manipulations as in Eqs. (\ref{uv15}) and (\ref{uv16}), we confirm that the moments of $\vare_{\rm H}^2$ and $(\vare_{\rm H}')^2$ agree, i.e.
\begin{align}
 \nonumber \int_{\vec{k}} (\vare_{\rm H}')^{2n} &= \mathscr{C}\Bigl[\Bigl(1+\frac{u^2}{v}+\frac{u}{v^2}\Bigr)^n\Bigl(1+\frac{v}{u^2}+\frac{v^2}{u}\Bigr)^n\Biggr]\\
 \label{uv53} &= \int_{\vec{k}} \vare_{\rm H}^{2n} = S_{2n}^{(\rm H)}.
\end{align}
We further define
\begin{align}
 \label{uv54} I_{n_1,n_2}:=\int_{\vec{k}} \vare_{\rm H}^{2n_1}(\vare_{\rm H}')^{2n_2}.
\end{align}
The generating functions for the R, ML, and SHD lattices can be written as
\begin{align}
 \label{uv55} \mathcal{G}(z,\vec{k}) = \frac{P_0(z)-P_1(z)\vare_{\rm H}^2-P_2(z)(\vare_{\rm H}')^2}{Q_0(z)-Q_1(z)\vare_{\rm H}^2-Q_2(z)(\vare_{\rm H}')^2-Q_3(z)\vare_{\rm H}^4},
\end{align}
with
\begin{equation}\label{uv56} 
\begin{aligned}
  P_0^{(\rm R)} &= 1-8z^2+z^3+12z^4,\ P_1^{(\rm R)} = \frac{1}{3}(3+4z)z^3,\\
  P_2^{(\rm R)} &= 0,\ Q_0^{(\rm R)} = 1 -12z^2 +2z^3+36z^4 +9z^6,\\
 Q_1^{(\rm R)} &= 2z^3(1+2z+3z^3),\ Q_2^{(\rm R)} = 4z^6,\ Q_3^{(\rm R)} =-z^6,
\end{aligned}
\end{equation}
and
\begin{equation}\label{uv57} 
\begin{aligned}
  P_0^{(\rm ML)} &= 1-10z^2-2z^3+21z^4+8z^5,\\
  P_1^{(\rm ML)} &= z^3(2+4z+z^2),\ P_2^{(\rm ML)} = \frac{1}{3}z^5,\\
  Q_0^{(\rm ML)} &= 1-15z^2-4z^3+63z^4+48z^5+4z^6,\\
 Q_1^{(\rm ML)} &= 2z^3(2+6z+3z^2+2z^3),\\
 Q_2^{(\rm ML)} &= 2z^5(1+3z),\ Q_3^{(\rm ML)} = -z^6,
\end{aligned}
\end{equation}
and
\begin{equation}\label{uv58} 
\begin{aligned}
 P_0^{(\rm SHD)} &= 1-15z^2+74z^4-131z^6+59z^8,\\
 P_1^{(\rm SHD)} &= z^6(3-10z^2+5z^4),\ P_2^{(\rm SHD)} = \frac{2}{3}z^{10},\\
 Q_0^{(\rm SHD)} &= 1 - 18z^2+111z^4 -262z^6+177z^8,\\
 Q_1^{(\rm SHD)} &= 6z^6(1-5z^2+5z^4),\\
 Q_2^{(\rm SHD)} &= 4z^{10},\ Q_3^{(\rm SHD)} = -z^{12}.
\end{aligned}
\end{equation}
Expanding the geometric series and using the binomial theorem, we can expand the denominator in powers of $\vare_{\rm H}^2$ and $(\vare_{\rm H}')^2$ according to
\begin{align}
\nonumber  &\frac{1}{Q_0-Q_1\vare_{\rm H}^2-Q_2(\vare_{\rm H}')^2-Q_3\vare_{\rm H}^4}\\
 \label{uv59} &=\sum_{\ell \geq 0} \frac{1}{Q_0^{\ell+1}}\Bigl(Q_1\vare_{\rm H}^2+Q_2(\vare_{\rm H}')^2+Q_3\vare_{\rm H}^4\Bigr)^\ell\\
 \nonumber &=\sum_{\ell \geq 0}\sum_{m_1=0}^\ell\sum_{m_2=0}^{m_1}\binom{\ell}{m_1}\binom{m_1}{m_2} \frac{Q_1^{m_2}Q_2^{\ell-{m_1}}Q_3^{m_1-m_2}}{Q_0^{\ell+1}}\\
 \label{uv60} &\times \vare_{\rm H}^{2(2m_1-m_2)}(\vare_{\rm H}')^{2(\ell-m_1)}.
\end{align}
Using $\mathcal{G}(z)=\int_{\vec{k}}\mathcal{G}(z,\vec{k})$  we arrive at
\begin{align}
 \nonumber \mathcal{G}(z) ={}&\sum_{\ell \geq 0}\sum_{m_1=0}^\ell\sum_{m_2=0}^{m_1}\binom{\ell}{m_1}\binom{m_1}{m_2} \frac{Q_1^{m_2}Q_2^{\ell-{m_1}}Q_3^{m_1-m_2}}{Q_0^{\ell+1}} \\
 \nonumber &\times \Bigl(P_0I_{2m_1-m_2,\ell-m_1}-P_1I_{2m_1-m_2+1,\ell-m_1}\\
 \label{uv61}&-P_2I_{2m_1-m_2,\ell-m_1+1}\Bigr),
\end{align}
with $I_{n_1,n_2}$ given in Eq. (\ref{uv54}). An explicit expression for $I_{n_1,n_2}$ is
\begin{equation}\label{uv62}
\begin{aligned}
 I_{n_1,n_2} ={}& \sum_{s_1=0}^{n_1}\sum_{\ell_1=0}^{n_2}\sum_{\ell_2=0}^{n_2} \sum_{m_1=0}^{\ell_1}\sum_{m_2=0}^{\ell_2}\binom{n_1}{s_1}\binom{n_2}{\ell_1}\binom{n_2}{\ell_2}\\
 &\times \binom{\ell_1}{m_1}\binom{n_1}{s_1+(m_1-m_2) -2(\ell_1-\ell_2)}\\
 &\times \binom{\ell_2}{m_2}\binom{2s_1+(m_1-m_2) -2(\ell_1-\ell_2)}{s_1-(m_1-m_2) -(\ell_1-\ell_2)},
\end{aligned}
\end{equation}
which can be obtained from writing the integrand in terms of $u$ and $v$ as in Eqs. (\ref{uv17}) and (\ref{uv56}) and subsequently taking the constant part.
With this, we confirm that
\begin{align}
 \nonumber \mathcal{G}_{\rm R}(z) ={}& 1 + 4 z^2 + 2 z^3 + 32 z^4 + 40 z^5 + 314 z^6 \\
 \label{uv63} &+ 616 z^7 + 3488 z^8 +  8864 z^9 +\dots\ \\
 \nonumber \mathcal{G}_{\rm ML}(z) ={}&1 + 5 z^2 + 8 z^3 + 57 z^4 + 180 z^5 + 907 z^6 \\
 \label{uv64} &+ 3612 z^7 +  16777 z^8 + 72896 z^9 +\dots\\
 \nonumber \mathcal{G}_{\rm SHD}(z) ={}& 1 + 3 z^2 + 17 z^4 + 113 z^6 + 809 z^8 \\
 \label{uv65} &+ 6063 z^{10} + 46835 z^{12} +\dots
\end{align}

\subsection{DOS from returning walks}\label{SecReturnDOS}

In this section, we apply the results on returning walks to determine the DOS $D(E)$ for seven Archimedean lattices. The DOS is defined for finite graphs with $N$ sites through the eigenvalues $E_i$ of the Hamiltonian matrix $H=-A$ by
\begin{align}
 \label{bloch26} D(E) = \frac{1}{N} \sum_{i=1}^N\delta(E-E_i).
\end{align}
For infinite systems with $N\to \infty$, this reproduces Eq. (\ref{bloch27}), whereas for finite graphs a discretization of momenta is required. 

To see appearance of the returning walk numbers, use Eq. (\ref{avgTr}) and the presentation $\delta(x-a)= \lim_{\eta\to 0}\frac{1}{\pi}\mathfrak{I}(\frac{1}{(x-a)-\rmi\eta })$ to obtain
\begin{align}
  \nonumber  D(E)&= \frac{1}{N}\mbox{Tr}\ \delta (E-H)\\
  \nonumber   &=\lim_{\eta\to 0}\frac{1}{\pi}\mathfrak{I} \Biggl\{\frac{1}{N}\mbox{Tr}\left(\frac{1}{(E-\rmi\eta)-H}\right)\Biggr\}\\
  \nonumber  &=\lim_{\eta\to 0}\frac{1}{\pi} \mathfrak{I} \sum_{n=0}^\infty \frac{\frac{1}{N}\mbox{Tr} (H^n)}{(E- \rmi\eta)^{n+1}}\\
  \label{def4}  &= \lim_{\eta\to 0} \frac{1}{\pi}\mathfrak{I}\sum_{n=0}^\infty (-1)^n\frac{S_{n}}{(E-\rmi \eta)^{n+1}}.
\end{align}
As an example, consider again the square lattice discussed in the introduction with $S^{(\square)}_{2n}=\binom{2n}{n}^2$. For $|E|<4$ we have
\begin{align}
 \nonumber D_\square(E) &= \lim_{\eta \to 0}\frac{1}{\pi}\mathfrak{I}\sum_{n=0}^\infty \frac{S^{(\square)}_{2n}}{(E-\rmi \eta)^{2n+1}}\\
\label{def5} &=\lim_{\eta \to 0} \frac{1}{\pi}\mathfrak{I}\Bigl[\frac{2}{\pi(E-\rmi \eta)} K \Bigl(\frac{16}{(E-\rmi \eta)^2}\Bigr)\Bigr],
 \end{align}
with complete elliptic integral of the first kind $K(k^2)$ defined in Eq. (\ref{bloch29}). Here we used
\begin{align}
 \label{def6} K(k^2) = \frac{\pi}{2}\sum_{n\geq 0}\binom{2n}{n}^2 \Bigl(\frac{k^2}{16}\Bigr)^n.
\end{align}
Now apply
\begin{align}
 \label{def7} \lim_{\eta \to 0} \mathfrak{I}\Bigl[\frac{1}{E-\rmi \eta} K \Bigl(\frac{16}{(E-\rmi \eta)^2}\Bigr)\Bigr] = \frac{1}{4}K\Bigl(1-\frac{E^2}{16}\Bigr)
\end{align}
 for $|E|<4$ to arrive at Eq. (\ref{intro2}).

A closed formula for $D(E)$ in terms of the Bloch generating function $\mathcal{G}(z,\vec{k})$ defined in Eqs. (\ref{gf3more}) and (\ref{gf3}) can be found by rearranging the sum in Eq. (\ref{def4}) as
\begin{equation}
    D(E)=\lim_{\eta\to 0}\frac{1}{\pi}\mathfrak{I}\left[\frac{1}{E-{\rm i} \eta}\sum_{n=0}^\infty S_{n}\left(\frac{-1}{E- {\rm i} \eta}\right)^n\right].
\end{equation}
We recognize that the remaining sum can be written as
\begin{align}
  \nonumber \sum_{n=0}^\infty S_{n}\left(\frac{-1}{E- {\rm i} \eta}\right)^n &=\sum_{n=0}^{\infty} \left(\int_{\bk}S_{n}(\bk)\right) \left(\frac{-1}{z}\right)^n\biggr|_{z=E- {\rm i\eta}}\\ \nonumber&=\int_{\bk}\overbrace{\left(\sum_{n=0}^{\infty} S_{n}(\bk) \left(\frac{-1}{z}\right)^n\right)}^{\mathcal{G}(-1/z,\bk)}\Biggr|_{z=E- {\rm i\eta}}\\ 
   &=\int_{\bk}\mathcal{G}\left(\frac{-1}{z},\bk\right)\biggr|_{z=E-{\rm i}\eta}.
\end{align}
Therefore, we arrive at
\begin{equation}
\label{DOSBOGF}
    D(E)=\lim_{\eta\to 0}\frac{1}{\pi}\mathfrak{{I}}\left[\int_{\bk}\frac{\mathcal{G}(\frac{-1}{z},\bk)}{z}\Biggr|_{z=E- {\rm i} \eta}\right].
\end{equation}
This connection is not entirely surprising, since the Bloch generating function $\mathcal{G}(z,\bk)$ may be understood as a rewriting of the momentum-space Green function of the Hamiltonian, derived from the resolvent operator $\hat{G}=(E\hat{\mathbb{1}}-\hat{H})^{-1}$.

As an application of Eq. (\ref{DOSBOGF}), we use the Bloch generating functions from Tab. \ref{TabGF} to compute the DOS of several Archimedean lattices. For the Triangular lattice, note that
\begin{equation}
    \frac{\mathcal{G}_\Delta(-1/z,\bk)}{z}= \frac{1}{z+2 \vare_{\Delta}(\bk)},
\end{equation}
hence
\begin{equation}
   D_\Delta(E) =  \frac{1}{\pi}\lim_{\eta\to 0} \mathfrak{I}\int_{\bk}\left\{\frac{1}{z+2\vare_{\Delta}(\bk)}\right\}.
\end{equation}
This yields $D_\Delta(E)$ as in Eq. (\ref{bloch32}) (see Ref. \cite{Kogan2021-fy} for the evaluation of the momentum integral). More generally, for any two polynomials $q(z),p(z)$ we have
\begin{equation}
\label{TrDOSId}
    \frac{1}{\pi}\lim_{\eta\to 0} \mathfrak{I}\int_{\bk}\left\{\frac{q(z)}{p(z)+2\vare_{\Delta}(\bk)}\right\}= |q(E)|D_\Delta(p(E)).
\end{equation}
For the Honeycomb lattice we have 
\begin{align}
    &\mathcal{G}_{\rm H}(z)= \frac{1}{1-z^2(3-\vare_\Delta(\bk))}\\
    \Rightarrow\  &\frac{\mathcal{G}_{\rm H}(-1/z,\bk)}{z}= \frac{-z}{(3-z^2)+2 \vare_\Delta(\bk)}.
\end{align} 
Using Eq. (\ref{TrDOSId})  with $p(z)=3-z^2$ and $q(z)=-z$, we arrive at
\begin{equation}
    D_{\rm H}(E)=|E|D_{\Delta}(3-E^2)
\end{equation}
as in Eq. (\ref{bloch34}). For the Kagome lattice, we have
\begin{align}
    &\mathcal{G}_{
    \rm K}(z,\bk)=\frac{1}{3}\Bigl(\frac{1}{1+2 z}-\frac{2 (1-z)}{1-2 z (1+z)+z^2 \varepsilon_{\Delta }(\vec{k})}\Bigr),\\
    \Rightarrow\ &\frac{\mathcal{G}_{\rm K}(\frac{-1}{z},\bk)}{z}= \frac{1}{3}\left(\frac{1}{z-2}+\frac{-2(1+z)}{(2-z(2+z)+2 \vare_\Delta(\bk)}\right).
\end{align}
Using $\frac{1}{\pi}\lim_{\eta\to 0}\int_{\bk} \frac{1}{E- \rmi \eta-2}=\int_{\bk}\delta(E-2)=\delta(E-2)$ and Eq. (\ref{TrDOSId}), we arrive at Eq. (\ref{bloch35}). For the Star lattice we have
\begin{align}
    \nonumber \mathcal{G}_{\rm St}(z,\bk)&=\frac{1}{3}\Bigl(\frac{1+z}{1+2 z}\\ &+\frac{(2-z)(1-z-3 z^2)}{(1-2 z^2) [1-2 z (1+z)]+z^4 \varepsilon _{\Delta }(\vec{k})}\Bigr).
\end{align} Then, with
    \begin{align}
        \nonumber\frac{\mathcal{G}_{\rm St}(-1/z,\bk)}{z}&=\frac{1}{3}\Biggl(\frac{1/2}{z-2}+\frac{1/2}{z}\\ &\phantom{ss}+ \frac{-(1+2z)(-3+z+z^2)}{(3-z^2)(z(2+z)-2)+2 \vare_\Delta(\bk)}\Biggr),
    \end{align}
we obtain Eq. (\ref{bloch36}). Explicit expressions of the functions $D_{\rm Ca}(E)$ [Eqs. (\ref{EDOS2}) and (\ref{EDOS2b})] and $D_{\rm Sr}(E)$ [Eqs. (\ref{EDOS3}) and (\ref{EDOS3b})] of the CaVO and SrCuBO lattices in terms of a one-dimensional integral are derived from the Bloch generating function in App. \ref{DOSExplicit}.

In App. \ref{AppCF}, we show how the returning walk numbers $S_n$ can be used to compute the coefficients $(a_n,b_n)$ in the continued fraction expansion of the DOS. The advantage of using the returning walk numbers is that it may avoid numerical rounding errors when compared to a straightforward computation of the Krylov basis through the Lanczos algorithm. For the Archimedean lattices, however, due to the presence of singularities in the function $D(E)$, the convergence properties of the continued fraction expansion are bad and the formulas quoted above describe the DOS better.

\subsection{Asymptotics of returning walk numbers}
In this section, we determine the asymptotics of the returning walks number $S_n=[{z^n}]\mathcal{G}(z)$ for the Archimedean lattices as $n\to \infty$. For the Archimedean lattices, each step provides $q$ choices and therefore $q^n$ is the total number of walks in $n$ steps. Hence
\begin{equation}
    \label{as0}
 p_n = \frac{S_n}{q^n}  
\end{equation}
is the return probability in $n$ steps of a random walker. We show in the following that $p_n \sim \alpha /n$ for large $n$, with a coefficient that depends on the particular lattice. The divergence of $\sum_{n\geq 0}p_n=\infty$ reflects the well-known fact a random walk in two dimensions always returns to its starting point.

Specifically, we compute in the following for the non-bipartite Archimedean lattices the coefficient
\begin{align}
 \alpha = \lim_{n\to \infty} \frac{n S_n}{q^n},
\end{align}
whereas for the bipartite lattices we compute
\begin{align}
 \alpha = \lim_{n\to\infty} \frac{2nS_{2n}}{q^{2n}}.
\end{align}
Given the exact expressions for $S_n$ or $\mathcal{G}(z)$ from Tab. \ref{TabWalks}, the coefficient $\alpha$ can be obtained numerically by computing the walk numbers for large $n$. Here we show that using the Bloch generating functions $\mathcal{G}(z,\vec{k})$ is an efficient alternative method to get an analytical answer. The expressions for all eleven lattices are summarized in Tab. \ref{Tabas}. These result have been confirmed numerically by comparing to the explicit expressions for $S_n$ for large $n$.

\renewcommand{\arraystretch}{2.5}
\begin{table}[t!]
\begin{tabular}{|c|c|}
\hline
 \multicolumn{2}{|c|}{\ Bipartite Archimedean lattices \ }\\
\hline
Lattice  & \ $\displaystyle S_{2n} \sim \alpha \frac{q^{2n}}{2n}$ \ \\
\hline
\hline
 Square  & $\dfrac{2}{\pi}\,\dfrac{4^{\,2n}}{2n}$\\ 
\hline
 \ Honeycomb \ & $\dfrac{3\sqrt{3}}{2\pi}\,\dfrac{3^{\,2n}}{2n}$	\\
\hline
 CaVO  & $\displaystyle \frac{3}{\pi}\,\frac{3^{\,2n}}{2n}$	\\
\hline
 SHD	  &	$\displaystyle \frac{2\sqrt{3}}{\pi\,}\frac{3^{\,2n}}{2n}$	\\
\hline
\hline
\multicolumn{2}{|c|}{\ Non-bipartite Archimedean lattices \ } \\
 \hline
 Lattice &  $\displaystyle S_n \sim \alpha \frac{q^{n}}{n}$\\
\hline
\hline
Triangular  & $\displaystyle \frac{\sqrt{3}}{2\pi}\,\frac{6^{\,n}}{n}$	\\
\hline
Kagome  & $\displaystyle \frac{2}{\pi\sqrt{3}}\,\frac{4^{\,n}}{n}$ \	\\
\hline
Trellis & $\displaystyle \frac{\sqrt{15}}{4\pi}\,\frac{5^{\,n}}{n}$ \\
\hline
Star  & $\displaystyle \frac{5\sqrt{3}}{4\pi}\,\frac{3^{n}}{n}$ \\
\hline
 \ SrCuBO & $\displaystyle \dfrac{15}{16\pi}\,\dfrac{5^{\,n}}{n}$ \\
\hline
Ruby & $\displaystyle \dfrac{2}{\pi\sqrt{3}}\,\dfrac{4^{\,n}}{n}$ \\
\hline
Maple-Leaf & $\displaystyle \frac{5\sqrt{3}}{8\pi}\,\frac{5^{\,n}}{n}$\\
\hline
\end{tabular}
\caption{The asymptotic number of returning walks for each of the eleven Archimedean lattices are shown. For bipartite lattices, the number of walks that return have an even number of steps, reflected in the formulas by its dependence on $2n$. The asymptotic return probability in $m$ steps is therefore given by $p_m=\frac{S_m}{q^m}\sim \frac{\alpha}{m}$, where $m=2n$ for bipartite and $m=n$ for non-bipartite lattices.}
\label{Tabas}
\end{table}
\renewcommand{\arraystretch}{1}

The derivation of the asymptotics of the returning walk numbers for the Archimedean lattices is presented in Appendix App. \ref{asympApp}. The idea is to write the generating function as
\begin{align}
 \label{Asym45} \mathcal{G}(z) = \int_{\bk} \mathcal{G}(z,\bk) = \int_{\bk} \frac{P(z,\bk)}{Q(z,\bk)}
\end{align}
and observe that $\mathcal{G}(z)$ is an analytical function for real $z$ such that $|z|<q^{-1}$, with coordination number $q$. The function diverges, however, at the critical value $z_{\rm c}=q^{-1}$. By expanding the denominator around its zero located at $(z_{\rm c},\bk_{\rm c})$ the asymptotics can be determined. (For non-bipartite lattices, $\bk_{\rm c}=\vec{0}$, but bipartite lattices may have several critical points $\bk_{\rm c}$.) Introducing the $2\times 2$ Hessian matrix
\begin{align}
 Q^{(2)}(z,\bk)_{ij} = \partial_{k_i}\partial_{k_j}Q(z,\bk),
\end{align}
we find for the seven non-bipartite Archimedean lattices that
\begin{align}
 \label{asym1} \alpha = \frac{P(z_{\rm c},\vec{0})}{2\pi\sqrt{\mbox{det} Q^{(2)}(z_{\rm c},\vec{0})}}. 
\end{align}
For the four bipartite Archimedean lattices, we discuss the proper expansion of $\mathcal{G}(z,\bk)$ in Eq. (\ref{Asym45}) in App. \ref{asympApp} and compute the associated coefficient $\alpha$. We find that $\alpha$ equals twice the value obtained from Eq. (\ref{asym1}) in these four cases. However, we were not able to derive the latter simple relation for general bipartite tessellations, and it may be specific to the four bipartite lattices considered in this work.

\section{Summary and Outlook}

In this work, we established a remarkable connection between the combinatorial problem of counting returning walks on Archimedean lattices and their physically relevant Bloch band theory. In particular, we have shown that the number of returning walks of length $n$, $S_n$, can be obtained from averaging the $n$-th power of the Bloch Hamiltonian matrix $H(\vec{k})=-A(\vec{k})$ over the Brillouin zone, see Eq. (\ref{intro7b}). For one, this is a numerically efficient way of computing $S_n$. On the other hand, it enabled us to derive explicit expressions for all eleven Archimedean lattices for either $S_n$ directly or their generating function $\mathcal{G}(z)=\sum_{n\geq 0}S_nz^n$, as summarized in Tab. \ref{TabWalks}. We further presented a general method to construct large flakes or clusters of Archimedean lattices with both open and periodic boundary conditions, which can be used to compute the numbers $S_n$ directly from powers of the adjacency matrix.

Our analysis recovered the known sequences of returning walk numbers $S_n$ for the Square, Honeycomb, Triangular, and Kagome lattices \cite{AllRoads,PhysRevE.105.014112}, corresponding to sequences A002894, A002893, A002898, A338672 in the Online Encyclopedia of Integers Sequences (OEIS) \cite{oeis}. On the other hand, the sequences for the remaining seven Archimedean lattices received less attention before and were not included in the OEIS at the time of writing. Furthermore, we applied the explicit expressions for the returning walk numbers and their generating functions to obtain closed formulas for the density of states of single-particle excitations on seven of the eleven Archimedean lattices, some of which do not appear to have been reported before (Star, CaVO, SrCuBO).

Archimedean lattices offer a rich enough structure to highlight the nontrivial problem of counting returning walks on lattices. However, the methods outlined in this work generalize to other (symmorphic) crystallographic tessellations in Euclidean space, both in two and higher dimensions, as long as they split into a unit cell and Bravais lattice. In particular, Eq. (\ref{gf0c}) for $S_n^{(i_{\rm u})}$ can be applied to compute the number of returning walks starting from any unit cell site $i_{\rm u}$. While there is no $i_{\rm u}$-dependence for Archimedean lattices, which are vertex-transitive and hence each lattice site has the same surroundings, the start site $i_{\rm u}$ plays a role for other lattices such as the Laves lattices \cite{Laves,Grunbaum2016-rs}. 

The master formula (\ref{mastereq1}) for constructing the adjacency matrices of finite graphs can be applied directly to other two-dimensional tessellations through specifying the unit cell $A_{\rm u}$ and connection matrices $\Gamma$. The construction can be generalized to periodic tiling of $\mathbb{R}^d$ with $d\geq 3$ by including additional tensor products involving the right (left) shift matrices $R_m\ (L_m)$. While the explicit construction of the flakes allowed us to compute the returning walk numbers here, their adjacency matrices are of relevance to other applications of tight-binding models in condensed matter physics.

Other combinatorial problems related to counting paths on lattices appear in physical and mathematical applications \cite{graphsBook} and lend themselves to be studied in the present framework. For instance, the problem of finding the number of returning walks on a lattice with prescribed area \cite{OUVRY2020115174,PhysRevE.105.014112,OuvryArea,PhysRevE.108.054104,Gan:2024rzb} is related to a charged particle hopping on a lattice in a uniform external magnetic field, i.e. the Hofstadter problem. Another example is counting the number of self-avoiding, non-backtracking returning walks on a planar lattice, which is related to finding the free energy of the classical two-dimensional Ising model on the lattice through the Feynman--Kac--Ward formula  \cite{Kac1952-ip,F0,vdovichenko1965calculation}. For an application to compute the critical temperature and free energy of two-dimensional Ising models on the Archimedean and Laves lattices, see Refs. \cite{codello2010exact,IsingArch}. 
Here the counting problem reduces to determining traces of powers of a Bloch-type Hamiltonian matrix of dimension $N_{\rm u}q\times N_{\rm u}q$, as will be presented elsewhere. 

In summary, we are convinced that the intersection of physics of lattice models, graph theory, and combinatorics offers many exciting avenues to be explored through the framework established in this work.

\section{Acknowledgments}
The authors thank Albion Arifi, Anffany Chen, Hanbang Lu, Canon Sun, and Connor Walsh for inspiring discussions. They acknowledge funding from the University of Alberta startup fund UOFAB Startup Boettcher and the Natural Sciences and Engineering Research Council of Canada (NSERC) Discovery Grants RGPIN-2021-02534 and DGECR2021-00043.

\begin{appendix}

\section{Explicit formulas for flakes and clusters of Archimedean lattices} \label{AppApq}
In this appendix, we present the explicit expressions for the adjacency matrices $A_{p,q}$ for flakes and clusters of all eleven Archimedean lattices. For this, we specify the unit cells by $A_\uc$ and the connection matrices $\Gamma_{(i,j)}$ with $(i,j)\in \mathcal{I}(\bd{d})$.  For brevity, we define $\Gamma (\bd{d}):=\sum_{(i,j)\in \mathcal{I}(\bd{d})}\Gamma_{(i,j)}$, which allows us to conveniently write Eq. (\ref{mastereq1}) as 
\begin{align}
    \nonumber A_{p,q}={}& (\mathbb{1}_p\otimes \mathbb{1}_q)\otimes A_{\rm u} \\
\nonumber &+ [(\mathbb{1}_p\otimes R_q)\otimes \Gamma(\bd{e}_1)+ \text{h.c.}]\\
\nonumber &+ [(R_p\otimes \mathbb{1}_q)\otimes \Gamma(\bd{e}_2)+\text{h.c.}]\\
\nonumber & + [(R_p\otimes R_q)\otimes \Gamma(\bd{e}_1+\bd{e}_2)+\text{h.c.}]\\
&+ [(L_p\otimes R_q)\otimes \Gamma(\bd{e}_1-\bd{e}_2)+\text{h.c.}].
\end{align} 
The unit cells, alongside the corresponding connection matrices $\Gamma(\bd{d})$ are shown in Tab. \ref{ALFlakes}. The sets $\mathcal{I}(\vec{d})$ can be read off from the $\Gamma(\vec{d})$ appearing for each lattice.

An alternative method to determine the $A_{\rm u}$ and $\mathcal{I}(\vec{d})$ is to start from a given Bloch adjacency matrix $A(\vec{k})$ and express it as a function of $1$, $e^{\pm \rmi k_1}$, and $e^{\pm \rmi k_2}$ according to
\begin{align}
 A(\vec{k}) = \bar{A}(1,e^{\rmi k_1},e^{-\rmi k_1},e^{\rmi k_2},e^{-\rmi k_2}).
\end{align}
Importantly, for lattices such as Trellis that contain $\cos k_1$ in the matrix components, we have to write $2\cos k_1 = e^{\rmi k_1}+e^{-\rmi k_1}$. The matrix $A_{\rm u}$ is obtain by erasing all non-unit entries of $A(\vec{k})$, i.e.
\begin{align}
A_{\rm u} = \bar{A}(1,0,0,0,0).
\end{align}
Since the complex phases in $A(\vec{k})$ correspond to translations along the Bravais lattice, we have
\begin{align}
 \nonumber \mathcal{I}(\vec{e}_1) &= \{(i,j):\ \bar{A}_{ij} = e^{-\rmi k_1} \},\\
 \nonumber \mathcal{I}(\vec{e}_2) &= \{(i,j):\ \bar{A}_{ij} = e^{-\rmi k_2} \},\\
 \nonumber \mathcal{I}(\vec{e}_1+\vec{e}_2) &= \{(i,j):\ \bar{A}_{ij} = e^{-\rmi (k_1+k_2)} \},\\
 \label{flake9} \mathcal{I}(\vec{e}_1-\vec{e}_2) &= \{(i,j):\ \bar{A}_{ij} = e^{-\rmi (k_1-k_2)} \}.
\end{align}
This allows us to read off the sets $\mathcal{I}(\vec{d})$ from the matrices $A(\vec{k})$ given in Sec. \ref{SecBloch}. They are consistent with the entries of Tab. \ref{ALFlakes}.

As an example of the alternative method, consider again the CaVO lattice. The Bloch adjacency matrix is given in Eq. (\ref{bloch21}), which is already in the form $A=\bar{A}$. By erasing all non-unit entries, we obtain the adjacency matrix for the unit cell as
\begin{align}
 \label{flake10} A_{\rm u }^{(\rm Ca)}=\begin{pmatrix}    0 & 1 & 0 & 1 \\ 1 & 0 & 1 & 0 \\ 0 & 1 & 0 & 1 \\ 1 & 0 & 1  & 0
\end{pmatrix}.
\end{align}
The edges that leave the unit cell in the $\boldsymbol{e}_1$- and $\vec{e}_2$-directions are then given by $\mathcal{I}(\vec{e}_1)=\{(1,3)\}$ and $\mathcal{I}(\vec{e}_2)=\{(2,4)\}$, respectively, since $A(\vec{k})_{13}=e^{-\rmi k_1}$ and $A(\vec{k})_{24}=e^{-\rmi k_2}$. There are no appearances of $e^{-\rmi (k_1\pm k_2)}$ in $A(\vec{k})$, hence no edges leave the unit cell in the $\vec{e}_1\pm\vec{e}_2$ directions, and $\mathcal{I}(\vec{e}_1\pm\vec{e}_2)=\emptyset$.

\begin{table*}[htp]
\centering
\setlength{\arraycolsep}{3pt}
\setlength{\extrarowheight}{2pt} 
\resizebox{\textwidth}{!}{\begin{tabular}{|c|c|c|c|c|c|c|c|c|} 
\hline
 Lattice & $N_\uc$ & $A_\uc$ & $\Gamma(\bd{e}_1)$ & 
 $\Gamma(\bd{e}_2)$ & 
 $\Gamma(\bd{e}_1+\bd{e}_2)$ & 
 $\Gamma(\bd{e}_1-\bd{e}_2)$ & Type \\
\toprule
Square & $1$ & $0$ & $1$ & $1$ & $0$ & $0$ & S\\ 
\hline
Triangular & $1$ & $0$ & $1$ & $1$ & $1$ &  $0$ & T\\
\hline
Honeycomb & $2$ & 
$\begin{pmatrix*}[r]
0 & 1 \\
1 & 0
\end{pmatrix*}$ & 
$\Gamma_{(1,2)}$ & $\Gamma_{(1,2)}$ & $0$ & $0$ & T\\
\hline
Trellis & $2$ & $\begin{pmatrix}
0 & 1 \\
1 & 0
\end{pmatrix}$ & $\Gamma_{(1,1)}+\Gamma_{(1,2)}+\Gamma_{(2,2)}$ & $\Gamma_{(1,2)}$ & $0$ & $0$ & S\\
\hline
Kagome & $3$ & 
$\begin{pmatrix}
0 & 1 & 1 \\
1 & 0 & 1 \\
1 & 1 & 0
\end{pmatrix}$ & 
$\Gamma_{(3,2)}$ & $\Gamma_{(1,2)}$ & $0$ & $\Gamma_{(3,1)}$ & T\\
\hline
CaVO & $4$ & $\left(
\begin{array}{cccc}
 0 & 1 & 0 & 1 \\
 1 & 0 & 1 & 0 \\
 0 & 1 & 0 & 1 \\
 1 & 0 & 1 & 0 \\
\end{array}
\right)$ & $\Gamma_{(1,3)}$ & $\Gamma_{(2,4)}$ & $0$ & $0$ & S\\
\hline
SrCuBO & $4$ & $\left(
\begin{array}{cccc}
 0 & 1 & 0 & 1 \\
 1 & 0 & 1 & 0 \\
 0 & 1 & 0 & 1 \\
 1 & 0 & 1 & 0 \\
\end{array}
\right)$ & $\Gamma_{(2,1)}+\Gamma_{(2,4)}+\Gamma_{(3,4)}$ & $\Gamma_{(3,1)}+\Gamma_{(4,1)}+\Gamma_{(3,2)}$ & $0$ & $0$ & S\\
\hline
Star & $6$ & $\left(
\begin{array}{cccccc}
 0 & 1 & 0 & 0 & 0 & 0 \\
 1 & 0 & 1 & 0 & 1 & 0 \\
 0 & 1 & 0 & 1 & 1 & 0 \\
 0 & 0 & 1 & 0 & 0 & 0 \\
 0 & 1 & 1 & 0 & 0 & 1 \\
 0 & 0 & 0 & 0 & 1 & 0 \\
\end{array}
\right)$ & $\Gamma_{(1,4)}$ & $\Gamma_{(4,6)}$ & $\Gamma_{(1,6)}$ & $0$ & T\\
\hline
Ruby & $6$ & $\left(
\begin{array}{cccccc}
 0 & 1 & 0 & 0 & 0 & 1 \\
 1 & 0 & 1 & 0 & 0 & 0 \\
 0 & 1 & 0 & 1 & 0 & 0 \\
 0 & 0 & 1 & 0 & 1 & 0 \\
 0 & 0 & 0 & 1 & 0 & 1 \\
 1 & 0 & 0 & 0 & 1 & 0 \\
\end{array}
\right)$ & $\Gamma_{(4,2)}+\Gamma_{(5,1)}$ & $\Gamma_{(3,1)}+\Gamma_{(4,6)}$ & $0$ & $\Gamma_{(5,3)}+\Gamma_{(6,2)}$ & T\\
\hline
Maple-Leaf & $6$ & $\left(
\begin{array}{cccccc}
 0 & 1 & 0 & 0 & 0 & 1 \\
 1 & 0 & 1 & 0 & 0 & 0 \\
 0 & 1 & 0 & 1 & 0 & 0 \\
 0 & 0 & 1 & 0 & 1 & 0 \\
 0 & 0 & 0 & 1 & 0 & 1 \\
 1 & 0 & 0 & 0 & 1 & 0 \\
\end{array}
\right)$ & $\Gamma_{(4,1)}+\Gamma_{(5,1)}+\Gamma_{(4,2)}$ & $\Gamma_{(5,2)}+\Gamma_{(5,3)}+\Gamma_{(6,2)}$ & $0$ & $\Gamma_{(3,1)}+\Gamma_{(3,6)}+\Gamma_{(4,6)}$ & T\\
\hline
SHD & $12$ & ${\small\left(
\begin{array}{cccccccccccc}
 0 & 1 & 0 & 0 & 0 & 0 & 0 & 0 & 0 & 0 & 0 & 1 \\
 1 & 0 & 1 & 0 & 0 & 0 & 0 & 0 & 0 & 0 & 0 & 0 \\
 0 & 1 & 0 & 1 & 0 & 0 & 0 & 0 & 0 & 0 & 0 & 0 \\
 0 & 0 & 1 & 0 & 1 & 0 & 0 & 0 & 0 & 0 & 0 & 0 \\
 0 & 0 & 0 & 1 & 0 & 1 & 0 & 0 & 0 & 0 & 0 & 0 \\
 0 & 0 & 0 & 0 & 1 & 0 & 1 & 0 & 0 & 0 & 0 & 0 \\
 0 & 0 & 0 & 0 & 0 & 1 & 0 & 1 & 0 & 0 & 0 & 0 \\
 0 & 0 & 0 & 0 & 0 & 0 & 1 & 0 & 1 & 0 & 0 & 0 \\
 0 & 0 & 0 & 0 & 0 & 0 & 0 & 1 & 0 & 1 & 0 & 0 \\
 0 & 0 & 0 & 0 & 0 & 0 & 0 & 0 & 1 & 0 & 1 & 0 \\
 0 & 0 & 0 & 0 & 0 & 0 & 0 & 0 & 0 & 1 & 0 & 1 \\
 1 & 0 & 0 & 0 & 0 & 0 & 0 & 0 & 0 & 0 & 1 & 0 \\
\end{array}
\right)}$ & $\Gamma_{(9,2)}+\Gamma_{(8,3)}$ & $\Gamma_{(6,1)}+\Gamma_{(7,12)}$ & $0$ & $\Gamma_{(10,5)}+\Gamma_{(11,4)}$ & T\\
\hline
\end{tabular}}
\caption{We show the unit cell adjacency matrices $A_\uc$ and connection matrices $\Gamma_{(i,j)}$ for all eleven Archimedean lattices, which allows us to compute arbitrarily large finite flakes or clusters. Specifically, the information is needed to construct the adjacency matrices $A_{p,q}$ in Eq. (\ref{mastereq1}). The corresponding connection matrices for each lattice is of size $N_\uc\times N_\uc$ and $0$ refers to $0\cdot \mathbb{1}_{N_\uc}$. The topologically distinct quotient graphs or Bravais lattices are type S and type T, with only type T having nonzero entries $\Gamma(\vec{e}_1\pm \vec{e}_2)$. We use the abbreviation $\Gamma(\bd{d})= \sum_{(i,j)\in\mathcal{I}(\bd{d})}\Gamma_{(i,j)}$.}
\label{ALFlakes}
\end{table*}

\section{Properties of Bloch adjacency matrix}\label{AppBAdj}

In this appendix, we derive Eq. (\ref{gf11}) for the components of the Bloch adjacency matrix $\hat{A}(\bk)$, that is
 \begin{equation}
  \label{ind0}  \Mel{i_\uc}{\hat{A}(\bk)}{j_\uc}= \sum_{\bd{v}} \Mel{i_{\uc}}{\hat{A}\hat{T}_{\bd{v}}}{j_{\uc}}e^{- \rmi \bd{v}\cdot \bk},
\end{equation}
and Eq. (\ref{gf12}) for the components of its moments, that is
\begin{equation}
 \label{ind1} \Mel{i_\uc}{\hat{A}(\vec{k})^n}{j_\uc}=\sum_{\bd{v}} \Mel{i_\uc}{\hat{A}^n \hat{T}_{\bd{v}}}{j_\uc} e^{-\rmi \bd{v}\cdot \bk }
\end{equation}
for any integer $n\geq 1$.

We start by showing Eq. (\ref{ind0}). Let $\hat{P}_\uc$ denote the projection operator onto the unit cell $\Lambda_\uc$ defined through the relation
\begin{equation}
    \hat{P}_{\uc}|i\rangle=
    \begin{cases}
    |i_\uc\rangle &\text{ if $|i\rangle=|i_\uc\rangle$ belongs to the unit cell}\\
    0 &\text{ otherwise}
\end{cases}.
\end{equation}
Let $|\bk\rangle$ denote an eigenstate of both  $\hat{A}$ and $\hat{T}_{\bd{v}}$, where $\bd{v}$ is a direction that leaves the system invariant under translation. The eigenvalue equation $\hat{A}|\bk\rangle=\lambda|\bk\rangle$ can be projected onto the unit cell to yield
\begin{align}
    \nonumber\lambda \hat{P}_\uc\ket{\bk}&=\hat{P}_{\uc}\left(\sum_{i,j}\Mel{i}{\hat{A}}{j}\ketbra{i}{j}\right)\ket{\bk}\\
    \nonumber&=\sum_{i,j}\Mel{i}{\hat{A}}{j}\braket{j}{\bk}(\hat{P}_\uc\ket{i})\\
    \nonumber&=\sum_{i_\uc,j}\Mel{i_\uc}{\hat{A}}{j}\braket{j}{\bk}\ket{i_\uc}\\
    &=\sum_{i_\uc}\left(\Mel{i_\uc}{\hat{A}\left(\sum_{j} \ketbra{j}{j}\right)}{\bk}\right)\ket{i_\uc}.
\end{align}
Using the unique split $|i\rangle=|i_{\rm u},\vec{v}\rangle$, the completeness relation reads
\begin{align}
 \label{ind3} \mathbb{1}_N = \sum_{i=1}^N|i\rangle\langle i| = \sum_{i_{\rm u}=1}^{N_{\rm u}}\sum_{\vec{v}} \hat{T}_{\vec{v}}|i_{\rm u}\rangle\langle i_{\rm u}|\hat{T}_{\vec{v}}^\dagger.
\end{align}
We arrive at
\begin{align}
\nonumber \lambda\hat{P}_{\uc}\ket{\bk}&=\sum_{i_\uc}\Mel{i_\uc}{\hat{A}\left(\sum_{\bd{v},j_\uc} \hat{T}_{\bd{v}}\ketbra{j_\uc}{j_\uc}\hat{T}^{\dagger}_{\bd{v}}\right)}{\bk}\ket{i_\uc}\\
\nonumber &=\sum_{i_\uc,j_\uc}\bra{i_\uc}\hat{A}\sum_{\bd{v}} \hat{T}_{\bd{v}}\ket{j_{\uc}}\bra{j_\uc}\underbrace{\hat{T}_{\bd{v}}^\dagger\ket{\bk}}_{=e^{-\rmi\bd{v}\cdot\bd{k}}\ket{\bk}}\ket{i_\uc}\\
 &=\left(\sum_{i_\uc,j_\uc}\left(\sum_{\bd{v}}\Mel{i_\uc}{\hat{A}\hat{T}_{\bd{v}}}{j_\uc}e^{-\rmi \bd{v}\cdot \bk}\right)\ketbra{i_\uc}{j_u}\right)\ket{\bk}.
\end{align}
Using $\bra{j_\uc}=\bra{j_\uc}\hat{P}_{\uc}$ we find
\begin{align}
\nonumber&\lambda \hat{P}_{\uc}\ket{\bk}\\ &=\left(\sum_{i_\uc,j_\uc}\left(\sum_{\bd{v}}\Mel{i_\uc}{\hat{A}\hat{T}_{\bd{v}}}{j_\uc}e^{-\rmi \bd{v}\cdot \bk}\right)\ketbra{i_\uc}{j_u}\right)\left(\hat{P}_{\uc}\ket{\bk}\right)
\end{align}
or, equivalently, 
\begin{align}
\lambda \ket{\bk_\uc}&=\overbrace{\left(\sum_{i_\uc,j_\uc}\left(\sum_{\bd{v}}\Mel{i_\uc}{\hat{A}\hat{T}_{\bd{v}}}{j_\uc}e^{-\rmi \bd{v}\cdot \bk}\right)\ketbra{i_\uc}{j_u}\right)}^{:=\hat{A}(\bk)}\ket{\bk_\uc
    }.
\end{align}
Here, $\ket{\bk_\uc}=\hat{P}_{\uc}\ket{\bk}$ is the Bloch state restricted to the unit cell. Since the choice of the unit cell $\Lambda_\uc$ was arbitrary due to translation symmetry, the spectrum of the infinite lattice is entirely described by solving the equation
\begin{equation}
    \hat{A}(\bk)\ket{k_\uc}=\lambda \ket{\bk_\uc}
\end{equation}
for $\bk\in [0,2\pi)\times[0,2\pi)$, where
\begin{equation}
 \Mel{i_\uc}{\hat{A}(\vec{k})}{j_\uc}=\sum_{\bd{v}} \Mel{i_\uc}{\hat{A}\hat{T}_{\bd{v}}}{j_\uc} e^{-\rmi \bd{v}\cdot \bk }.
\end{equation}
This proves Eqs. (\ref{ind0}) and (\ref{gf12}) $\square.$

We next derive Eq. (\ref{ind1}). For $n=1$, the result follows from the definition of the Bloch adjacency matrix in Eq. (\ref{gf11}). Let $\mathbb{N}\ni n>1$ and assume that Eq. (\ref{ind1}) is true for $n-1$. We then have
\begin{align}
    \nonumber &\Mel{i_\uc}{\hat{A}(\bk)^{n}}{j_\uc}=\sum_{m_\uc}\Mel{i_\uc}{\hat{A}(\bk)^{n-1}}{m_\uc}\Mel{m_\uc}{\hat{A}(\bk)}{j_\uc}\\
    \nonumber &=\sum_{m_\uc}\sum_{\bd{v},\vec{w}}\langle i_{\rm u}|\hat{A}^{n-1}\hat{T}_{\bd{v}}|m_{\rm u}\rangle \langle m_{\rm u}|
    \hat{A}\hat{T}_{\bd{w}}|j_{\rm u}\rangle e^{-\rmi (\vec{v}+\bd{w})\cdot \vec{k}}\\
    \label{ind2} &=\sum_{\vec{v},\vec{w}} \langle i_{\rm u}|\hat{A}^{n-1}\Biggl(\sum_{m_{\rm u}}\hat{T}_{\vec{v}}|m_{\rm u}\rangle\langle m_{\rm u}|\hat{T}_{\vec{w}}\Biggr)\hat{A}|i_{\rm u}\rangle e^{-\rmi (\vec{v}+\bd{w})\cdot \vec{k}}.
\end{align}
In the last line we used that the lattice is invariant under translations, so that $\hat{A}$ commutes with all $\hat{T}_{\vec{v}}$. 
Note that $\hat{T}_{-\vec{v}}=\hat{T}^\dagger_{\vec{v}}$ and $\hat{T}_{\vec{v}+\vec{w}}=\hat{T}_{\vec{v}}\hat{T}_{\vec{w}}$. 
 Thus the term in brackets in Eq. (\ref{ind2}) reads
\begin{align}
 \label{ind4} \Biggl(\sum_{m_{\rm u}}\hat{T}_{\vec{v}}|m_{\rm u}\rangle\langle m_{\rm u}|\hat{T}_{\vec{w}}\Biggr) &=  \Biggl(\sum_{m_{\rm u}}\hat{T}_{\vec{v}}|m_{\rm u}\rangle\langle m_{\rm u}|\hat{T}_{\vec{v}}^\dagger\Biggr)\hat{T}_{\vec{v}+\vec{w}}.
\end{align}
Consequently, using that $\hat{A}$ commutes with all translations and the completeness relation in Eq. (\ref{ind3}), we arrive at
\begin{align}
 \nonumber \Mel{i_\uc}{\hat{A}(\bk)^{n}}{j_\uc} ={}& \langle i_{\rm u}|\hat{A}^{n-1}\Biggl(\sum_{m_{\rm u}}\sum_{\vec{v}}\hat{T}_{\vec{v}}|m_{\rm u}\rangle\langle m_{\rm u}|\hat{T}_{\vec{v}}^\dagger\Biggr) \\
 \nonumber &\times \sum_{\vec{w}}\hat{T}_{\vec{v}+\vec{w}}\hat{A}|i_{\rm u}\rangle e^{-\rmi (\vec{v}+\bd{w})\cdot \vec{k}}\\
 \nonumber ={}& \langle i_{\rm u}|\hat{A}^{n-1}\sum_{\vec{z}}\hat{T}_{\vec{z}}\hat{A}|i_{\rm u}\rangle e^{-\rmi \vec{z}\cdot \vec{k}}\\
 \label{ind5} ={}& \sum_{\vec{z}} \langle i_{\rm u}|\hat{A}^n\hat{T}_{\vec{z}}|i_{\rm u}\rangle e^{-\rmi \vec{z}\cdot \vec{k}}.
\end{align}
Here we shifted the summation variables $\vec{w}\to \vec{z}=\vec{v}+\vec{w}$. This completes the proof of Eqs. (\ref{ind1}) and (\ref{gf12}). $\square$

\section{Derivation of Eqs. (\ref{gf5}) and (\ref{gf5b})}\label{Trid}

In this appendix, we establish the identities
\begin{equation}
  \label{Tra0}  \lim_{N\to\infty}\frac{{\rm Tr}(\hat{A}^n)}{N}=\frac{1}{N_\uc}\int_{\bk} {\rm Tr}(\hat{A}(\bk)^n)
\end{equation}
and
\begin{equation}
\label{Tra-1}
    \lim_{N\to\infty}\Mel{i_\uc}{\hat{A}}{i_{\uc}}=\int_{\bk} \Mel{i_\uc}{\hat    A(\bk)^n}{i_\uc}
\end{equation}
for infinite lattices, i.e. Eqs. (\ref{gf5}) and (\ref{gf5b}). Let $p\in\mathbb{N}$. For simplicity, assume that the lattice $\Lambda$ is a periodic tiling of $p\times p$ unit cells $\Lambda_\uc$. Each $\Lambda_\uc$ consists of $N_\uc$ sites labeled by $i_{\uc}$, therefore the fixed total number of sites is $N= p^2 N_\uc$.

Now consider the lattice translation vectors $\bd{v},\bd{w}\in \mathbb{Z}^2_{p}$, i.e $\bd{v}=(v_1,v_2)^T$ and $\bd{w}=(w_1,w_2)^T$ where $ v_1,v_2,w_1,w_2\in\mathbb{Z}_p$. Our definitions imply
\begin{align}
    \sum_{\bd{v}}1=\sum_{\bd{w}}1=p^2.
\end{align}
Since there are $p^2$ unit cells, we define $p^2$ reciprocal lattice vectors $\bd{k}\in \mathcal{K}$, where 
\begin{equation}
\mathcal{K}=\left\{\left( \frac{2\pi n}{p},\frac{2\pi m}{p}\right)^T: n,m=0,1,2,\dots,p-1\right\}.    
\end{equation}
Every site can be uniquely identified via the split $|i\rangle=|i_\uc,\bd{v}\rangle$ for some $\bd{v}\in\mathbb{Z}_p^2$. The returning walk numbers may depend on $i_{\rm u}$, but not $\vec{v}$ because of the periodicity of the lattice. Thus we have
\begin{equation}
\label{Snk1}
\sum_{\bd{v}}\Mel{i_\uc,\bd{v}}{\hat{A}^n}{i_\uc,\bd{v}}=p^2\Mel{i_\uc,\vec{0}}{ \hat{A}^n}{i_\uc,\vec{0}}.
\end{equation}
This implies
\begin{align}
\label{Snk2}
    S_{n}^{(i_\uc)} &= \frac{1}{p^2}\sum_{\bd{v}}\Mel{i_\uc,\bd{v}}{\hat{A}^n}{i_\uc,\bd{v}}\\
    &=\sum_{\bd{v}}\frac{\Mel{i_\uc}{\hat{T}^{\dagger}_{\bd{v}}\hat{A}^n\hat{T}_{\bd{v}}}{i_\uc}}{p^2},
\end{align}
where we used $T_{\bd{v}}\ket{i_{\uc},\bd{0}}=\ket{i_{\uc},\bd{v}}$ and defined $\ket{i_\uc,\bd{0}}=\ket{i_\uc}$.
Using the identity
\begin{align}
\nonumber\frac{1}{p^2}\sum_{\bd{k}\in\mathcal{K}}e^{i (\bd{w}+\bd{v})\cdot \bd{k}}&=\frac{1}{p^2}\sum_{n=0}^{p-1}\sum_{m=0}^{p-1}e^{i (w_1+v_1)\frac{2\pi n}{p}} e^{i (w_2+v_2) \frac{2\pi m}{p}}\\ &=\delta_{\bd{w},-\bd{v}},
\end{align} 
we find
\begin{align}
    \nonumber&\sum_{\bd{v}}\frac{\Mel{i_\uc}{\hat{T}^{\dagger}_{\bd{v}}\hat{A}^n\hat{T}_{\bd{v}}}{i_\uc}}{p^2}\\
    \nonumber&=\sum_{\bd{v},\bd{w}}\sum_{\bk \in \mathcal{{K}}}\frac{e^{i (\bd{w}+\bd{v})\cdot \bd{k}}}{ p^2}\Mel{i_\uc}{\hat
    T_{-\bd{w}}^{\dagger}\hat{A}^n\hat{T}_{\bd{v}}}{i_\uc}\\
   \label{Tra2} &=\sum_{\bk \in \mathcal{{K}}}\sum_{\bd{v},\bd{w}}\frac{e^{i (\bd{w}+\bd{v})\cdot \bd{k}}}{p^2}\Mel{i_\uc}{\hat
    T_{-\bd{w}}^{\dagger}\hat{A}^n\hat{T}_{\bd{v}}}{i_\uc}.
\end{align}
Since $[\hat{T}_{\bd{v}},\hat{A}]=0$ for $\bd{v}\in\mathbb{Z}^2_p$ due to discrete translation symmetry of the lattice, we have
\begin{align}
    \Mel{i_\uc}{\hat
    T_{-\bd{w}}^{\dagger}\hat{A}^n\hat{T}_{\bd{v}}}{i_\uc}&=\Mel{i_\uc}{\hat{A}^n\hat{T}_{\bd{v}+\bd{w}}}{i_\uc}.
\end{align}
Under the change of variables $\bd{z}:=\bd{w}+\bd{v}\in\mathbb{Z}^2_p$, observe that for fixed $\vec{k}$ we have
\begin{align}
    \nonumber&\sum_{\bd{v},\bd{w}}\frac{e^{i (\bd{w}+\bd{v})\cdot \bd{k}}}{p^2}\Mel{i_\uc}{\hat{A}^n\hat{T}_{\bd{w}+\bd{v}}}{i_\uc}\\
    \nonumber&=\left(\sum_{\bd{z}}\frac{e^{i \bd{z}\cdot \bd{k}}}{p^2}\Mel{i_\uc}{\hat{A}^n\hat{T}_{\bd{z}}}{i_\uc}\right)\left(\sum_{\bd{v}} 1\right)  \\ 
    \nonumber&=\sum_{\bd{z}}\frac{e^{i \bd{z}\cdot \bd{k}}}{p^2}\Mel{i_\uc}{\hat{A}^n\hat{T}_{\bd{z}}}{i_\uc}\cdot p^2 \\  
    \label{Tra3} &=\sum_{\bd{z}}e^{i \bd{z}\cdot \bd{k}}\Mel{i_\uc}{\hat{A}^n\hat{T}_{\bd{z}}}{i_\uc}.
\end{align} 
Combining Eqs. (\ref{Tra2}), (\ref{Tra3}), and (\ref{ind1}) we arrive at
 \begin{align}
     \nonumber S_{n}^{(i_{\uc})}&=\sum_{\bd{v}}\frac{\Mel{i_\uc}{\hat{T}^{\dagger}_{\bd{v}}\hat{A}^n\hat{T}_{\bd{v}}}{i_\uc}}{p^2}\\
     \nonumber&=\sum_{\bk\in \mathcal{K}}\sum_{\bd{z}}\frac{e^{i \bd{z}\cdot \bd{k}}}{p^2}\Mel{i_\uc}{\hat{A}^n\hat{T}_{\bd{z}}}{i_\uc}\\ 
     &= \frac{1}{(2\pi)^2}\sum_{\bk\in\mathcal{K}}\left(\frac{2\pi}{p}\right)\left(\frac{2\pi}{p}\right) \Mel{i_\uc}{\hat{A}(\bk)^n}{i_\uc}.
 \end{align}
 Since the spacing $\Delta k_1,\Delta k_2$ between each $k_1,k_2$ of the reciprocal vector is $2\pi/p$, we have 
\begin{align}
\label{Sndelk}
    S_{n}^{(i_{\uc})}&=\frac{1}{(2\pi)^2}\sum_{\bd{k}\in \mathcal{K}}\Delta k_1 \Delta k_2\Mel{i_\uc}{\hat{A}(\bk)^n}{i_\uc},
\end{align}
which we recognize as a Riemann sum. In the thermodynamic limit $N\to \infty$, the spacing $\Delta k_1,\Delta k_2 \to 0$ and we arrive at
\begin{align}
     \nonumber S_{n}^{(i_{\uc})}&= \frac{1}{(2\pi)^2}\int_{0}^{2\pi}\int_{0}^{2\pi} \mbox{d}k_1  \mbox{d} k_2\Mel{i_\uc}{\hat{A}(\bk)^n}{i_\uc}\\
     \nonumber&= \int_{\bk} \Mel{i_\uc}{\hat{A}(\bk)^n}{i_\uc}\\
     &:=\int_{\bk} S_{n}^{(i_{\uc})}(\bk).
\end{align}
This proves Eqs. (\ref{Tra-1}) and (\ref{gf5b}) $\square$.

To show Eqs. (\ref{Tra0}) and (\ref{gf5}), we observe that
\begin{equation}
    {\rm Tr}(\hat{A}^n)= \sum_{i_{\rm u}=1}^{N_{\rm u}}\sum_{\bd{v}}\Mel{i_\uc,\bd{v}}{\hat{A}^n}{i_\uc,\bd{v}} = p^2\sum_{i_{\rm u}=1}^{N_{\rm u}}S_{n}^{(i_\uc)}.
\end{equation}
Using the fact that $N=p^2 N_{\uc}$ and Eq. (\ref{Sndelk}), we obtain
\begin{align}
    \nonumber \frac{{{\rm Tr}(\hat{A}^n})}{N}&=\sum_{i_\uc}\frac{1}{N_{\uc}} \frac{1}{(2\pi)^2}\sum_{\bd{k}\in \mathcal{K}}\Delta k_1 \Delta k_2\Mel{i_\uc}{\hat{A}(\bk)^n}{i_\uc}\\
    &= \frac{1}{N_{\uc}} \frac{1}{(2\pi)^2}\sum_{\bd{k}\in \mathcal{K}}\Delta k_1 \Delta k_2{\rm Tr}(\hat{A}(\bk)^n),
\end{align}
which, in the thermodynamic limit $N\to\infty$ yields
\begin{equation}
    \lim_{N\to\infty} \frac{{{\rm Tr}(\hat{A}^n})}{N}=\frac{1}{N_{\uc}} \int_{\bk}{\rm Tr}(\hat{A}(\bk)^n).\qquad \square
\end{equation}

\section{Continued fraction expansion and $S_n$}\label{AppCF}

In this section, we derive a recursive method to compute the coefficients $(a_n,b_n)$ of the continued fraction (CF) expansion of the DOS from the returning walks numbers $S_n$. Throughout the rest of this section, we refer to $(a_n,b_n)$ as the CF coefficients. 

The central idea behind the CF expansion is to choose a basis that allows us to write the Hamiltonian matrix $H=-A$ as a symmetric tri-diagonal matrix in the form
\begin{equation}
    H=\begin{pmatrix}
    a_0 & b_1 & 0 & 0 & 0 & \cdots \\
    b_1 & a_1 & b_2 & 0 & 0 & \\
    0 & b_2 & a_2 & b_3 & 0 & \\
    0 & 0 & b_3 & a_3 & b_4 & \\
    0 & 0 & 0 & b_4 & a_4 & \\
    \vdots & & & & & \ddots
\end{pmatrix}.
\end{equation}
The associated basis is constructed by choosing a fixed site $i$ and the state $|i\rangle$, such that
\begin{align}
\Mel{i}{\hat{H}}{i}=H_{ii}=a_0,
\end{align}
while the remaining basis states are linear combination of the position states $|i'\rangle$. For $\eta>0$, the $(i,i)$ component of the Green operator $\hat{G}=([E- \rmi\eta] \hat{\mathbb{1}}-\hat{H})^{-1}$ then has the form 
\begin{align}
\label{CF1}
    \langle i|\hat{G}|i\rangle = \frac{1}{E - \rmi\eta + a_0 - \cfrac{b_1^2}{E - \rmi\eta + a_1 - \cfrac{b_2^2}{\ddots}}}.
\end{align}
Taking the imaginary part, we find that the local density of states (LDOS) at site $i$ is given by
\begin{equation}\label{CF2}
    \rho_i(E)=\frac{1}{\pi} \lim_{\eta\to 0 }\mathfrak{I} \Mel{i}{\hat
    G}{i}.
\end{equation}
Note that the sum of all the LDOS over all sites $i$ gives the full DOS, i.e
\begin{equation}
    D(E)=  \frac{1}{N}\frac{1}{\pi}\lim_{\eta\to 0}\mathfrak{I}\  {\rm Tr}(\hat{G}) = \frac{1}{N}\sum_{i=1}^{N}\rho_i(E),
\end{equation}
with $N\to \infty$ in the thermodynamic limit. 
Since the choice of the  state $\ket{i}$ corresponds to an arbitrary site $i$, we may instead choose all the sites in the unit cell $i_\uc=1,2,\dots,N_\uc$ and exploit translation symmetry. Thus the full DOS is given by
\begin{align}
\label{CF3}
 D(E)  &=\frac{1}{N_\uc}\sum_{i_u=1}^{N_\uc}\rho_{i_\uc}(E).
\end{align}

The construction of the orthonormal basis $\{|u_n\rangle\}_{n=1,2,3\dots}$, called Krylov basis, where the operator $\hat{H}$ is represented by a tri-diagonal matrix $H$ is accomplished through the Lanczos algorithm. For this, we define 
\begin{align}
|u_0\rangle=|i\rangle.
\end{align}
For $n\geq 1$, we impose the Gram–Schmidt relation
\begin{equation}\label{CF4}
    \ket{\tilde{u}_{n+1}}=\hat{H}\ket{u_{n}}-\Mel{u_{n}}{\hat{H}}{u_{n}}\ket{u_{n}}- \Mel{u_{n-1}}{\hat{H}}{u_{n}}\ket{u_{n-1}}
\end{equation}
and obtain $|u_n\rangle = |\tilde{u}_n\rangle/\sqrt{\langle \tilde{u}_n|\tilde{u}_n\rangle}$ after normalization. The CF coefficients are defined for $n\geq 0$ by 
\begin{align}
    \label{CF5}
    a_n&:= \Mel{u_n}{\hat{H}}{u_n},\\
    b_{n+1}&:=\Mel{u_{n}}{\hat{H}}{u_{n+1}}= \sqrt{\braket{\tilde{u}_{n+1}}{\tilde{u}_{n+1}}}.
\end{align}
Note that $b_n^2=\langle \tilde{u}_n|\tilde{u}_n\rangle$ and $|\tilde{u}_n\rangle= b_n|u_n\rangle$. For $n<0$ we set
\begin{align}
 \label{CFneg} a_{n<0}=b_{n<0}&=0.
\end{align}    
With the CF coefficients, Eq. (\ref{CF4}) can be rewritten in the form 
\begin{equation} \label{CF6}
\begin{aligned}
 |\tilde{u}_{n+1}\rangle &= \hat{H}|u_n\rangle - a_n |u_n\rangle -b_{n}|u_{n-1}\rangle\\
 \Rightarrow\ \hat{H}|u_n\rangle  &=  a_n |u_n\rangle +b_{n}|u_{n-1}\rangle+|\tilde{u}_{n+1}\rangle\\
  &= a_n |u_n\rangle +b_{n}|u_{n-1}\rangle+b_{n+1} |u_{n+1}\rangle
\end{aligned}
\end{equation}
As a result, the CF coefficients $(a_n,b_n)$ are obtained through repeated matrix multiplication with $\hat{H}$ and orthogonalization of the Krylov basis. However, due to numerical errors that may pile up, orthogonality between states $\braket{u_n}{u_{n'}}$ is typically lost for large $n,n' \gg 1$ in practice, and only the first few $(a_n,b_n)$ may be reliable. Here we propose an alternative method to circumvent this issue by using the returning walk numbers $S_n$.

We define the Lanczos polynomials $\mathcal{P}_n(\hat{H})$ as the polynomial of order $n$ in the variable $\hat{H}$ such that 
\begin{equation}
    \label{CF7}
    \ket{u_n}=\mathcal{P}_n(\hat{H}) \ket{u_0}.
\end{equation}
By construction of the Krylov basis, the Lanczos polynomial assumes the form
\begin{equation}
\label{CF8}
    \mathcal{P}_n(\hat{H})= \sum_{j=0}^{n} c_{n,j} \hat{H}^{j}.
\end{equation}
We set
\begin{equation}
\label{CF9}
c_{n,j} =0\ \text{for}\ n-j\notin[0,n].
\end{equation} 
Equation (\ref{CF8}) allows us to express the CF coefficients in terms of $\mathcal{P}_n(\hat{H})$ through
\begin{align}
    a_n&= \Mel{u_0}{\mathcal{P}_n(\hat{H})\hat{H}\mathcal{P}_n(\hat{H})}{u_0},\label{CF10}\\
    b_n^2&=\Mel{u_0}{\mathcal{P}_{n-1}(\hat{H})\hat{H}^2\mathcal{P}_{n-1}(\hat{H)}}{u_0}- (a_{n-1}^2+b_{n-1}^2),\label{CF11}
\end{align}
where Eq. (\ref{CF10}) follows from Eq. (\ref{CF5}), and Eq. (\ref{CF11}) is obtained by using Eq. (\ref{CF6}) to compute the norm-squared of $|\psi\rangle=\hat{H}|u_{n-1}\rangle$  and isolating for $b_{n}^2$.
Due to Eq. (\ref{CF6}), the Lanczos polynomials satisfy the recursion
\begin{equation}
\label{CF12}
    b_{n+2}\mathcal{P}_{n+2}(\hat{H})= (\hat{H}- a_{n+1}\hat{\mathbb{1}})\mathcal{P}_{n+1}(\hat{H})- b_{n+1} \mathcal{P}_{n} (\hat{H}),
\end{equation}
where we define
\begin{align}
\label{CF13}
    P_0(\hat{H})&= \hat{\mathbb{1}}\\
    P_{1}(\hat{H})&= \frac{1}{b_1}(\hat{H}- a_0 \hat{\mathbb{1}}).
\end{align}
From this, we read off the first three coefficients $c_{n,j}$ as
\begin{align}
\label{CF14}
    c_{0,0}&= 1,\ c_{1,0}= -\frac{a_0}{b_1},\ c_{1,1}= \frac{1}{b_1}.
\end{align}

In order to express $a_{n}$ and $b_{n}$ in terms of the coefficients $c_{n,j}$, we define
\begin{equation}
\label{CF15}
    \kappa_{n,j}= \sum_{l=0}^{\infty} c_{n,l} c_{n,j-l},
\end{equation}
where the sum only contains a finite number of terms because of Eq. (\ref{CF9}). Using Eq. (\ref{CF10}), we write $a_n$ for $n\geq 1$ in terms of the moments of $H$ as 
\begin{align}
\label{CF16}
    \nonumber a_{n}&=\Mel{u_0}{\sum_{j=0}^{2n}\kappa_{n,j}\hat{H}^{j+1}}{u_0}\\
    &= \sum_{j=0}^{2n} \kappa_{n,j} (H^{j+1})_{ii}.
\end{align}
Note that $a_0$ is given by
\begin{equation}
   a_0= \Mel{u_0}{\hat{H}}{u_0} = H_{ii},
\end{equation} 
and for adjacency matrices $A=-H$ that correspond to lattices without self-loops, we have $a_0=0$. Similarly, using Eq. (\ref{CF11}), we rewrite $b_n$ for $n\geq 2$ as 
\begin{align}
\label{CF17}
    \nonumber b_n^2&= \Mel{u_0}{\sum_{j=0}^{2(n-1)} \kappa_{n-1,j}\hat{H}^{j+2}}{u_0}-(a_{n-1}^2+b_{n-1}^2)\\ &=\sum_{j=0}^{2(n-1)} \kappa_{n-1,j} (H^{j+2})_{ii}- (a_{n-1}^2+b_{n-1}^2),
\end{align}
where $b_1$ is given by
\begin{equation}
    b_1=\sqrt{\Mel{u_0}{\hat{H}^2}{u_0}-a_0^2}=\sqrt{(H^2)_{ii}-a_0^2}.
\end{equation}
Equations (\ref{CF16}) and (\ref{CF17}) specify the CF coefficients if the $c_{n,j}$ are known. To find a recursion relation for $c_{n,j}$, we plug the parametrization (\ref{CF8}) of $P_n(\hat{H})$ into Eq. (\ref{CF12}) to arrive at
 \begin{align}
     \nonumber c_{n+2,m}&=\Biggl( - \frac{b_{n+1}}{b_{n+2}}c_{n,m}- \frac{a_{n+1}}{b_{n+2}} c_{n+1,m}\\ \nonumber &\phantom{sss}+\frac{c_{n+1,m-1}}{b_{n+2}} (1- \delta_{m,0})\Biggr) \Theta(n+1-m) \\ &\phantom{sss}+ \frac{c_{n+1,n+1}}{b_{n+2}}\delta_{n+2,m},
\end{align}
 where $\Theta(x)$ is the Heaviside step function defined such that $\Theta(x)=1 \iff x\geq 0$ else  $\Theta(x)=0$. The step function and Kronecker delta ensure that Eq. (\ref{CF9}) is satisfied.
 
To find the relation to the returning walk numbers, we insert $H=- A$ so that
 \begin{align}
 (H^j)_{ii} = (-1)^jS_j^{(i)} = (-1)^j S_j.
 \end{align}
Here we assumed that the returning walk numbers are independent of the site $i$ chosen, $S_n^{(i)}=S_n$, as is true for the Archimedean lattices. Hence the CF coefficients for $n\geq 0$ are given by
\begin{align}
a_{n+1}&=\sum_{j=0}^{2(n+1)} \kappa_{n+1,j} (-1)^{j+1}S_{j+1}, \\
b_{n+2}&= \Biggl(\sum_{j=0}^{2(n+1)}\kappa_{n+1,j}(-1)^{j+2} S_{j+2}-(a_{n+1}^2+b_{n+1}^2)\Biggr)^{1/2}.
\end{align}
Using these relations, we can start form the (known) returning walk numbers $\{S_n\}$ and the coefficients $c_{0,0}$, $c_{1,0}$, $c_{1,1}$ given in Eq. (\ref{CF14}), to recursively compute the numbers $\kappa_{n,j}$, $a_n$, $b_n$ for any $n$.

\section{DOS formula from returning walks for CaVO and SrCuBO lattices}\label{DOSExplicit}

In this appendix, we derive explicit expressions for the DOS $D(E)$ of the CaVO and SrCuBO lattices from the Bloch generating function of returning walks $\mathcal{G}(z,\bk)$ using Equation (\ref{DOSBOGF}).

For the CaVO lattice, we use from Tab. \ref{TabGF} that
\begin{equation}
    \mathcal{G}_{\rm Ca}(z,\bk)=\frac{1-3z^2-z^3(c_1+c_2)}{1-6z^2+z^4-4z^3(c_1+c_2)-4z^4c_1c_2},
\end{equation}
where $c_1:=\cos(k_1)$ and $c_2:=\cos(k_2)$. This yields
\begin{equation}
    \frac{\mathcal{G}_{\rm Ca}(-1/z,\bk)}{z}= \frac{c_1+c_2+z^3-3z}{4 z(c_1+c_2)-4 c_1 c_2+z^4-6z^2+1}.
    \label{EDOS1}
\end{equation}
Equation (\ref{EDOS1}) can be rewritten in the form  
\begin{equation}
    \frac{c_1+\alpha}{4 z(c_1+c_2)-4 c_1 c_2+\beta }+\frac{c_2+\alpha}{4 z(c_1+c_2)-4 c_1 c_2+\beta },
\end{equation}
where $\alpha= \frac{z^3-3z}{2}$ and $\beta=z^4-6z^2+1$. Exploiting the fact that $\int_{\bk}\mathcal{G}_{\rm Ca}(-1/z,\bk)/z$ is symmetric under the exchange $c_1\leftrightarrow c_2$, we have 
\begin{align}
   \nonumber &\frac{1}{\pi}\int_{\bk} \frac{\mathcal{G}_{\rm Ca}(-1/z,\bk)}{z}\\
   \nonumber &=\frac{2}{\pi} \int_{\bk} \frac{c_2+\frac{z^3-3z}{2}}{4z(c_1+c_2)-4c_1c_2+z^4-6z^2+1}\\ 
    &=\frac{2}{\pi} \int_{\bk} \frac{c_2+\frac{z^3-3z}{2}}{(1-6 z^2+z^4+4 z c_2)+4(z-c_2)c_1}.
\end{align}
Making use of Eq. (\ref{uv39}), we integrate over $k_1$ to obtain 
\begin{equation}
    \frac{2}{2\pi^2}\int_{0}^{2\pi} \mbox{d}k_2\frac{c_2+\frac{z^3-3z}{2}}{\sqrt{(1-6z^2+z^4+4z c_2)^2-16(z-c_2)^2}}.
\end{equation}
Using the change of variables $u=c_1\implies \mbox{d}k_2= \frac{-1}{\sqrt{1-u^2}}\mbox{d}u$ combined with the fact that $\cos(-x)=\cos(x)$, we evaluate the integral over the interval $[-\pi,\pi]$ to get
\begin{equation}
\frac{2}{\pi^2}\int_{-1}^{1}\frac{\mbox{d}u}{\sqrt{1-u^2}}\frac{u+\frac{z^3-3z}{2}}{\sqrt{(1-6z^2+z^4+4z\ u)^2-16(z-u)^2}}.
\end{equation}
Now we set $z=E- \ii \eta$ and extract the imaginary part when $\eta\to 0$ whilst making use of the identity 
\begin{align}
\label{imId}
    \nonumber&\int_{-1}^{1} \mbox{d}u\lim_{\eta\to 0}\mathfrak{I} \frac{P(z,u)}{\sqrt{Q(z,u)}}\Biggr|_{z=E-\ii \eta}\\
    \nonumber &=\int_{-1}^1\mbox{d}u \frac{|P(E,u)|}{\sqrt{|Q(E,u)|}}\sin\left(\frac{\arg(Q(E,u))}{2}\right)\\
    &=\int_{-1}^1\mbox{d}u \frac{|P(E,u)|}{\sqrt{|Q(E,u)|}} \Theta(- Q(E,u))
\end{align}
and use the fact that the argument of a real number is either $0$ or $\pi$. Therefore,
\begin{equation}
    \sin\left(\frac{\arg(x)}{2}\right)=\begin{cases}
    0 & x>0 \\ 
    1 & x<0
\end{cases}:=\Theta(-x).
\end{equation}
Thus, we arrive at the CaVO-lattice DOS
\begin{equation}
    \label{EDOS2}    D_{\rm Ca}(E)=\frac{2}{\pi^2}\int_{-1}^{1}\mbox{d}u \frac{|P(E,u)|}{\sqrt{|Q(E,u)|}} \Theta(- Q(E,u)),
\end{equation} 
with
\begin{align}
    \nonumber P(z,u)&=u+\frac{z^3-3z}{2},\\
    \label{EDOS2b} Q(z,u)&=(1-u^2)((1-6z^2+z^4+4z\ u)^2-16(z-u)^2).
\end{align}

For the SrCuBO lattice, we start with $\mathcal{G}_{\rm sr}(z,\bk)$ from Tab. \ref{TabGF} and transform the integrals over $\mbox{d}k_i$ in $\int_{\bk}\mathcal{G}_{\rm Sr}(-1/z,\bk)/z$ into the form 
\begin{equation}
    I(a,b,c):=\frac{\alpha}{2\pi}\int_{0}^{2\pi}\frac{a+\cos(k_1)}{c+b\cos (k_1)+\cos(k_1)^2} \mbox{d}k_1.
\end{equation}
This transformation allows us to use partial fraction decomposition to formally express the integral as 
\begin{equation}
    I(a,b,c)= \frac{\alpha}{2\pi}\int_{0}^{2\pi}\left( \frac{\frac{r_++a}{r_+-r_-}}{\cos(k_1)-r_+}- \frac{\frac{r_-+a}{r_+-r_-}}{\cos(k_1)-r_-}\right) \mbox{d}k_1,
\end{equation}
which, through Eq. (\ref{uv39}), becomes
\begin{equation}
    I(a,b,c)= -\alpha\left(\frac{J(a,b,r_+)-J(a,b,r_-)}{r_+-r_-}\right).
\end{equation}
Here we define
\begin{align}
    \nonumber J(a,b,r_\pm)&:= \frac{r_\pm+a}{\sqrt{r_\pm^2-1}},\\
    \nonumber r_{\pm}&=\frac{-b\pm 2\sqrt{\Delta}}{2},\\
    \Delta&= \frac{b^2- 4 c}{4}.
\end{align}
In our case, we identify
\begin{align}
    \nonumber &\alpha=\frac{1}{2} \left(c_2-z+1\right),\\
    \nonumber &a=\frac{z^3-2 c_2 (z-1)-5 z+2}{2 c_2-2 z+2},\\
    \nonumber &b=2 c_2 (z-2)-(z-1)^2,\\
    \nonumber &c=c_2^2-c_2 (z-1)^2+\frac{1}{4} \left(z \left(z^3-10 z+8\right)+1\right),\\
    \nonumber &\Delta =(1+c_2)(c_2-z)(z-3)(z-1),\\
    &r_+-r_-= 2\sqrt{\Delta}.
\end{align}
This yields 
\begin{align}
    \nonumber\alpha \frac{J(a,b,r_+)}{r_+-r_-}&=\frac{1}{4}\frac{X_{+}}{\sqrt{Y_{+}}},\\
    \alpha \frac{J(a,b,r_-)}{r_+-r_-}&=\frac{1}{4} \frac{X_-}{\sqrt{Y_{-}}},
\end{align}
where
\begin{align}
  \nonumber X_{\pm}&=\left(c_2+1\right) \left(-2 c_2 (z-2)+3 z^2-8 z+3\right)\\\nonumber  &\phantom{-}\pm2\sqrt{\Delta } \left( c_2- z+1\right),\\
    \nonumber Y_{\pm}&=\Delta\left(-2 c_2 (z-2)\pm2 \sqrt{\Delta }+(z-2) z-1\right)\\ \nonumber &\phantom{\Delta}\times \left(-2 c_2 (z-2)\pm2 \sqrt{\Delta }+(z-2) z+3\right).
\end{align}
Under the change of variables $u=c_2$ and using the fact that cosine is even, we use Eq. (\ref{imId}) and set $z=E$ to deduce
\begin{align}
    \nonumber D_{\rm Sr}(E)=\frac{1}{\pi^2}\Biggl({}&\int_{-1}^1 \mbox{d}u \frac{|P_{+}(E,u)|}{\sqrt{|Q_{+}(E,u)|}}\Theta\left(-Q_+(E,u)\right)\\ 
    \label{EDOS3} &+ \int_{-1}^1 \mbox{d}u \frac{|P_{-}(E,u)|}{\sqrt{|Q_{-}(E,u)|}}\Theta\left(-Q_-(E,u)\right)\Biggr), 
\end{align}
where 
\begin{equation}\label{EDOS3b}
\begin{aligned}
 P_\pm(E,u)&=-\frac{1}{4}X_{\pm}\Bigr|_{z\to E,c_2\to u},\\
 Q_{\pm}(E,u)&=(1-u^2)Y_{\pm}\Bigr|_{z\to E,c_2\to u}.
\end{aligned}
\end{equation}

\section{Derivation of asymptotic expansion coefficient $\alpha$}\label{asympApp}

\subsection{Non-bipartite lattices}

We first consider non-bipartite Archimedean lattices. We assume real $z$ such that $|z|<q^{-1}$. In order to obtain the numerator and denominator in
\begin{align}
 \mathcal{G}(z,\bk) = \frac{P(z,\bk)}{Q(z,\bk)},
\end{align}
we use Eq. (\ref{gf3}) to write
\begin{equation}
    \label{as5}
    \G(z,\bk)=\frac{1}{N_\uc}\tr{\frac{1}{\hat{\mathbbm{1}}-z \hat{A}(\bk)}}= \frac{\tr{{\rm adj}(\hat{\mathbbm{1}}-z \hat{A}(\bk))}}{N_\uc\det(\hat{\mathbbm{1}}-z \hat{A}(\bk))}
\end{equation}
and identify
\begin{align}
 Q(z,\bk)= \det\Bigl(\hat{\mathbbm{1}}- z \hat{A}(\bk)\Bigr).
\end{align}
The function $Q(z,\bk)$ is positive, 
\begin{align}
\text{(i)}:\ Q(z,\bk)\geq 0,\label{as4a}
\end{align}
as is shown in Sec. \ref{asympIneq}, and it vanishes for $(z_{\rm c},\bk_{\rm c})$ with $z_{\rm c}=q^{-1}$ and $\bk_{\rm c}=\vec{0}$. At this point, we have
\begin{align}
    \text{(ii)}:\ &Q(z_{\rm c},\vec{0})=0,\label{as4b}\\
    \text{(iii)}:\ &\partial_zQ(z_{\rm c},\vec{0})<0, \label{as4c}
\end{align}
as shown in Sec. \ref{asympIneq}. For non-bipartite Archimedean lattices, $z_{\rm c}$ is the only point in the interval $|z|\leq q^{-1}$ where $\mathcal{G}(z)$ diverges, and constitutes the radius of convergence.

For $z$ close to the critical point, we have
\begin{equation}
    \label{as1}
    \mathcal{G}(z) = \int_{\bk}\mathcal{G}(z,\bk)\sim -\alpha\ \ln\left(1-\frac{z}{z_{\rm c}}\right).
\end{equation}
Here  $f(z)\sim g(z)$ for two functions $f(z)$ and $g(z)$ means that they are equal up to analytic additions, i.e. have the same singular behavior. The function
\begin{align}
 f(z) = -\alpha\ \ln\left(1-\frac{z}{z_{\rm c}}\right)
\end{align}
is analytical in the interval $|z|<q^{-1}$, with the the power series expansion
\begin{align}
 -\log\Bigl(1-\frac{z}{z_{\rm c}}\Bigr) &= \sum_{n\geq 1} \frac{1}{n} \Bigl(\frac{z}{z_{\rm c}}\Bigr)^n,\\
 \Rightarrow\ 
 -[z^n]\log\Bigl(1-\frac{z}{z_{\rm c}}\Bigr) &= \frac{1}{nz_{\rm c}^n}.
 \end{align}
Hence Eq. (\ref{as1}) implies that the asymptotics for the returning walk numbers of non-bipartite lattices are governed by 
\begin{equation}
\label{as2}
    S_n \sim  \frac{\alpha}{n z_{\rm c}^n}
\end{equation}
for large $n$.

To show Eq. (\ref{as1}), we expand $Q(z,\bk)$ about its minimum located at $z=z_{\rm c}$ and $\bk=\0$ and use the fact that the gradient term $\nabla_\bk Q(z_c,\0)$ vanishes at an extremum. We find that 
\begin{equation}
    \label{as6}
    \int_{\bk}\G(z,\bk)\sim \int_{\bk} \frac{P(z_{\rm c},\0)}{\partial_zQ(z_{\rm c},\0)(z-z_{\rm c})+\frac{1}{2}\bk^T Q^{(2)}(z_{\rm c},\0)\bk}.
\end{equation}
Since $(z_{\rm c},\0)$ is a minimum, the Hessian matrix is positive semi-definite. This allows us to perform the coordinate change $\bh= \sqrt{Q^{(2)}(z_{\rm c},\0))}\bk$, which yields
\begin{equation}
\label{as7}
    \frac{1}{2}\bk^T Q^{(2)}(z_{\rm c},\0)\bk=\frac{1}{2}\bh^T \bh=\frac{1}{2}|\bh|^2:=\frac{1}{2}h^2.
\end{equation}
Integrating with respect to $\bh$ in a disk of sufficiently small radius $\Lambda$ around $\bk=0$ using polar coordinates, we uncover the logarithmic divergence in Eq. (\ref{as1}):
\begin{align}
\label{as8}
\nonumber \int_{\bk}\G(z,\bk)&\sim \int_{\bh}\frac{P(z_{\rm c},\0)}{\mbox{det}(\sqrt{Q^{(2)}})}\frac{1}{\zeta(z)+\frac{1}{2}|\bh|^2}\\  \nonumber 
\nonumber &\sim \frac{P(z_{\rm c},\0)}{\sqrt{\mbox{det}Q^{(2)}}} \frac{2\pi}{(2\pi)^2}\int_{0}^{\Lambda}\mbox{d}h  \frac{h}{\zeta(z)+\frac{1}{2}h^2}\\
\nonumber &= \frac{P(z_{\rm c},\0)}{2\pi\sqrt{\mbox{det}Q^{(2)}}}\int_{0}^{\Lambda^2/2}\mbox{d}\eta  \frac{1}{\zeta(z)+\eta}\\
 &= \frac{P(z_{\rm c},\0)}{2\pi\sqrt{\mbox{det}Q^{(2)}}}\ln \left(\frac{\zeta(z)+\Lambda^2/2}{\zeta(z)}\right).
\end{align}
Here we abbreviate $\zeta(z)=(z-z_{\rm c}) \partial_zQ(z_{\rm c},\0)$ and $Q^{(2)}=Q^{(2)}(z_{\rm c},\0)$. Since the Hessian is positive semi-definite, we have $\det{(\sqrt{Q^{(2)}})}=\sqrt{\det Q^{(2)}}$. We further have $\partial_z Q(z_{\rm c},\0)<0$, which together with $z\leq z_{\rm c}$ yields $\zeta(z)\geq 0$.  We arrive at
\begin{align}
\nonumber \int_{\bk}\G(z,\bk) &\sim \frac{P(z_{\rm c},\0)}{2\pi \sqrt{\det Q^{(2)}}}\ln\left( \frac{\Lambda^2}{|z-z_{\rm c}|\cdot|\partial_z Q(z_{\rm c},\0)|}\right)\\
& \sim -\frac{P(z_{\rm c},\0)}{2\pi \sqrt{\det Q^{(2)}}} \ln\left(1-\frac{z}{z_{\rm c}}\right).
\end{align} 
This shows the validity of Eq. (\ref{as1}).

\subsection{Bipartite lattices}

In this section, we consider bipartite Archimedean lattices. There, $S_n$ vanishes for odd $n$ because $\mathcal{G}(z)$ is an analytic function of $z^2$, not $z$.

As an example, consider the bipartite Honeycomb lattice with
\begin{align}
 \mathcal{G}(z) = \int_{\bk} \frac{1}{1-z^2(3-\vare_\Delta(\bk))}.
\end{align}
The generating function is a function of $z^2$. The integrand diverges for $(z^2,\bk)=(z_{\rm c}^2,\vec{0})$ with $z_{\rm c}=\frac{1}{q}=\frac{1}{3}$. Assuming $z\lesssim z_{\rm c}$, we expand the denominator about this point to obtain
\begin{align}
 \mathcal{G}(z\to z_{\rm c}) &\sim \int_{\bk} \frac{1}{1-9z^2 + \frac{1}{2}\bk^TQ^{(2)}\bk}\\
 &\sim\int_{\vec{h}} \frac{1}{\sqrt{\mbox{det} Q^{(2)}}} \frac{1}{1-\frac{z^2}{z_{\rm c}^2}+\frac{1}{2}\vec{h}^2}\\
  &\sim \frac{1}{2\pi \sqrt{\mbox{det}Q^{(2)}}} \ln\Bigl(\frac{1-\frac{z^2}{z_{\rm c}^2}+\Lambda^2/2}{1-\frac{z^2}{z_{\rm c}^2}}\Bigr)\\
  &\sim - \frac{1}{2\pi \sqrt{\mbox{det}Q^{(2)}}}  \log\Bigl(1-\frac{z^2}{z_{\rm c}^2}\Bigr).
\end{align}
Using the power series representation of the logarithm
\begin{align}
 -\log\Bigl(1-\frac{z^2}{z_{\rm c}^2}\Bigr) &= \sum_{n\geq 1} \frac{1}{n} \Bigl(\frac{z^2}{z_{\rm c}^2}\Bigr)^n,\\
 \label{logexp} \Rightarrow\ -[z^{2n}]\log\Bigl(1-\frac{z^2}{z_{\rm c}^2}\Bigr) &= \frac{1}{nz_{\rm c}^{2n}} = \frac{2}{(2n)z_{\rm c}^{2n}},
\end{align}
we arrive at
\begin{align}
 S_{2n} = [z^{2n}]\mathcal{G}(z) &\sim \frac{2}{2\pi \sqrt{\mbox{det}Q^{(2)}}} \frac{q^{2n}}{2n} = \frac{3\sqrt{3}}{2\pi} \cdot \frac{q^{2n}}{2n}.
\end{align}

As another example, consider the bipartite Square lattice with
\begin{align}
 \label{sq1xx} \mathcal{G}(z) = \int_{\bk} \frac{1}{1+z\vare_\square(\bk)} = \int_{\bk} \frac{P(z,\bk)}{Q(z,\bk)}.
\end{align}
We have $Q(z,\bk)=1+z\vare_{\square}(\bk)$, which yields $\mbox{det} Q^{(2)}=\frac{1}{4}$ and
\begin{align}
  \alpha=2\times \frac{1}{2\pi\sqrt{\mbox{det} Q^{(2)}}} = \frac{2}{\pi}.
\end{align}
This is the correct coefficient of $S_{2n} \sim \frac{2}{\pi n} \frac{q^{2n}}{2n}$, which can be obtained readily using $S_{2n} = \binom{2n}{n}^2$ and Stirling's formula. However, in the representation of Eq. (\ref{sq1xx}), the dependence of $\mathcal{G}(z)$ on $z^2$ is not manifest. To achieve this, we symmetrize the integrand in $z$ and find
\begin{align}
 \label{sq2xx} \mathcal{G}(z) = \int_{\bk} \frac{1}{1-z^2\vare_\square(\bk)^2} = \int_{\bk} \frac{\bar{P}(z^2,\bk)}{\bar{Q}(z^2,\bk)}.
\end{align}
The denominator $\bar{Q}(z,\bk)=1-z^2\vare_\square^2(\vec{k})$ not only vanishes for $(z_{\rm c},\vec{0})$ where $\vare_{\square}(\vec{0})=-q$, but also for $(z_{\rm c},\vec{\pi})$ with $\vec{\pi}=(\pi,\pi)$, since $\vare_\square(\vec{\pi})=+q$. Thus the integrand receives divergent contributions at both critical values $\vec{k}_{\rm c}=\vec{0},\vec{\pi}$ for $z\to z_{\rm c}$. For $z\to z_{\rm c}$, we have
\begin{align}
 \nonumber \mathcal{G}(z) &\sim \sum_{\vec{k}_{\rm c}=\vec{0},\vec{\pi}} \int_{\bk\approx \vec{0}} \frac{1}{1-16z^2+\frac{1}{2}\vec{k}^T\bar{Q}^{(2)}(z_{\rm c}^2,\vec{k}_{\rm c})\vec{k}}\\
 &\sim\sum_{\vec{k}_{\rm c}=\vec{0},\vec{\pi}} \frac{-1}{2\pi\sqrt{\mbox{det} \bar{Q}^{(2)}(z_{\rm c}^2,\bk_{\rm c})}} \log\Bigl(1-\frac{z^2}{z_{\rm c}^2}\Bigr).
\end{align} 
We have
\begin{align}
\mbox{det} \bar{Q}^{(2)}(z_{\rm c}^2,\vec{0}) = \mbox{det} \bar{Q}^{(2)}(z_{\rm c}^2,\vec{\pi}) =1
\end{align}
and thus arrive at
\begin{align}
 \nonumber \mathcal{G}(z) &\sim\sum_{\vec{k}_{\rm c}=\vec{0},\vec{\pi}} \frac{-1}{2\pi} \log\Bigl(1-\frac{z^2}{z_{\rm c}^2}\Bigr)\\
  &= 2\times \frac{-1}{2\pi} \log\Bigl(1-\frac{z^2}{z_{\rm c}^2}\Bigr).
\end{align} 
Using the power series of the logarithm according to Eq. (\ref{logexp}) eventually gives the result $S_{2n} \sim \frac{2}{\pi } \frac{q^{2n}}{2n}$ derived before.

For general bipartite lattices, we write
\begin{align}
 \nonumber \mathcal{G}(z)  &=\sum_{n\geq 0} S_{2n} z^{2n}\\
 \nonumber &= \frac{1}{2}\int_{\bk}\Bigl( \mathcal{G}(z,\bk)+\mathcal{G}(-z,\bk)\Bigr)\\
 \nonumber &= \frac{1}{2N_{\uc}} \int_{\bk} \Biggl[\tr{\frac{1}{\mathbb{1}-z A(\bk)}}+\tr{\frac{1}{\mathbb{1}+z A(\bk)}}\Biggr]\\
 \nonumber &= \frac{1}{N_{\uc}} \int_{\bk} \tr{\frac{1}{\mathbb{1}-z^2 A(\bk)^2}}\\
 &= \int_{\vec{k}} \frac{\bar{P}(z^2,\bk)}{\bar{Q}(z^2,\bk)},
\end{align}
and assume that the denominator vanishes at the points $(z_{\rm c}^2,\vec{k}_{\rm c})$. We then expand
\begin{align}
 \mathcal{G}(z) &\sim \sum_{\vec{k}_{\rm c}} \int_{\bk\approx \vec{0}} \frac{\bar{P}(z_{\rm c}^2,\bk_{\rm c})}{\bar{Q}(z_{\rm c}^2,\bk_{\rm c})+\frac{1}{2}\vec{k}^T\bar{Q}^{(2)}(z_{\rm c}^2,\vec{k}_{\rm c})\vec{k}}\\
 &\sim\sum_{\vec{k}_{\rm c}} \frac{-\bar{P}(z_{\rm c}^2,\bk)}{2\pi\sqrt{\mbox{det} \bar{Q}^{(2)}(z_{\rm c}^2,\bk_{\rm c})}} \log\Bigl(1-\frac{z^2}{z_{\rm c}^2}\Bigr).
\end{align} 
This gives the asymptotic coefficient of
\begin{align}
 S_{2n} \sim \alpha \frac{q^{2n}}{2n}
\end{align}
as
\begin{align}
 \alpha = \sum_{\vec{k}_{\rm c}}\frac{\bar{P}(z_{\rm c}^2,\bk)}{\pi\sqrt{\mbox{det} \bar{Q}^{(2)}(z_{\rm c}^2,\bk_{\rm c})}}.
\end{align}
We confirm for the four bipartite Archimedean lattices that this agrees with
\begin{align}
 \alpha = 2\times \frac{P(z,\vec{0})}{2\pi\sqrt{\mbox{det} Q^{(2)}(z_{\rm c},\vec{0})}},
\end{align}
although this agreement might be peculiar to the Archimedean lattices and might not necessarily generalize to other periodic tilings. We note that for the CaVO lattice we have $\vec{k}_{\rm c}=\vec{0},\vec{\pi}$, whereas for the SHD lattice $\vec{k}_{\rm c}=\vec{0}$.

\subsection{Inequalities}\label{asympIneq}

In this section we prove Eqs. (\ref{as4a},\ref{as4b},\ref{as4c}). First, we prove non-negativity of $Q(z,\bk)$. Without loss of generality assume that $0\leq z \leq 1/q$. Since  
\begin{equation}
    Q(z,\bk)= N_{\uc}\det(\mathbbm{1}- z \hat{A}(\bk)),
\end{equation}
we can rewrite the determinant in terms of the eigenvalues $\{\lambda_{i_\uc}(\bk)\}$ of $\hat{A}(\bk)$ as
\begin{equation}
\label{asapp1}
    \det(\hat{\mathbbm{1}}- z \hat{A}(\bk))=\prod_{i_\uc=1}^{N_\uc}(1-z \lambda_{i_\uc}(\bk)).
\end{equation}
Now since $\lambda_{i_\uc}(\bk)$ is an eigenvalue which lies within the spectral radius $q$, we have for all $\bk$ in the Brillouin Zone,
\begin{equation}
-q\leq \lambda_{i_\uc}(\bk)\leq q,
\end{equation}
where $\lambda_{i_\uc}(\bk) \leq q$ is a tight upper bound. This implies that 
\begin{align}
    &-q \leq \lambda_{i_\uc}(\bk) \leq q\\
    &z q \geq -z \lambda_{i_\uc}(\bk) \geq -z q\\
    &1+ z q \geq 1-z \lambda_{i_\uc}(\bk) \geq 1- zq\geq 0,
\end{align}
where we used $z \leq \frac{1}{q}\implies 1-z q\geq 0$ in the last line. Since each $1-z \lambda_{i_\uc}(\bk)$ is non-negative (even for $-1/q \geq z > 0$ where the proof follows identically), their product is as well and therefore we obtain $Q(z,\bk)\geq 0$ for $|z|<1/q \quad \square$.

To prove Eq. (\ref{as4b}), we observe that the matrix $A(\0)$ is the adjacency matrix of the unit cell with PBC. Here, if the non-zero entries of $A(\0)_{ij}=m$ where $m\geq 2$, then the site $i$ connects to $j$ in two different ways (as it is in the case of Trellis, SrCuBO, Kagome and Honeycomb in Sec. \ref{SecBlochA}). Since each site connects to $q$ neighbors, the vector $\textbf{1}\in\mathbb{R}^{N_\uc}$ which consists of entries $\textbf{1}_{i}=\frac{1}{\sqrt{N_{\uc}}}$ for $i=1,2,3,\dots N_{\uc}$ is an eigenvector of $A(\0)$ with eigenvalue $q$ due to the fact that the sum of the rows adds up to the coordination number. That is, 
\begin{equation}
    A(\0)\textbf{1}=q \textbf{1}.
\end{equation}
This means that there exists one $\lambda_{i_\uc^{\prime}}(\0)=q$, allowing us to rewrite Eq. (\ref{asapp1}) as 
\begin{equation}
    \det(\hat{\mathbbm{1}}-z \hat{A}(\0))= (1-z q)\prod_{i_{\uc}\neq i^{\prime}_{\uc}} (1-z \lambda_{i_\uc}(\0)),
\end{equation}
which implies that $Q\left(z=\frac{1}{q},\0\right)=0 \quad \square$.

To prove the last equality Eq. (\ref{as7}), we use the Jacobi formula given by
\begin{equation}
\label{jacobi1}
    \mbox{d}(\det(M))=\det(M) \tr{M^{-1} \mbox{d}M}.
\end{equation}
Here \begin{equation}
    \mbox{d}M(z)\equiv \frac{\partial}{\partial z}M(z),
\end{equation}
and
\begin{equation}
\label{jacobi2}
    M(z)= \mathbbm{1}-z A(\0).
\end{equation}
Equations (\ref{jacobi1}) and (\ref{jacobi2}) imply that
\begin{equation}
\label{jacobi3}
    \partial_z Q(z,\0)= - Q(z,\0) \tr{[\mathbbm{1}-z A(\0)]^{-1} A(\0)}.
\end{equation}
Expanding Eq. (\ref{jacobi3}) in the eigenbasis of $A(\0)$ within the unit cell and using the fact that $\lambda_{i_\uc^{\prime}}(\0)=q$ is an eigenvalue, we have
\begin{align}
    \nonumber
    \partial_z Q(z,\0)&= - (1-z q) \frac{q}{(1-z q)}\prod_{i_{\uc}\neq i^{\prime}_{\uc}} (1-z\lambda_{i_{\uc}}(\0))\\
    &\phantom{=}+(1-z q) \Psi(z), \label{jacobi4}
\end{align}
where $|\Psi\left(z=\frac{1}{q}\right)|<\infty$ and $\prod_{i_{\uc}\neq i^{\prime}_{\uc}} (1-\frac{1}{q}\lambda_{i_{\uc}}(\0))> 0$. Performing the limit $z\to 1/q$, we obtain
\begin{equation}
    \partial_z Q(z_c,\0)< 0 \quad \square.
\end{equation}

\end{appendix}

\bibliography{refs_walks}

\begin{thebibliography}{79}%
\makeatletter
\providecommand \@ifxundefined [1]{%
 \@ifx{#1\undefined}
}%
\providecommand \@ifnum [1]{%
 \ifnum #1\expandafter \@firstoftwo
 \else \expandafter \@secondoftwo
 \fi
}%
\providecommand \@ifx [1]{%
 \ifx #1\expandafter \@firstoftwo
 \else \expandafter \@secondoftwo
 \fi
}%
\providecommand \natexlab [1]{#1}%
\providecommand \enquote  [1]{``#1''}%
\providecommand \bibnamefont  [1]{#1}%
\providecommand \bibfnamefont [1]{#1}%
\providecommand \citenamefont [1]{#1}%
\providecommand \href@noop [0]{\@secondoftwo}%
\providecommand \href [0]{\begingroup \@sanitize@url \@href}%
\providecommand \@href[1]{\@@startlink{#1}\@@href}%
\providecommand \@@href[1]{\endgroup#1\@@endlink}%
\providecommand \@sanitize@url [0]{\catcode `\\12\catcode `\$12\catcode
  `\&12\catcode `\#12\catcode `\^12\catcode `\_12\catcode `\%12\relax}%
\providecommand \@@startlink[1]{}%
\providecommand \@@endlink[0]{}%
\providecommand \url  [0]{\begingroup\@sanitize@url \@url }%
\providecommand \@url [1]{\endgroup\@href {#1}{\urlprefix }}%
\providecommand \urlprefix  [0]{URL }%
\providecommand \Eprint [0]{\href }%
\providecommand \doibase [0]{https://doi.org/}%
\providecommand \selectlanguage [0]{\@gobble}%
\providecommand \bibinfo  [0]{\@secondoftwo}%
\providecommand \bibfield  [0]{\@secondoftwo}%
\providecommand \translation [1]{[#1]}%
\providecommand \BibitemOpen [0]{}%
\providecommand \bibitemStop [0]{}%
\providecommand \bibitemNoStop [0]{.\EOS\space}%
\providecommand \EOS [0]{\spacefactor3000\relax}%
\providecommand \BibitemShut  [1]{\csname bibitem#1\endcsname}%
\let\auto@bib@innerbib\@empty
\bibitem [{\citenamefont {Wallace}(1947)}]{Wallace1947}%
  \BibitemOpen
  \bibfield  {author} {\bibinfo {author} {\bibfnamefont {P.~R.}\ \bibnamefont
  {Wallace}},\ }\bibfield  {title} {\bibinfo {title} {{The Band Theory of
  Graphite}},\ }\href {https://doi.org/10.1103/PhysRev.71.622} {\bibfield
  {journal} {\bibinfo  {journal} {Phys. Rev.}\ }\textbf {\bibinfo {volume}
  {71}},\ \bibinfo {pages} {622} (\bibinfo {year} {1947})}\BibitemShut
  {NoStop}%
\bibitem [{\citenamefont {Boehm}\ \emph {et~al.}(1962)\citenamefont {Boehm},
  \citenamefont {Clauss}, \citenamefont {Fischer},\ and\ \citenamefont
  {Hofmann}}]{Boehm1962}%
  \BibitemOpen
  \bibfield  {author} {\bibinfo {author} {\bibfnamefont {H.~P.}\ \bibnamefont
  {Boehm}}, \bibinfo {author} {\bibfnamefont {A.}~\bibnamefont {Clauss}},
  \bibinfo {author} {\bibfnamefont {G.~O.}\ \bibnamefont {Fischer}},\ and\
  \bibinfo {author} {\bibfnamefont {U.}~\bibnamefont {Hofmann}},\ }\bibfield
  {title} {\bibinfo {title} {{Das Adsorptionsverhalten sehr dünner
  Kohlenstoff-Folien}},\ }\href
  {https://doi.org/https://doi.org/10.1002/zaac.19623160303} {\bibfield
  {journal} {\bibinfo  {journal} {{Zeitschrift für anorganische und allgemeine
  Chemie}}\ }\textbf {\bibinfo {volume} {316}},\ \bibinfo {pages} {119}
  (\bibinfo {year} {1962})}\BibitemShut {NoStop}%
\bibitem [{\citenamefont {Boehm}\ \emph {et~al.}(1986)\citenamefont {Boehm},
  \citenamefont {Setton},\ and\ \citenamefont {Stumpp}}]{Boem1986}%
  \BibitemOpen
  \bibfield  {author} {\bibinfo {author} {\bibfnamefont {H.}~\bibnamefont
  {Boehm}}, \bibinfo {author} {\bibfnamefont {R.}~\bibnamefont {Setton}},\ and\
  \bibinfo {author} {\bibfnamefont {E.}~\bibnamefont {Stumpp}},\ }\bibfield
  {title} {\bibinfo {title} {Nomenclature and terminology of graphite
  intercalation compounds},\ }\href
  {https://doi.org/https://doi.org/10.1016/0008-6223(86)90126-0} {\bibfield
  {journal} {\bibinfo  {journal} {Carbon}\ }\textbf {\bibinfo {volume} {24}},\
  \bibinfo {pages} {241} (\bibinfo {year} {1986})}\BibitemShut {NoStop}%
\bibitem [{\citenamefont {Saito}\ \emph {et~al.}(1992)\citenamefont {Saito},
  \citenamefont {Fujita}, \citenamefont {Dresselhaus},\ and\ \citenamefont
  {Dresselhaus}}]{Saito1992}%
  \BibitemOpen
  \bibfield  {author} {\bibinfo {author} {\bibfnamefont {R.}~\bibnamefont
  {Saito}}, \bibinfo {author} {\bibfnamefont {M.}~\bibnamefont {Fujita}},
  \bibinfo {author} {\bibfnamefont {G.}~\bibnamefont {Dresselhaus}},\ and\
  \bibinfo {author} {\bibfnamefont {M.~S.}\ \bibnamefont {Dresselhaus}},\
  }\bibfield  {title} {\bibinfo {title} {Electronic structure of chiral
  graphene tubules},\ }\href {https://doi.org/10.1063/1.107080} {\bibfield
  {journal} {\bibinfo  {journal} {Appl. Phys. Lett.}\ }\textbf {\bibinfo
  {volume} {60}},\ \bibinfo {pages} {2204} (\bibinfo {year}
  {1992})}\BibitemShut {NoStop}%
\bibitem [{\citenamefont {Shioyama}(2001)}]{Shioyama2001}%
  \BibitemOpen
  \bibfield  {author} {\bibinfo {author} {\bibfnamefont {H.}~\bibnamefont
  {Shioyama}},\ }\bibfield  {title} {\bibinfo {title} {Cleavage of graphite to
  graphene},\ }\href {https://doi.org/10.1023/a:1010907928709} {\bibfield
  {journal} {\bibinfo  {journal} {Journal of Materials Science Letters}\
  }\textbf {\bibinfo {volume} {20}},\ \bibinfo {pages} {499–500} (\bibinfo
  {year} {2001})}\BibitemShut {NoStop}%
\bibitem [{\citenamefont {Reich}\ \emph {et~al.}(2002)\citenamefont {Reich},
  \citenamefont {Maultzsch}, \citenamefont {Thomsen},\ and\ \citenamefont
  {Ordej\'on}}]{Reich2002}%
  \BibitemOpen
  \bibfield  {author} {\bibinfo {author} {\bibfnamefont {S.}~\bibnamefont
  {Reich}}, \bibinfo {author} {\bibfnamefont {J.}~\bibnamefont {Maultzsch}},
  \bibinfo {author} {\bibfnamefont {C.}~\bibnamefont {Thomsen}},\ and\ \bibinfo
  {author} {\bibfnamefont {P.}~\bibnamefont {Ordej\'on}},\ }\bibfield  {title}
  {\bibinfo {title} {Tight-binding description of graphene},\ }\href
  {https://doi.org/10.1103/PhysRevB.66.035412} {\bibfield  {journal} {\bibinfo
  {journal} {Phys. Rev. B}\ }\textbf {\bibinfo {volume} {66}},\ \bibinfo
  {pages} {035412} (\bibinfo {year} {2002})}\BibitemShut {NoStop}%
\bibitem [{\citenamefont {Novoselov}\ \emph {et~al.}(2004)\citenamefont
  {Novoselov}, \citenamefont {Geim}, \citenamefont {Morozov}, \citenamefont
  {Jiang}, \citenamefont {Zhang}, \citenamefont {Dubonos}, \citenamefont
  {Grigorieva},\ and\ \citenamefont {Firsov}}]{Geim2004}%
  \BibitemOpen
  \bibfield  {author} {\bibinfo {author} {\bibfnamefont {K.~S.}\ \bibnamefont
  {Novoselov}}, \bibinfo {author} {\bibfnamefont {A.~K.}\ \bibnamefont {Geim}},
  \bibinfo {author} {\bibfnamefont {S.~V.}\ \bibnamefont {Morozov}}, \bibinfo
  {author} {\bibfnamefont {D.}~\bibnamefont {Jiang}}, \bibinfo {author}
  {\bibfnamefont {Y.}~\bibnamefont {Zhang}}, \bibinfo {author} {\bibfnamefont
  {S.~V.}\ \bibnamefont {Dubonos}}, \bibinfo {author} {\bibfnamefont {I.~V.}\
  \bibnamefont {Grigorieva}},\ and\ \bibinfo {author} {\bibfnamefont {A.~A.}\
  \bibnamefont {Firsov}},\ }\bibfield  {title} {\bibinfo {title} {{Electric
  Field Effect in Atomically Thin Carbon Films}},\ }\href
  {https://doi.org/10.1126/science.1102896} {\bibfield  {journal} {\bibinfo
  {journal} {Science}\ }\textbf {\bibinfo {volume} {306}},\ \bibinfo {pages}
  {666} (\bibinfo {year} {2004})}\BibitemShut {NoStop}%
\bibitem [{\citenamefont {Geim}\ and\ \citenamefont
  {Novoselov}(2007)}]{Geim2007}%
  \BibitemOpen
  \bibfield  {author} {\bibinfo {author} {\bibfnamefont {A.~K.}\ \bibnamefont
  {Geim}}\ and\ \bibinfo {author} {\bibfnamefont {K.~S.}\ \bibnamefont
  {Novoselov}},\ }\bibfield  {title} {\bibinfo {title} {{The rise of
  graphene}},\ }\href {https://doi.org/10.1038/nmat1849} {\bibfield  {journal}
  {\bibinfo  {journal} {Nat. Mat.}\ }\textbf {\bibinfo {volume} {6}},\ \bibinfo
  {pages} {183–191} (\bibinfo {year} {2007})}\BibitemShut {NoStop}%
\bibitem [{\citenamefont {Mecklenburg}\ and\ \citenamefont
  {Regan}(2011)}]{Mecklenburg2011}%
  \BibitemOpen
  \bibfield  {author} {\bibinfo {author} {\bibfnamefont {M.}~\bibnamefont
  {Mecklenburg}}\ and\ \bibinfo {author} {\bibfnamefont {B.~C.}\ \bibnamefont
  {Regan}},\ }\bibfield  {title} {\bibinfo {title} {{Spin and the Honeycomb
  Lattice: Lessons from Graphene}},\ }\href
  {https://doi.org/10.1103/PhysRevLett.106.116803} {\bibfield  {journal}
  {\bibinfo  {journal} {Phys. Rev. Lett.}\ }\textbf {\bibinfo {volume} {106}},\
  \bibinfo {pages} {116803} (\bibinfo {year} {2011})}\BibitemShut {NoStop}%
\bibitem [{\citenamefont {Yang}\ \emph {et~al.}(2018)\citenamefont {Yang},
  \citenamefont {Li}, \citenamefont {Lee},\ and\ \citenamefont
  {Ng}}]{Yang2018}%
  \BibitemOpen
  \bibfield  {author} {\bibinfo {author} {\bibfnamefont {G.}~\bibnamefont
  {Yang}}, \bibinfo {author} {\bibfnamefont {L.}~\bibnamefont {Li}}, \bibinfo
  {author} {\bibfnamefont {W.~B.}\ \bibnamefont {Lee}},\ and\ \bibinfo {author}
  {\bibfnamefont {M.~C.}\ \bibnamefont {Ng}},\ }\bibfield  {title} {\bibinfo
  {title} {Structure of graphene and its disorders: a review},\ }\href
  {https://doi.org/10.1080/14686996.2018.1494493} {\bibfield  {journal}
  {\bibinfo  {journal} {Science and Technology of Advanced Materials}\ }\textbf
  {\bibinfo {volume} {19}},\ \bibinfo {pages} {613–648} (\bibinfo {year}
  {2018})}\BibitemShut {NoStop}%
\bibitem [{\citenamefont {{Sriram Shastry}}\ and\ \citenamefont
  {Sutherland}(1981)}]{SRIRAMSHASTRY19811069}%
  \BibitemOpen
  \bibfield  {author} {\bibinfo {author} {\bibfnamefont {B.}~\bibnamefont
  {{Sriram Shastry}}}\ and\ \bibinfo {author} {\bibfnamefont {B.}~\bibnamefont
  {Sutherland}},\ }\bibfield  {title} {\bibinfo {title} {Exact ground state of
  a quantum mechanical antiferromagnet},\ }\href
  {https://doi.org/https://doi.org/10.1016/0378-4363(81)90838-X} {\bibfield
  {journal} {\bibinfo  {journal} {Physica B+C}\ }\textbf {\bibinfo {volume}
  {108}},\ \bibinfo {pages} {1069} (\bibinfo {year} {1981})}\BibitemShut
  {NoStop}%
\bibitem [{\citenamefont {Kitaev}(2006)}]{Kitaev2006}%
  \BibitemOpen
  \bibfield  {author} {\bibinfo {author} {\bibfnamefont {A.}~\bibnamefont
  {Kitaev}},\ }\bibfield  {title} {\bibinfo {title} {Anyons in an exactly
  solved model and beyond},\ }\href {https://doi.org/10.1016/j.aop.2005.10.005}
  {\bibfield  {journal} {\bibinfo  {journal} {Annals of Phys.}\ }\textbf
  {\bibinfo {volume} {321}},\ \bibinfo {pages} {2–111} (\bibinfo {year}
  {2006})}\BibitemShut {NoStop}%
\bibitem [{\citenamefont {Song}\ \emph {et~al.}(2016)\citenamefont {Song},
  \citenamefont {You},\ and\ \citenamefont {Balents}}]{Song2016}%
  \BibitemOpen
  \bibfield  {author} {\bibinfo {author} {\bibfnamefont {X.~Y.}\ \bibnamefont
  {Song}}, \bibinfo {author} {\bibfnamefont {Y.~Z.}\ \bibnamefont {You}},\ and\
  \bibinfo {author} {\bibfnamefont {L.}~\bibnamefont {Balents}},\ }\bibfield
  {title} {\bibinfo {title} {{Low-Energy Spin Dynamics of the Honeycomb Spin
  Liquid Beyond the Kitaev Limit}},\ }\href
  {https://doi.org/10.1103/PhysRevLett.117.037209} {\bibfield  {journal}
  {\bibinfo  {journal} {Phys. Rev. Lett.}\ }\textbf {\bibinfo {volume} {117}},\
  \bibinfo {pages} {037209} (\bibinfo {year} {2016})}\BibitemShut {NoStop}%
\bibitem [{\citenamefont {Kos}\ and\ \citenamefont {Punk}(2017)}]{Kos2017}%
  \BibitemOpen
  \bibfield  {author} {\bibinfo {author} {\bibfnamefont {P.}~\bibnamefont
  {Kos}}\ and\ \bibinfo {author} {\bibfnamefont {M.}~\bibnamefont {Punk}},\
  }\bibfield  {title} {\bibinfo {title} {{Quantum spin liquid ground states of
  the Heisenberg-Kitaev model on the triangular lattice}},\ }\href
  {https://doi.org/10.1103/PhysRevB.95.024421} {\bibfield  {journal} {\bibinfo
  {journal} {Phys. Rev. B}\ }\textbf {\bibinfo {volume} {95}},\ \bibinfo
  {pages} {024421} (\bibinfo {year} {2017})}\BibitemShut {NoStop}%
\bibitem [{\citenamefont {Takagi}\ \emph {et~al.}(2019)\citenamefont {Takagi},
  \citenamefont {Takayama}, \citenamefont {Jackeli}, \citenamefont
  {Khaliullin},\ and\ \citenamefont {Nagler}}]{Takagi2019}%
  \BibitemOpen
  \bibfield  {author} {\bibinfo {author} {\bibfnamefont {H.}~\bibnamefont
  {Takagi}}, \bibinfo {author} {\bibfnamefont {T.}~\bibnamefont {Takayama}},
  \bibinfo {author} {\bibfnamefont {G.}~\bibnamefont {Jackeli}}, \bibinfo
  {author} {\bibfnamefont {G.}~\bibnamefont {Khaliullin}},\ and\ \bibinfo
  {author} {\bibfnamefont {S.~E.}\ \bibnamefont {Nagler}},\ }\bibfield  {title}
  {\bibinfo {title} {{Concept and realization of Kitaev quantum spin
  liquids}},\ }\href {https://doi.org/10.1038/s42254-019-0038-2} {\bibfield
  {journal} {\bibinfo  {journal} {Nat. Rev. Phys.}\ }\textbf {\bibinfo {volume}
  {1}},\ \bibinfo {pages} {264–280} (\bibinfo {year} {2019})}\BibitemShut
  {NoStop}%
\bibitem [{\citenamefont {Verresen}\ and\ \citenamefont
  {Vishwanath}(2022)}]{Verresen2022}%
  \BibitemOpen
  \bibfield  {author} {\bibinfo {author} {\bibfnamefont {R.}~\bibnamefont
  {Verresen}}\ and\ \bibinfo {author} {\bibfnamefont {A.}~\bibnamefont
  {Vishwanath}},\ }\bibfield  {title} {\bibinfo {title} {{Unifying Kitaev
  Magnets, Kagom\'e Dimer Models, and Ruby Rydberg Spin Liquids}},\ }\href
  {https://doi.org/10.1103/PhysRevX.12.041029} {\bibfield  {journal} {\bibinfo
  {journal} {Phys. Rev. X}\ }\textbf {\bibinfo {volume} {12}},\ \bibinfo
  {pages} {041029} (\bibinfo {year} {2022})}\BibitemShut {NoStop}%
\bibitem [{\citenamefont {Sonnenschein}\ \emph {et~al.}(2024)\citenamefont
  {Sonnenschein}, \citenamefont {Maity}, \citenamefont {Liu}, \citenamefont
  {Thomale}, \citenamefont {Ferrari},\ and\ \citenamefont
  {Iqbal}}]{PhysRevB.110.014414}%
  \BibitemOpen
  \bibfield  {author} {\bibinfo {author} {\bibfnamefont {J.}~\bibnamefont
  {Sonnenschein}}, \bibinfo {author} {\bibfnamefont {A.}~\bibnamefont {Maity}},
  \bibinfo {author} {\bibfnamefont {C.}~\bibnamefont {Liu}}, \bibinfo {author}
  {\bibfnamefont {R.}~\bibnamefont {Thomale}}, \bibinfo {author} {\bibfnamefont
  {F.}~\bibnamefont {Ferrari}},\ and\ \bibinfo {author} {\bibfnamefont
  {Y.}~\bibnamefont {Iqbal}},\ }\bibfield  {title} {\bibinfo {title}
  {{Candidate quantum spin liquids on the maple-leaf lattice}},\ }\href
  {https://doi.org/10.1103/PhysRevB.110.014414} {\bibfield  {journal} {\bibinfo
   {journal} {Phys. Rev. B}\ }\textbf {\bibinfo {volume} {110}},\ \bibinfo
  {pages} {014414} (\bibinfo {year} {2024})}\BibitemShut {NoStop}%
\bibitem [{\citenamefont {Koll{\'a}r}\ \emph {et~al.}(2019)\citenamefont
  {Koll{\'a}r}, \citenamefont {Fitzpatrick},\ and\ \citenamefont
  {Houck}}]{kollar2019hyperbolic}%
  \BibitemOpen
  \bibfield  {author} {\bibinfo {author} {\bibfnamefont {A.~J.}\ \bibnamefont
  {Koll{\'a}r}}, \bibinfo {author} {\bibfnamefont {M.}~\bibnamefont
  {Fitzpatrick}},\ and\ \bibinfo {author} {\bibfnamefont {A.~A.}\ \bibnamefont
  {Houck}},\ }\bibfield  {title} {\bibinfo {title} {{Hyperbolic lattices in
  circuit quantum electrodynamics}},\ }\href
  {https://doi.org/10.1038/s41586-019-1348-3} {\bibfield  {journal} {\bibinfo
  {journal} {Nature}\ }\textbf {\bibinfo {volume} {571}},\ \bibinfo {pages}
  {45} (\bibinfo {year} {2019})}\BibitemShut {NoStop}%
\bibitem [{\citenamefont {Koll{\'a}r}\ \emph {et~al.}(2020)\citenamefont
  {Koll{\'a}r}, \citenamefont {Fitzpatrick}, \citenamefont {Sarnak},\ and\
  \citenamefont {Houck}}]{kollar2019line}%
  \BibitemOpen
  \bibfield  {author} {\bibinfo {author} {\bibfnamefont {A.~J.}\ \bibnamefont
  {Koll{\'a}r}}, \bibinfo {author} {\bibfnamefont {M.}~\bibnamefont
  {Fitzpatrick}}, \bibinfo {author} {\bibfnamefont {P.}~\bibnamefont
  {Sarnak}},\ and\ \bibinfo {author} {\bibfnamefont {A.~A.}\ \bibnamefont
  {Houck}},\ }\bibfield  {title} {\bibinfo {title} {{Line-graph lattices:
  Euclidean and non-Euclidean flat bands, and implementations in circuit
  quantum electrodynamics}},\ }\href
  {https://doi.org/10.1007/s00220-019-03645-8} {\bibfield  {journal} {\bibinfo
  {journal} {Commun. Math. Phys.}\ }\textbf {\bibinfo {volume} {376}},\
  \bibinfo {pages} {1909} (\bibinfo {year} {2020})}\BibitemShut {NoStop}%
\bibitem [{\citenamefont {Boyle}\ \emph {et~al.}(2020)\citenamefont {Boyle},
  \citenamefont {Dickens},\ and\ \citenamefont {Flicker}}]{PhysRevX.10.011009}%
  \BibitemOpen
  \bibfield  {author} {\bibinfo {author} {\bibfnamefont {L.}~\bibnamefont
  {Boyle}}, \bibinfo {author} {\bibfnamefont {M.}~\bibnamefont {Dickens}},\
  and\ \bibinfo {author} {\bibfnamefont {F.}~\bibnamefont {Flicker}},\
  }\bibfield  {title} {\bibinfo {title} {Conformal quasicrystals and
  holography},\ }\href {https://doi.org/10.1103/PhysRevX.10.011009} {\bibfield
  {journal} {\bibinfo  {journal} {Phys. Rev. X}\ }\textbf {\bibinfo {volume}
  {10}},\ \bibinfo {pages} {011009} (\bibinfo {year} {2020})}\BibitemShut
  {NoStop}%
\bibitem [{\citenamefont {Yu}\ \emph {et~al.}(2020)\citenamefont {Yu},
  \citenamefont {Piao},\ and\ \citenamefont {Park}}]{Yu2020}%
  \BibitemOpen
  \bibfield  {author} {\bibinfo {author} {\bibfnamefont {S.}~\bibnamefont
  {Yu}}, \bibinfo {author} {\bibfnamefont {X.}~\bibnamefont {Piao}},\ and\
  \bibinfo {author} {\bibfnamefont {N.}~\bibnamefont {Park}},\ }\bibfield
  {title} {\bibinfo {title} {{{Topological Hyperbolic Lattices}}},\ }\href
  {https://link.aps.org/doi/10.1103/PhysRevLett.125.053901} {\bibfield
  {journal} {\bibinfo  {journal} {Phys. Rev. Lett.}\ }\textbf {\bibinfo
  {volume} {125}},\ \bibinfo {pages} {053901} (\bibinfo {year}
  {2020})}\BibitemShut {NoStop}%
\bibitem [{\citenamefont {Asaduzzaman}\ \emph {et~al.}(2020)\citenamefont
  {Asaduzzaman}, \citenamefont {Catterall}, \citenamefont {Hubisz},
  \citenamefont {Nelson},\ and\ \citenamefont
  {Unmuth-Yockey}}]{PhysRevD.102.034511}%
  \BibitemOpen
  \bibfield  {author} {\bibinfo {author} {\bibfnamefont {M.}~\bibnamefont
  {Asaduzzaman}}, \bibinfo {author} {\bibfnamefont {S.}~\bibnamefont
  {Catterall}}, \bibinfo {author} {\bibfnamefont {J.}~\bibnamefont {Hubisz}},
  \bibinfo {author} {\bibfnamefont {R.}~\bibnamefont {Nelson}},\ and\ \bibinfo
  {author} {\bibfnamefont {J.}~\bibnamefont {Unmuth-Yockey}},\ }\bibfield
  {title} {\bibinfo {title} {{Holography on tessellations of hyperbolic
  space}},\ }\href {https://doi.org/10.1103/PhysRevD.102.034511} {\bibfield
  {journal} {\bibinfo  {journal} {Phys. Rev. D}\ }\textbf {\bibinfo {volume}
  {102}},\ \bibinfo {pages} {034511} (\bibinfo {year} {2020})}\BibitemShut
  {NoStop}%
\bibitem [{\citenamefont {Boettcher}\ \emph {et~al.}(2020)\citenamefont
  {Boettcher}, \citenamefont {Bienias}, \citenamefont {Belyansky},
  \citenamefont {Koll\'ar},\ and\ \citenamefont
  {Gorshkov}}]{PhysRevA.102.032208}%
  \BibitemOpen
  \bibfield  {author} {\bibinfo {author} {\bibfnamefont {I.}~\bibnamefont
  {Boettcher}}, \bibinfo {author} {\bibfnamefont {P.}~\bibnamefont {Bienias}},
  \bibinfo {author} {\bibfnamefont {R.}~\bibnamefont {Belyansky}}, \bibinfo
  {author} {\bibfnamefont {A.~J.}\ \bibnamefont {Koll\'ar}},\ and\ \bibinfo
  {author} {\bibfnamefont {A.~V.}\ \bibnamefont {Gorshkov}},\ }\bibfield
  {title} {\bibinfo {title} {{Quantum simulation of hyperbolic space with
  circuit quantum electrodynamics: From graphs to geometry}},\ }\href
  {https://doi.org/10.1103/PhysRevA.102.032208} {\bibfield  {journal} {\bibinfo
   {journal} {Phys. Rev. A}\ }\textbf {\bibinfo {volume} {102}},\ \bibinfo
  {pages} {032208} (\bibinfo {year} {2020})}\BibitemShut {NoStop}%
\bibitem [{\citenamefont {Jahn}\ \emph {et~al.}(2020)\citenamefont {Jahn},
  \citenamefont {Zimbor{\'{a}}s},\ and\ \citenamefont {Eisert}}]{Jahn2020}%
  \BibitemOpen
  \bibfield  {author} {\bibinfo {author} {\bibfnamefont {A.}~\bibnamefont
  {Jahn}}, \bibinfo {author} {\bibfnamefont {Z.}~\bibnamefont
  {Zimbor{\'{a}}s}},\ and\ \bibinfo {author} {\bibfnamefont {J.}~\bibnamefont
  {Eisert}},\ }\bibfield  {title} {\bibinfo {title} {{Central charges of
  aperiodic holographic tensor-network models}},\ }\href
  {https://doi.org/10.1103/PhysRevA.102.042407} {\bibfield  {journal} {\bibinfo
   {journal} {Phys. Rev. A}\ }\textbf {\bibinfo {volume} {102}},\ \bibinfo
  {pages} {042407} (\bibinfo {year} {2020})}\BibitemShut {NoStop}%
\bibitem [{\citenamefont {Maciejko}\ and\ \citenamefont
  {Rayan}(2021)}]{maciejko2020hyperbolic}%
  \BibitemOpen
  \bibfield  {author} {\bibinfo {author} {\bibfnamefont {J.}~\bibnamefont
  {Maciejko}}\ and\ \bibinfo {author} {\bibfnamefont {S.}~\bibnamefont
  {Rayan}},\ }\bibfield  {title} {\bibinfo {title} {{Hyperbolic band theory}},\
  }\href {https://www.science.org/doi/10.1126/sciadv.abe9170} {\bibfield
  {journal} {\bibinfo  {journal} {Sci. Adv.}\ }\textbf {\bibinfo {volume} {7}}
  (\bibinfo {year} {2021})}\BibitemShut {NoStop}%
\bibitem [{\citenamefont {Brower}\ \emph {et~al.}(2021)\citenamefont {Brower},
  \citenamefont {Cogburn}, \citenamefont {Fitzpatrick}, \citenamefont
  {Howarth},\ and\ \citenamefont {Tan}}]{brower2021lattice}%
  \BibitemOpen
  \bibfield  {author} {\bibinfo {author} {\bibfnamefont {R.~C.}\ \bibnamefont
  {Brower}}, \bibinfo {author} {\bibfnamefont {C.~V.}\ \bibnamefont {Cogburn}},
  \bibinfo {author} {\bibfnamefont {A.~L.}\ \bibnamefont {Fitzpatrick}},
  \bibinfo {author} {\bibfnamefont {D.}~\bibnamefont {Howarth}},\ and\ \bibinfo
  {author} {\bibfnamefont {C.-I.}\ \bibnamefont {Tan}},\ }\bibfield  {title}
  {\bibinfo {title} {{Lattice setup for quantum field theory in
  ${\mathrm{AdS}}_{2}$}},\ }\href {https://doi.org/10.1103/PhysRevD.103.094507}
  {\bibfield  {journal} {\bibinfo  {journal} {Phys. Rev. D}\ }\textbf {\bibinfo
  {volume} {103}},\ \bibinfo {pages} {094507} (\bibinfo {year}
  {2021})}\BibitemShut {NoStop}%
\bibitem [{\citenamefont {Zhu}\ \emph {et~al.}(2021)\citenamefont {Zhu},
  \citenamefont {Guo}, \citenamefont {Breuckmann}, \citenamefont {Guo},\ and\
  \citenamefont {Feng}}]{Zhu:2021}%
  \BibitemOpen
  \bibfield  {author} {\bibinfo {author} {\bibfnamefont {X.}~\bibnamefont
  {Zhu}}, \bibinfo {author} {\bibfnamefont {J.}~\bibnamefont {Guo}}, \bibinfo
  {author} {\bibfnamefont {N.~P.}\ \bibnamefont {Breuckmann}}, \bibinfo
  {author} {\bibfnamefont {H.}~\bibnamefont {Guo}},\ and\ \bibinfo {author}
  {\bibfnamefont {S.}~\bibnamefont {Feng}},\ }\bibfield  {title} {\bibinfo
  {title} {Quantum phase transitions of interacting bosons on hyperbolic
  lattices},\ }\href {https://doi.org/10.1088/1361-648X/ac0a1a} {\bibfield
  {journal} {\bibinfo  {journal} {J. Phys.: Condens. Matter}\ }\textbf
  {\bibinfo {volume} {33}},\ \bibinfo {pages} {335602} (\bibinfo {year}
  {2021})}\BibitemShut {NoStop}%
\bibitem [{\citenamefont {Maciejko}\ and\ \citenamefont
  {Rayan}(2022)}]{Maciejko2022}%
  \BibitemOpen
  \bibfield  {author} {\bibinfo {author} {\bibfnamefont {J.}~\bibnamefont
  {Maciejko}}\ and\ \bibinfo {author} {\bibfnamefont {S.}~\bibnamefont
  {Rayan}},\ }\bibfield  {title} {\bibinfo {title} {{{Automorphic Bloch
  theorems for hyperbolic lattices}}},\ }\href
  {https://doi.org/10.1073/pnas.2116869119} {\bibfield  {journal} {\bibinfo
  {journal} {Proc. Natl. Acad. Sci. U.S.A.}\ }\textbf {\bibinfo {volume}
  {119}},\ \bibinfo {pages} {e2116869119} (\bibinfo {year} {2022})}\BibitemShut
  {NoStop}%
\bibitem [{\citenamefont {Stegmaier}\ \emph {et~al.}(2022)\citenamefont
  {Stegmaier}, \citenamefont {Upreti}, \citenamefont {Thomale},\ and\
  \citenamefont {Boettcher}}]{Stegmaier2022}%
  \BibitemOpen
  \bibfield  {author} {\bibinfo {author} {\bibfnamefont {A.}~\bibnamefont
  {Stegmaier}}, \bibinfo {author} {\bibfnamefont {L.~K.}\ \bibnamefont
  {Upreti}}, \bibinfo {author} {\bibfnamefont {R.}~\bibnamefont {Thomale}},\
  and\ \bibinfo {author} {\bibfnamefont {I.}~\bibnamefont {Boettcher}},\
  }\bibfield  {title} {\bibinfo {title} {{Universality of Hofstadter
  Butterflies on Hyperbolic Lattices}},\ }\href
  {https://doi.org/10.1103/PhysRevLett.128.166402} {\bibfield  {journal}
  {\bibinfo  {journal} {Phys. Rev. Lett.}\ }\textbf {\bibinfo {volume} {128}},\
  \bibinfo {pages} {166402} (\bibinfo {year} {2022})}\BibitemShut {NoStop}%
\bibitem [{\citenamefont {Zhang}\ \emph {et~al.}(2022)\citenamefont {Zhang},
  \citenamefont {Yuan}, \citenamefont {Sun}, \citenamefont {Sun},\ and\
  \citenamefont {Zhang}}]{zhang2022observation}%
  \BibitemOpen
  \bibfield  {author} {\bibinfo {author} {\bibfnamefont {W.}~\bibnamefont
  {Zhang}}, \bibinfo {author} {\bibfnamefont {H.}~\bibnamefont {Yuan}},
  \bibinfo {author} {\bibfnamefont {N.}~\bibnamefont {Sun}}, \bibinfo {author}
  {\bibfnamefont {H.}~\bibnamefont {Sun}},\ and\ \bibinfo {author}
  {\bibfnamefont {X.}~\bibnamefont {Zhang}},\ }\bibfield  {title} {\bibinfo
  {title} {Observation of novel topological states in hyperbolic lattices},\
  }\href {https://doi.org/10.1038/s41467-022-30631-x} {\bibfield  {journal}
  {\bibinfo  {journal} {Nat. Commun.}\ }\textbf {\bibinfo {volume} {13}},\
  \bibinfo {pages} {2937} (\bibinfo {year} {2022})}\BibitemShut {NoStop}%
\bibitem [{\citenamefont {Lenggenhager}\ \emph {et~al.}(2022)\citenamefont
  {Lenggenhager}, \citenamefont {Stegmaier}, \citenamefont {Upreti},
  \citenamefont {Hofmann}, \citenamefont {Helbig}, \citenamefont {Vollhardt},
  \citenamefont {Greiter}, \citenamefont {Lee}, \citenamefont {Imhof},
  \citenamefont {Brand}, \citenamefont {Kie{\ss}ling}, \citenamefont
  {Boettcher}, \citenamefont {Neupert}, \citenamefont {Thomale},\ and\
  \citenamefont {Bzdu\v{s}ek}}]{Lenggenhager2022}%
  \BibitemOpen
  \bibfield  {author} {\bibinfo {author} {\bibfnamefont {P.~M.}\ \bibnamefont
  {Lenggenhager}}, \bibinfo {author} {\bibfnamefont {A.}~\bibnamefont
  {Stegmaier}}, \bibinfo {author} {\bibfnamefont {L.~K.}\ \bibnamefont
  {Upreti}}, \bibinfo {author} {\bibfnamefont {T.}~\bibnamefont {Hofmann}},
  \bibinfo {author} {\bibfnamefont {T.}~\bibnamefont {Helbig}}, \bibinfo
  {author} {\bibfnamefont {A.}~\bibnamefont {Vollhardt}}, \bibinfo {author}
  {\bibfnamefont {M.}~\bibnamefont {Greiter}}, \bibinfo {author} {\bibfnamefont
  {C.~H.}\ \bibnamefont {Lee}}, \bibinfo {author} {\bibfnamefont
  {S.}~\bibnamefont {Imhof}}, \bibinfo {author} {\bibfnamefont
  {H.}~\bibnamefont {Brand}}, \bibinfo {author} {\bibfnamefont
  {T.}~\bibnamefont {Kie{\ss}ling}}, \bibinfo {author} {\bibfnamefont
  {I.}~\bibnamefont {Boettcher}}, \bibinfo {author} {\bibfnamefont
  {T.}~\bibnamefont {Neupert}}, \bibinfo {author} {\bibfnamefont
  {R.}~\bibnamefont {Thomale}},\ and\ \bibinfo {author} {\bibfnamefont
  {T.}~\bibnamefont {Bzdu\v{s}ek}},\ }\bibfield  {title} {\bibinfo {title}
  {Simulating hyperbolic space on a circuit board},\ }\href
  {https://doi.org/10.1038/s41467-022-32042-4} {\bibfield  {journal} {\bibinfo
  {journal} {Nat. Commun.}\ }\textbf {\bibinfo {volume} {13}},\ \bibinfo
  {pages} {4373} (\bibinfo {year} {2022})}\BibitemShut {NoStop}%
\bibitem [{\citenamefont {Boettcher}\ \emph {et~al.}(2022)\citenamefont
  {Boettcher}, \citenamefont {Gorshkov}, \citenamefont {Kollár}, \citenamefont
  {Maciejko}, \citenamefont {Rayan},\ and\ \citenamefont
  {Thomale}}]{Boettcher2022}%
  \BibitemOpen
  \bibfield  {author} {\bibinfo {author} {\bibfnamefont {I.}~\bibnamefont
  {Boettcher}}, \bibinfo {author} {\bibfnamefont {A.~V.}\ \bibnamefont
  {Gorshkov}}, \bibinfo {author} {\bibfnamefont {A.~J.}\ \bibnamefont
  {Kollár}}, \bibinfo {author} {\bibfnamefont {J.}~\bibnamefont {Maciejko}},
  \bibinfo {author} {\bibfnamefont {S.}~\bibnamefont {Rayan}},\ and\ \bibinfo
  {author} {\bibfnamefont {R.}~\bibnamefont {Thomale}},\ }\bibfield  {title}
  {\bibinfo {title} {{Crystallography of Hyperbolic Lattices}},\ }\href
  {https://doi.org/https://doi.org/10.1103/PhysRevB.105.125118} {\bibfield
  {journal} {\bibinfo  {journal} {Phys. Rev. B}\ }\textbf {\bibinfo {volume}
  {105}},\ \bibinfo {pages} {125118} (\bibinfo {year} {2022})}\BibitemShut
  {NoStop}%
\bibitem [{\citenamefont {Lenggenhager}\ \emph {et~al.}(2023)\citenamefont
  {Lenggenhager}, \citenamefont {Maciejko},\ and\ \citenamefont
  {Bzdu\v{s}ek}}]{LenggenhagerJsp2023}%
  \BibitemOpen
  \bibfield  {author} {\bibinfo {author} {\bibfnamefont {P.~M.}\ \bibnamefont
  {Lenggenhager}}, \bibinfo {author} {\bibfnamefont {J.}~\bibnamefont
  {Maciejko}},\ and\ \bibinfo {author} {\bibfnamefont {T.}~\bibnamefont
  {Bzdu\v{s}ek}},\ }\bibfield  {title} {\bibinfo {title} {{Non-Abelian
  Hyperbolic Band Theory from Supercells}},\ }\href
  {https://doi.org/10.1103/PhysRevLett.131.226401} {\bibfield  {journal}
  {\bibinfo  {journal} {Phys. Rev. Lett.}\ }\textbf {\bibinfo {volume} {131}},\
  \bibinfo {pages} {226401} (\bibinfo {year} {2023})}\BibitemShut {NoStop}%
\bibitem [{\citenamefont {Dey}\ \emph {et~al.}(2024)\citenamefont {Dey},
  \citenamefont {Chen}, \citenamefont {Basteiro}, \citenamefont {Fritzsche},
  \citenamefont {Greiter}, \citenamefont {Kaminski}, \citenamefont
  {Lenggenhager}, \citenamefont {Meyer}, \citenamefont {Sorbello},
  \citenamefont {Stegmaier}, \citenamefont {Thomale}, \citenamefont
  {Erdmenger},\ and\ \citenamefont {Boettcher}}]{PhysRevLett.133.061603}%
  \BibitemOpen
  \bibfield  {author} {\bibinfo {author} {\bibfnamefont {S.}~\bibnamefont
  {Dey}}, \bibinfo {author} {\bibfnamefont {A.}~\bibnamefont {Chen}}, \bibinfo
  {author} {\bibfnamefont {P.}~\bibnamefont {Basteiro}}, \bibinfo {author}
  {\bibfnamefont {A.}~\bibnamefont {Fritzsche}}, \bibinfo {author}
  {\bibfnamefont {M.}~\bibnamefont {Greiter}}, \bibinfo {author} {\bibfnamefont
  {M.}~\bibnamefont {Kaminski}}, \bibinfo {author} {\bibfnamefont {P.~M.}\
  \bibnamefont {Lenggenhager}}, \bibinfo {author} {\bibfnamefont
  {R.}~\bibnamefont {Meyer}}, \bibinfo {author} {\bibfnamefont
  {R.}~\bibnamefont {Sorbello}}, \bibinfo {author} {\bibfnamefont
  {A.}~\bibnamefont {Stegmaier}}, \bibinfo {author} {\bibfnamefont
  {R.}~\bibnamefont {Thomale}}, \bibinfo {author} {\bibfnamefont
  {J.}~\bibnamefont {Erdmenger}},\ and\ \bibinfo {author} {\bibfnamefont
  {I.}~\bibnamefont {Boettcher}},\ }\bibfield  {title} {\bibinfo {title}
  {{Simulating Holographic Conformal Field Theories on Hyperbolic Lattices}},\
  }\href {https://doi.org/10.1103/PhysRevLett.133.061603} {\bibfield  {journal}
  {\bibinfo  {journal} {Phys. Rev. Lett.}\ }\textbf {\bibinfo {volume} {133}},\
  \bibinfo {pages} {061603} (\bibinfo {year} {2024})}\BibitemShut {NoStop}%
\bibitem [{\citenamefont {Gluscevich}\ \emph {et~al.}(2025)\citenamefont
  {Gluscevich}, \citenamefont {Samanta}, \citenamefont {Manna},\ and\
  \citenamefont {Roy}}]{PhysRevB.111.L121108}%
  \BibitemOpen
  \bibfield  {author} {\bibinfo {author} {\bibfnamefont {N.}~\bibnamefont
  {Gluscevich}}, \bibinfo {author} {\bibfnamefont {A.}~\bibnamefont {Samanta}},
  \bibinfo {author} {\bibfnamefont {S.}~\bibnamefont {Manna}},\ and\ \bibinfo
  {author} {\bibfnamefont {B.}~\bibnamefont {Roy}},\ }\bibfield  {title}
  {\bibinfo {title} {Dynamic mass generation on two-dimensional electronic
  hyperbolic lattices},\ }\href {https://doi.org/10.1103/PhysRevB.111.L121108}
  {\bibfield  {journal} {\bibinfo  {journal} {Phys. Rev. B}\ }\textbf {\bibinfo
  {volume} {111}},\ \bibinfo {pages} {L121108} (\bibinfo {year}
  {2025})}\BibitemShut {NoStop}%
\bibitem [{\citenamefont {Kasteleyn}(1967)}]{graphsBook}%
  \BibitemOpen
  \bibfield  {author} {\bibinfo {author} {\bibfnamefont {P.~W.}\ \bibnamefont
  {Kasteleyn}},\ }\bibfield  {title} {\bibinfo {title} {{Graph Theory and
  Crystal Physics}},\ }in\ \href@noop {} {\emph {\bibinfo {booktitle} {Graph
  Theory and Theoretical Physics}}},\ \bibinfo {editor} {edited by\ \bibinfo
  {editor} {\bibfnamefont {F.}~\bibnamefont {Harary}}}\ (\bibinfo  {publisher}
  {London, Academic Press},\ \bibinfo {year} {1967})\ p.~\bibinfo {pages}
  {44}\BibitemShut {NoStop}%
\bibitem [{\citenamefont {Haydock}\ \emph {et~al.}(1972)\citenamefont
  {Haydock}, \citenamefont {Heine},\ and\ \citenamefont {Kelly}}]{Haydock1972}%
  \BibitemOpen
  \bibfield  {author} {\bibinfo {author} {\bibfnamefont {R.}~\bibnamefont
  {Haydock}}, \bibinfo {author} {\bibfnamefont {V.}~\bibnamefont {Heine}},\
  and\ \bibinfo {author} {\bibfnamefont {M.~J.}\ \bibnamefont {Kelly}},\
  }\bibfield  {title} {\bibinfo {title} {{Electronic structure based on the
  local atomic environment for tight-binding bands}},\ }\href
  {https://doi.org/10.1088/0022-3719/5/20/004} {\bibfield  {journal} {\bibinfo
  {journal} {J. Phys. C: Solid State Phys.}\ }\textbf {\bibinfo {volume} {5}},\
  \bibinfo {pages} {2845–2858} (\bibinfo {year} {1972})}\BibitemShut
  {NoStop}%
\bibitem [{\citenamefont {Haydock}\ \emph {et~al.}(1975)\citenamefont
  {Haydock}, \citenamefont {Heine},\ and\ \citenamefont {Kelly}}]{Haydock1975}%
  \BibitemOpen
  \bibfield  {author} {\bibinfo {author} {\bibfnamefont {R.}~\bibnamefont
  {Haydock}}, \bibinfo {author} {\bibfnamefont {V.}~\bibnamefont {Heine}},\
  and\ \bibinfo {author} {\bibfnamefont {M.~J.}\ \bibnamefont {Kelly}},\
  }\bibfield  {title} {\bibinfo {title} {{Electronic structure based on the
  local atomic environment for tight-binding bands. II}},\ }\href
  {https://doi.org/10.1088/0022-3719/8/16/011} {\bibfield  {journal} {\bibinfo
  {journal} {J. Phys. C: Solid State Phys.}\ }\textbf {\bibinfo {volume} {8}},\
  \bibinfo {pages} {2591–2605} (\bibinfo {year} {1975})}\BibitemShut
  {NoStop}%
\bibitem [{\citenamefont {Mosseri}\ and\ \citenamefont
  {Vidal}(2023)}]{Mosseri2023}%
  \BibitemOpen
  \bibfield  {author} {\bibinfo {author} {\bibfnamefont {R.}~\bibnamefont
  {Mosseri}}\ and\ \bibinfo {author} {\bibfnamefont {J.}~\bibnamefont
  {Vidal}},\ }\bibfield  {title} {\bibinfo {title} {{Density of states of
  tight-binding models in the hyperbolic plane}},\ }\href
  {https://doi.org/10.1103/PhysRevB.108.035154} {\bibfield  {journal} {\bibinfo
   {journal} {Phys. Rev. B}\ }\textbf {\bibinfo {volume} {108}},\ \bibinfo
  {pages} {035154} (\bibinfo {year} {2023})}\BibitemShut {NoStop}%
\bibitem [{\citenamefont {Shankar}\ and\ \citenamefont
  {Maciejko}(2024)}]{PhysRevLett.133.146601}%
  \BibitemOpen
  \bibfield  {author} {\bibinfo {author} {\bibfnamefont {G.}~\bibnamefont
  {Shankar}}\ and\ \bibinfo {author} {\bibfnamefont {J.}~\bibnamefont
  {Maciejko}},\ }\bibfield  {title} {\bibinfo {title} {{Hyperbolic Lattices and
  Two-Dimensional Yang-Mills Theory}},\ }\href
  {https://doi.org/10.1103/PhysRevLett.133.146601} {\bibfield  {journal}
  {\bibinfo  {journal} {Phys. Rev. Lett.}\ }\textbf {\bibinfo {volume} {133}},\
  \bibinfo {pages} {146601} (\bibinfo {year} {2024})}\BibitemShut {NoStop}%
\bibitem [{\citenamefont {Mn\"{e}v}(2007)}]{MnevPath}%
  \BibitemOpen
  \bibfield  {author} {\bibinfo {author} {\bibfnamefont {P.}~\bibnamefont
  {Mn\"{e}v}},\ }\bibfield  {title} {\bibinfo {title} {{Discrete Path Integral
  Approach to the Selberg Trace Formula for Regular Graphs}},\ }\href
  {https://doi.org/10.1007/s00220-007-0257-8} {\bibfield  {journal} {\bibinfo
  {journal} {Commun. Math. Phys.}\ }\textbf {\bibinfo {volume} {274}},\
  \bibinfo {pages} {233} (\bibinfo {year} {2007})}\BibitemShut {NoStop}%
\bibitem [{\citenamefont {Contreras}\ \emph {et~al.}(2024)\citenamefont
  {Contreras}, \citenamefont {Kandel}, \citenamefont {Mnev},\ and\
  \citenamefont {Wernli}}]{MnevArxiv}%
  \BibitemOpen
  \bibfield  {author} {\bibinfo {author} {\bibfnamefont {I.}~\bibnamefont
  {Contreras}}, \bibinfo {author} {\bibfnamefont {S.}~\bibnamefont {Kandel}},
  \bibinfo {author} {\bibfnamefont {P.}~\bibnamefont {Mnev}},\ and\ \bibinfo
  {author} {\bibfnamefont {K.}~\bibnamefont {Wernli}},\ }\bibfield  {title}
  {\bibinfo {title} {{Combinatorial QFT on graphs: First quantization
  formalism}},\ }\bibfield  {journal} {\bibinfo  {journal} {Ann. Inst. Henri
  Poincar\'{e} Comb. Phys. Interact. (published online first)}\ }\href
  {https://doi.org/10.4171/AIHPD/194} {10.4171/AIHPD/194} (\bibinfo {year}
  {2024})\BibitemShut {NoStop}%
\bibitem [{\citenamefont {P\'{o}lya}(1921)}]{Polya1921}%
  \BibitemOpen
  \bibfield  {author} {\bibinfo {author} {\bibfnamefont {G.}~\bibnamefont
  {P\'{o}lya}},\ }\bibfield  {title} {\bibinfo {title} {{\"{U}ber eine Aufgabe
  der Wahrscheinlichkeitsrechnung betreffend die Irrfahrt im
  Stra{\ss}ennetz}},\ }\href {https://doi.org/10.1007/BF01458701} {\bibfield
  {journal} {\bibinfo  {journal} {Math. Ann.}\ }\textbf {\bibinfo {volume}
  {84}},\ \bibinfo {pages} {149} (\bibinfo {year} {1921})}\BibitemShut
  {NoStop}%
\bibitem [{\citenamefont {Montroll}(1956)}]{Montroll}%
  \BibitemOpen
  \bibfield  {author} {\bibinfo {author} {\bibfnamefont {E.~W.}\ \bibnamefont
  {Montroll}},\ }\bibfield  {title} {\bibinfo {title} {{Random Walks in
  Multidimensional Spaces, Especially on Periodic Lattices}},\ }\href
  {http://www.jstor.org/stable/2098789} {\bibfield  {journal} {\bibinfo
  {journal} {Journal of the Society for Industrial and Applied Mathematics}\
  }\textbf {\bibinfo {volume} {4}},\ \bibinfo {pages} {241} (\bibinfo {year}
  {1956})}\BibitemShut {NoStop}%
\bibitem [{\citenamefont {Hughes}(1986)}]{HUGHES1986443}%
  \BibitemOpen
  \bibfield  {author} {\bibinfo {author} {\bibfnamefont {B.~D.}\ \bibnamefont
  {Hughes}},\ }\bibfield  {title} {\bibinfo {title} {On returns to the starting
  site in lattice random walks},\ }\href
  {https://doi.org/https://doi.org/10.1016/0378-4371(86)90058-0} {\bibfield
  {journal} {\bibinfo  {journal} {Physica A: Statistical Mechanics and its
  Applications}\ }\textbf {\bibinfo {volume} {134}},\ \bibinfo {pages} {443}
  (\bibinfo {year} {1986})}\BibitemShut {NoStop}%
\bibitem [{\citenamefont {{de Haro}}\ and\ \citenamefont
  {Tierz}(2004)}]{DEHARO2004201}%
  \BibitemOpen
  \bibfield  {author} {\bibinfo {author} {\bibfnamefont {S.}~\bibnamefont {{de
  Haro}}}\ and\ \bibinfo {author} {\bibfnamefont {M.}~\bibnamefont {Tierz}},\
  }\bibfield  {title} {\bibinfo {title} {{Brownian motion, Chern–Simons
  theory, and 2d Yang–Mills}},\ }\href
  {https://doi.org/https://doi.org/10.1016/j.physletb.2004.09.033} {\bibfield
  {journal} {\bibinfo  {journal} {Physics Letters B}\ }\textbf {\bibinfo
  {volume} {601}},\ \bibinfo {pages} {201} (\bibinfo {year}
  {2004})}\BibitemShut {NoStop}%
\bibitem [{\citenamefont {Manai}\ \emph {et~al.}(2015)\citenamefont {Manai},
  \citenamefont {Clément}, \citenamefont {Chicireanu}, \citenamefont
  {Hainaut}, \citenamefont {Garreau}, \citenamefont {Szriftgiser},\ and\
  \citenamefont {Delande}}]{Manai2015}%
  \BibitemOpen
  \bibfield  {author} {\bibinfo {author} {\bibfnamefont {I.}~\bibnamefont
  {Manai}}, \bibinfo {author} {\bibfnamefont {J.~F.}\ \bibnamefont {Clément}},
  \bibinfo {author} {\bibfnamefont {R.}~\bibnamefont {Chicireanu}}, \bibinfo
  {author} {\bibfnamefont {C.}~\bibnamefont {Hainaut}}, \bibinfo {author}
  {\bibfnamefont {J.~C.}\ \bibnamefont {Garreau}}, \bibinfo {author}
  {\bibfnamefont {P.}~\bibnamefont {Szriftgiser}},\ and\ \bibinfo {author}
  {\bibfnamefont {D.}~\bibnamefont {Delande}},\ }\bibfield  {title} {\bibinfo
  {title} {{Experimental Observation of Two-Dimensional Anderson Localization
  with the Atomic Kicked Rotor}},\ }\href
  {https://doi.org/10.1103/PhysRevLett.115.240603} {\bibfield  {journal}
  {\bibinfo  {journal} {Phys. Rev. Lett.}\ }\textbf {\bibinfo {volume} {115}},\
  \bibinfo {pages} {240603} (\bibinfo {year} {2015})}\BibitemShut {NoStop}%
\bibitem [{\citenamefont {Thiel}\ \emph {et~al.}(2018)\citenamefont {Thiel},
  \citenamefont {Kessler},\ and\ \citenamefont {Barkai}}]{PhysRevA.97.062105}%
  \BibitemOpen
  \bibfield  {author} {\bibinfo {author} {\bibfnamefont {F.}~\bibnamefont
  {Thiel}}, \bibinfo {author} {\bibfnamefont {D.~A.}\ \bibnamefont {Kessler}},\
  and\ \bibinfo {author} {\bibfnamefont {E.}~\bibnamefont {Barkai}},\
  }\bibfield  {title} {\bibinfo {title} {{Spectral dimension controlling the
  decay of the quantum first-detection probability}},\ }\href
  {https://doi.org/10.1103/PhysRevA.97.062105} {\bibfield  {journal} {\bibinfo
  {journal} {Phys. Rev. A}\ }\textbf {\bibinfo {volume} {97}},\ \bibinfo
  {pages} {062105} (\bibinfo {year} {2018})}\BibitemShut {NoStop}%
\bibitem [{\citenamefont {Tikhonov}\ and\ \citenamefont
  {Mirlin}(2021)}]{Tikhonov2021}%
  \BibitemOpen
  \bibfield  {author} {\bibinfo {author} {\bibfnamefont {K.}~\bibnamefont
  {Tikhonov}}\ and\ \bibinfo {author} {\bibfnamefont {A.}~\bibnamefont
  {Mirlin}},\ }\bibfield  {title} {\bibinfo {title} {{From Anderson
  localization on random regular graphs to many-body localization}},\ }\href
  {https://doi.org/10.1016/j.aop.2021.168525} {\bibfield  {journal} {\bibinfo
  {journal} {Annals of Phys.}\ }\textbf {\bibinfo {volume} {435}},\ \bibinfo
  {pages} {168525} (\bibinfo {year} {2021})}\BibitemShut {NoStop}%
\bibitem [{\citenamefont {Afek}\ \emph {et~al.}(2023)\citenamefont {Afek},
  \citenamefont {Davidson}, \citenamefont {Kessler},\ and\ \citenamefont
  {Barkai}}]{RevModPhys.95.031003}%
  \BibitemOpen
  \bibfield  {author} {\bibinfo {author} {\bibfnamefont {G.}~\bibnamefont
  {Afek}}, \bibinfo {author} {\bibfnamefont {N.}~\bibnamefont {Davidson}},
  \bibinfo {author} {\bibfnamefont {D.~A.}\ \bibnamefont {Kessler}},\ and\
  \bibinfo {author} {\bibfnamefont {E.}~\bibnamefont {Barkai}},\ }\bibfield
  {title} {\bibinfo {title} {{Colloquium: Anomalous statistics of laser-cooled
  atoms in dissipative optical lattices}},\ }\href
  {https://doi.org/10.1103/RevModPhys.95.031003} {\bibfield  {journal}
  {\bibinfo  {journal} {Rev. Mod. Phys.}\ }\textbf {\bibinfo {volume} {95}},\
  \bibinfo {pages} {031003} (\bibinfo {year} {2023})}\BibitemShut {NoStop}%
\bibitem [{\citenamefont {Chen}\ \emph {et~al.}(2024)\citenamefont {Chen},
  \citenamefont {Maciejko},\ and\ \citenamefont {Boettcher}}]{Chen2024aug}%
  \BibitemOpen
  \bibfield  {author} {\bibinfo {author} {\bibfnamefont {A.}~\bibnamefont
  {Chen}}, \bibinfo {author} {\bibfnamefont {J.}~\bibnamefont {Maciejko}},\
  and\ \bibinfo {author} {\bibfnamefont {I.}~\bibnamefont {Boettcher}},\
  }\bibfield  {title} {\bibinfo {title} {{Anderson Localization Transition in
  Disordered Hyperbolic Lattices}},\ }\href
  {https://doi.org/10.1103/PhysRevLett.133.066101} {\bibfield  {journal}
  {\bibinfo  {journal} {Phys. Rev. Lett.}\ }\textbf {\bibinfo {volume} {133}},\
  \bibinfo {pages} {066101} (\bibinfo {year} {2024})}\BibitemShut {NoStop}%
\bibitem [{\citenamefont {Vidakovic}(1994)}]{AllRoads}%
  \BibitemOpen
  \bibfield  {author} {\bibinfo {author} {\bibfnamefont {B.}~\bibnamefont
  {Vidakovic}},\ }\bibfield  {title} {\bibinfo {title} {{All Roads Lead to
  Rome-Even in the Honeycomb World}},\ }\href {https://doi.org/10.2307/2684723}
  {\bibfield  {journal} {\bibinfo  {journal} {The American Statistician}\
  }\textbf {\bibinfo {volume} {48}},\ \bibinfo {pages} {234} (\bibinfo {year}
  {1994})}\BibitemShut {NoStop}%
\bibitem [{\citenamefont {Ouvry}\ and\ \citenamefont
  {Polychronakos}(2020)}]{OUVRY2020115174}%
  \BibitemOpen
  \bibfield  {author} {\bibinfo {author} {\bibfnamefont {S.}~\bibnamefont
  {Ouvry}}\ and\ \bibinfo {author} {\bibfnamefont {A.~P.}\ \bibnamefont
  {Polychronakos}},\ }\bibfield  {title} {\bibinfo {title} {{Lattice walk area
  combinatorics, some remarkable trigonometric sums and Ap\'{e}ry-like
  numbers}},\ }\href
  {https://doi.org/https://doi.org/10.1016/j.nuclphysb.2020.115174} {\bibfield
  {journal} {\bibinfo  {journal} {Nuclear Physics B}\ }\textbf {\bibinfo
  {volume} {960}},\ \bibinfo {pages} {115174} (\bibinfo {year}
  {2020})}\BibitemShut {NoStop}%
\bibitem [{\citenamefont {Gan}\ \emph {et~al.}(2022)\citenamefont {Gan},
  \citenamefont {Ouvry},\ and\ \citenamefont
  {Polychronakos}}]{PhysRevE.105.014112}%
  \BibitemOpen
  \bibfield  {author} {\bibinfo {author} {\bibfnamefont {L.}~\bibnamefont
  {Gan}}, \bibinfo {author} {\bibfnamefont {S.}~\bibnamefont {Ouvry}},\ and\
  \bibinfo {author} {\bibfnamefont {A.~P.}\ \bibnamefont {Polychronakos}},\
  }\bibfield  {title} {\bibinfo {title} {Algebraic area enumeration of random
  walks on the honeycomb lattice},\ }\href
  {https://doi.org/10.1103/PhysRevE.105.014112} {\bibfield  {journal} {\bibinfo
   {journal} {Phys. Rev. E}\ }\textbf {\bibinfo {volume} {105}},\ \bibinfo
  {pages} {014112} (\bibinfo {year} {2022})}\BibitemShut {NoStop}%
\bibitem [{\citenamefont {Ouvry}\ and\ \citenamefont
  {Polychronakos}(2022)}]{OuvryArea}%
  \BibitemOpen
  \bibfield  {author} {\bibinfo {author} {\bibfnamefont {S.}~\bibnamefont
  {Ouvry}}\ and\ \bibinfo {author} {\bibfnamefont {A.~P.}\ \bibnamefont
  {Polychronakos}},\ }\bibfield  {title} {\bibinfo {title} {{Algebraic area
  enumeration for open lattice walk}},\ }\href
  {https://doi.org/10.1088/1751-8121/aca573} {\bibfield  {journal} {\bibinfo
  {journal} {J. Phys. A: Math. Theor.}\ }\textbf {\bibinfo {volume} {55}},\
  \bibinfo {pages} {485005} (\bibinfo {year} {2022})}\BibitemShut {NoStop}%
\bibitem [{\citenamefont {Gan}(2023)}]{PhysRevE.108.054104}%
  \BibitemOpen
  \bibfield  {author} {\bibinfo {author} {\bibfnamefont {L.}~\bibnamefont
  {Gan}},\ }\bibfield  {title} {\bibinfo {title} {Algebraic area of cubic
  lattice walks},\ }\href {https://doi.org/10.1103/PhysRevE.108.054104}
  {\bibfield  {journal} {\bibinfo  {journal} {Phys. Rev. E}\ }\textbf {\bibinfo
  {volume} {108}},\ \bibinfo {pages} {054104} (\bibinfo {year}
  {2023})}\BibitemShut {NoStop}%
\bibitem [{\citenamefont {Gan}(2025)}]{Gan:2024rzb}%
  \BibitemOpen
  \bibfield  {author} {\bibinfo {author} {\bibfnamefont {L.}~\bibnamefont
  {Gan}},\ }\bibfield  {title} {\bibinfo {title} {{Lattice random walks and
  quantum A-period conjecture}},\ }\href
  {https://doi.org/10.21468/SciPostPhys.19.2.053} {\bibfield  {journal}
  {\bibinfo  {journal} {SciPost Phys.}\ }\textbf {\bibinfo {volume} {19}},\
  \bibinfo {pages} {053} (\bibinfo {year} {2025})}\BibitemShut {NoStop}%
\bibitem [{\citenamefont {Cheng}\ \emph {et~al.}(2024)\citenamefont {Cheng},
  \citenamefont {Shu}, \citenamefont {Zhang}, \citenamefont {Mao},\ and\
  \citenamefont {Sun}}]{PhysRevLett.133.216401}%
  \BibitemOpen
  \bibfield  {author} {\bibinfo {author} {\bibfnamefont {N.}~\bibnamefont
  {Cheng}}, \bibinfo {author} {\bibfnamefont {C.}~\bibnamefont {Shu}}, \bibinfo
  {author} {\bibfnamefont {K.}~\bibnamefont {Zhang}}, \bibinfo {author}
  {\bibfnamefont {X.}~\bibnamefont {Mao}},\ and\ \bibinfo {author}
  {\bibfnamefont {K.}~\bibnamefont {Sun}},\ }\bibfield  {title} {\bibinfo
  {title} {{Universal Spectral Moment Theorem and Its Applications in
  Non-Hermitian Systems}},\ }\href
  {https://doi.org/10.1103/PhysRevLett.133.216401} {\bibfield  {journal}
  {\bibinfo  {journal} {Phys. Rev. Lett.}\ }\textbf {\bibinfo {volume} {133}},\
  \bibinfo {pages} {216401} (\bibinfo {year} {2024})}\BibitemShut {NoStop}%
\bibitem [{\citenamefont {Farnell}\ \emph {et~al.}(2018)\citenamefont
  {Farnell}, \citenamefont {G\"{o}tze}, \citenamefont {Schulenburg},
  \citenamefont {Zinke}, \citenamefont {Bishop},\ and\ \citenamefont
  {Li}}]{Farnell}%
  \BibitemOpen
  \bibfield  {author} {\bibinfo {author} {\bibfnamefont {D.}~\bibnamefont
  {Farnell}}, \bibinfo {author} {\bibfnamefont {O.}~\bibnamefont {G\"{o}tze}},
  \bibinfo {author} {\bibfnamefont {J.}~\bibnamefont {Schulenburg}}, \bibinfo
  {author} {\bibfnamefont {R.}~\bibnamefont {Zinke}}, \bibinfo {author}
  {\bibfnamefont {R.}~\bibnamefont {Bishop}},\ and\ \bibinfo {author}
  {\bibfnamefont {P.}~\bibnamefont {Li}},\ }\bibfield  {title} {\bibinfo
  {title} {{Interplay between lattice topology, frustration, and spin quantum
  number in quantum antiferromagnets on Archimedean lattices}},\ }\href
  {https://doi.org/10.1103/PhysRevB.98.224402} {\bibfield  {journal} {\bibinfo
  {journal} {Phys. Rev. B}\ }\textbf {\bibinfo {volume} {98}} (\bibinfo {year}
  {2018})}\BibitemShut {NoStop}%
\bibitem [{\citenamefont {Betts}(1995)}]{Betts1995}%
  \BibitemOpen
  \bibfield  {author} {\bibinfo {author} {\bibfnamefont {D.}~\bibnamefont
  {Betts}},\ }\bibfield  {title} {\bibinfo {title} {{A new two-dimensional
  lattice of coordination number five}},\ }\href
  {https://doi.org/http://hdl.handle.net/10222/35332} {\bibfield  {journal}
  {\bibinfo  {journal} {Proc. N. S. Inst. Sci.}\ }\textbf {\bibinfo {volume}
  {40}},\ \bibinfo {pages} {95} (\bibinfo {year} {1995})}\BibitemShut {NoStop}%
\bibitem [{\citenamefont {Hu}\ \emph {et~al.}(2011)\citenamefont {Hu},
  \citenamefont {Kargarian},\ and\ \citenamefont {Fiete}}]{PhysRevB.84.155116}%
  \BibitemOpen
  \bibfield  {author} {\bibinfo {author} {\bibfnamefont {X.}~\bibnamefont
  {Hu}}, \bibinfo {author} {\bibfnamefont {M.}~\bibnamefont {Kargarian}},\ and\
  \bibinfo {author} {\bibfnamefont {G.~A.}\ \bibnamefont {Fiete}},\ }\bibfield
  {title} {\bibinfo {title} {{Topological insulators and fractional quantum
  Hall effect on the ruby lattice}},\ }\href
  {https://doi.org/10.1103/PhysRevB.84.155116} {\bibfield  {journal} {\bibinfo
  {journal} {Phys. Rev. B}\ }\textbf {\bibinfo {volume} {84}},\ \bibinfo
  {pages} {155116} (\bibinfo {year} {2011})}\BibitemShut {NoStop}%
\bibitem [{\citenamefont {Lukin}\ and\ \citenamefont
  {Sotnikov}(2024)}]{PhysRevE.109.045305}%
  \BibitemOpen
  \bibfield  {author} {\bibinfo {author} {\bibfnamefont {I.~V.}\ \bibnamefont
  {Lukin}}\ and\ \bibinfo {author} {\bibfnamefont {A.~G.}\ \bibnamefont
  {Sotnikov}},\ }\bibfield  {title} {\bibinfo {title} {{Corner transfer matrix
  renormalization group approach in the zoo of Archimedean lattices}},\ }\href
  {https://doi.org/10.1103/PhysRevE.109.045305} {\bibfield  {journal} {\bibinfo
   {journal} {Phys. Rev. E}\ }\textbf {\bibinfo {volume} {109}},\ \bibinfo
  {pages} {045305} (\bibinfo {year} {2024})}\BibitemShut {NoStop}%
\bibitem [{\citenamefont {Codello}(2010)}]{codello2010exact}%
  \BibitemOpen
  \bibfield  {author} {\bibinfo {author} {\bibfnamefont {A.}~\bibnamefont
  {Codello}},\ }\bibfield  {title} {\bibinfo {title} {{Exact Curie temperature
  for the Ising model on Archimedean and Laves lattices}},\ }\href
  {https://doi.org/10.1088/1751-8113/43/38/385002} {\bibfield  {journal}
  {\bibinfo  {journal} {J. Phys. A: Math. Theor.}\ }\textbf {\bibinfo {volume}
  {43}},\ \bibinfo {pages} {385002} (\bibinfo {year} {2010})}\BibitemShut
  {NoStop}%
\bibitem [{\citenamefont {Gr{\"u}nbaum}\ and\ \citenamefont
  {Shephard}(2016)}]{Grunbaum2016-rs}%
  \BibitemOpen
  \bibfield  {author} {\bibinfo {author} {\bibfnamefont {B.}~\bibnamefont
  {Gr{\"u}nbaum}}\ and\ \bibinfo {author} {\bibfnamefont {G.~C.}\ \bibnamefont
  {Shephard}},\ }\href@noop {} {\emph {\bibinfo {title} {{Tilings and
  Patterns}}}},\ Dover Books on Mathematics\ (\bibinfo  {publisher} {Dover
  Publications},\ \bibinfo {address} {Mineola, NY},\ \bibinfo {year}
  {2016})\BibitemShut {NoStop}%
\bibitem [{\citenamefont {Crasto~de Lima}\ \emph {et~al.}(2019)\citenamefont
  {Crasto~de Lima}, \citenamefont {Ferreira},\ and\ \citenamefont
  {Miwa}}]{CrastodeLima2019}%
  \BibitemOpen
  \bibfield  {author} {\bibinfo {author} {\bibfnamefont {F.}~\bibnamefont
  {Crasto~de Lima}}, \bibinfo {author} {\bibfnamefont {G.~J.}\ \bibnamefont
  {Ferreira}},\ and\ \bibinfo {author} {\bibfnamefont {R.~H.}\ \bibnamefont
  {Miwa}},\ }\bibfield  {title} {\bibinfo {title} {{Topological flat band,
  Dirac fermions and quantum spin Hall phase in 2D Archimedean lattices}},\
  }\href {https://doi.org/10.1039/c9cp04760c} {\bibfield  {journal} {\bibinfo
  {journal} {Phys. Chem. Chem. Phys.}\ }\textbf {\bibinfo {volume} {21}},\
  \bibinfo {pages} {22344} (\bibinfo {year} {2019})}\BibitemShut {NoStop}%
\bibitem [{\citenamefont {Pierre}\ \emph {et~al.}(2025)\citenamefont {Pierre},
  \citenamefont {Bernu},\ and\ \citenamefont {Messio}}]{IsingArch}%
  \BibitemOpen
  \bibfield  {author} {\bibinfo {author} {\bibfnamefont {L.}~\bibnamefont
  {Pierre}}, \bibinfo {author} {\bibfnamefont {B.}~\bibnamefont {Bernu}},\ and\
  \bibinfo {author} {\bibfnamefont {L.}~\bibnamefont {Messio}},\ }\bibfield
  {title} {\bibinfo {title} {{Derivation of free energy, entropy and specific
  heat for planar Ising models: Application to Archimedean lattices and their
  duals}},\ }\href {https://doi.org/10.21468/SciPostPhys.19.1.025} {\bibfield
  {journal} {\bibinfo  {journal} {SciPost Phys.}\ }\textbf {\bibinfo {volume}
  {19}},\ \bibinfo {pages} {025} (\bibinfo {year} {2025})}\BibitemShut
  {NoStop}%
\bibitem [{\citenamefont {Ortiz}\ \emph {et~al.}(2021)\citenamefont {Ortiz},
  \citenamefont {Sarte}, \citenamefont {Kenney}, \citenamefont {Graf},
  \citenamefont {Teicher}, \citenamefont {Seshadri},\ and\ \citenamefont
  {Wilson}}]{Ortiz2021}%
  \BibitemOpen
  \bibfield  {author} {\bibinfo {author} {\bibfnamefont {B.~R.}\ \bibnamefont
  {Ortiz}}, \bibinfo {author} {\bibfnamefont {P.~M.}\ \bibnamefont {Sarte}},
  \bibinfo {author} {\bibfnamefont {E.~M.}\ \bibnamefont {Kenney}}, \bibinfo
  {author} {\bibfnamefont {M.~J.}\ \bibnamefont {Graf}}, \bibinfo {author}
  {\bibfnamefont {S.~M.~L.}\ \bibnamefont {Teicher}}, \bibinfo {author}
  {\bibfnamefont {R.}~\bibnamefont {Seshadri}},\ and\ \bibinfo {author}
  {\bibfnamefont {S.~D.}\ \bibnamefont {Wilson}},\ }\bibfield  {title}
  {\bibinfo {title} {{Superconductivity in the $\mathbb{Z}_{2}$ kagome metal
  $\mathrm{KV}_{3}\mathrm{Sb}_{5}$}},\ }\href
  {https://doi.org/10.1103/PhysRevMaterials.5.034801} {\bibfield  {journal}
  {\bibinfo  {journal} {Phys. Rev. Mater.}\ }\textbf {\bibinfo {volume} {5}},\
  \bibinfo {pages} {034801} (\bibinfo {year} {2021})}\BibitemShut {NoStop}%
\bibitem [{\citenamefont {Kato}\ \emph {et~al.}(2024)\citenamefont {Kato},
  \citenamefont {Narumi}, \citenamefont {Morita}, \citenamefont {Matsushita},
  \citenamefont {Fukuoka}, \citenamefont {Yamashita}, \citenamefont {Nakazawa},
  \citenamefont {Oda}, \citenamefont {Hayashi}, \citenamefont {Yamaura},
  \citenamefont {Hagiwara},\ and\ \citenamefont {Yoshida}}]{Kato2024}%
  \BibitemOpen
  \bibfield  {author} {\bibinfo {author} {\bibfnamefont {M.}~\bibnamefont
  {Kato}}, \bibinfo {author} {\bibfnamefont {Y.}~\bibnamefont {Narumi}},
  \bibinfo {author} {\bibfnamefont {K.}~\bibnamefont {Morita}}, \bibinfo
  {author} {\bibfnamefont {Y.}~\bibnamefont {Matsushita}}, \bibinfo {author}
  {\bibfnamefont {S.}~\bibnamefont {Fukuoka}}, \bibinfo {author} {\bibfnamefont
  {S.}~\bibnamefont {Yamashita}}, \bibinfo {author} {\bibfnamefont
  {Y.}~\bibnamefont {Nakazawa}}, \bibinfo {author} {\bibfnamefont
  {M.}~\bibnamefont {Oda}}, \bibinfo {author} {\bibfnamefont {H.}~\bibnamefont
  {Hayashi}}, \bibinfo {author} {\bibfnamefont {K.}~\bibnamefont {Yamaura}},
  \bibinfo {author} {\bibfnamefont {M.}~\bibnamefont {Hagiwara}},\ and\
  \bibinfo {author} {\bibfnamefont {H.~K.}\ \bibnamefont {Yoshida}},\
  }\bibfield  {title} {\bibinfo {title} {{One-third magnetization plateau in
  Quantum Kagome antiferromagnet}},\ }\href
  {https://doi.org/10.1038/s42005-024-01922-0} {\bibfield  {journal} {\bibinfo
  {journal} {Commun. Phys.}\ }\textbf {\bibinfo {volume} {7}},\ \bibinfo
  {pages} {424} (\bibinfo {year} {2024})}\BibitemShut {NoStop}%
\bibitem [{\citenamefont {Pratt}\ \emph {et~al.}(2025)\citenamefont {Pratt},
  \citenamefont {L\'opez-Alcal\'a}, \citenamefont {Garcia-Lopez}, \citenamefont
  {Clemente-Le\'on}, \citenamefont {Baldov\'{\i}},\ and\ \citenamefont
  {Coronado}}]{Pratt2025}%
  \BibitemOpen
  \bibfield  {author} {\bibinfo {author} {\bibfnamefont {F.~L.}\ \bibnamefont
  {Pratt}}, \bibinfo {author} {\bibfnamefont {D.}~\bibnamefont
  {L\'opez-Alcal\'a}}, \bibinfo {author} {\bibfnamefont {V.}~\bibnamefont
  {Garcia-Lopez}}, \bibinfo {author} {\bibfnamefont {M.}~\bibnamefont
  {Clemente-Le\'on}}, \bibinfo {author} {\bibfnamefont {J.~J.}\ \bibnamefont
  {Baldov\'{\i}}},\ and\ \bibinfo {author} {\bibfnamefont {E.}~\bibnamefont
  {Coronado}},\ }\bibfield  {title} {\bibinfo {title} {{Properties of the
  frustrated spin 1/2 kagome material
  ${\mathrm{Cu}}_{3}{(\mathrm{HOTP})}_{2}$}},\ }\href
  {https://doi.org/10.1103/PhysRevResearch.7.023007} {\bibfield  {journal}
  {\bibinfo  {journal} {Phys. Rev. Res.}\ }\textbf {\bibinfo {volume} {7}},\
  \bibinfo {pages} {023007} (\bibinfo {year} {2025})}\BibitemShut {NoStop}%
\bibitem [{\citenamefont {Cai}\ \emph {et~al.}(2025)\citenamefont {Cai},
  \citenamefont {Yoon}, \citenamefont {Sheng}, \citenamefont {Zhao},
  \citenamefont {Seewald}, \citenamefont {Ghosh}, \citenamefont {Ingham},
  \citenamefont {Pasupathy}, \citenamefont {Queiroz}, \citenamefont {Lei},
  \citenamefont {Xie}, \citenamefont {Dai}, \citenamefont {Ito}, \citenamefont
  {Ke}, \citenamefont {Cava}, \citenamefont {Sharma}, \citenamefont {Pula},
  \citenamefont {Luke}, \citenamefont {Kojima},\ and\ \citenamefont
  {Uemura}}]{Cai2025}%
  \BibitemOpen
  \bibfield  {author} {\bibinfo {author} {\bibfnamefont {Y.}~\bibnamefont
  {Cai}}, \bibinfo {author} {\bibfnamefont {S.}~\bibnamefont {Yoon}}, \bibinfo
  {author} {\bibfnamefont {Q.}~\bibnamefont {Sheng}}, \bibinfo {author}
  {\bibfnamefont {G.}~\bibnamefont {Zhao}}, \bibinfo {author} {\bibfnamefont
  {E.~F.}\ \bibnamefont {Seewald}}, \bibinfo {author} {\bibfnamefont
  {S.}~\bibnamefont {Ghosh}}, \bibinfo {author} {\bibfnamefont
  {J.}~\bibnamefont {Ingham}}, \bibinfo {author} {\bibfnamefont {A.~N.}\
  \bibnamefont {Pasupathy}}, \bibinfo {author} {\bibfnamefont {R.}~\bibnamefont
  {Queiroz}}, \bibinfo {author} {\bibfnamefont {H.}~\bibnamefont {Lei}},
  \bibinfo {author} {\bibfnamefont {Y.}~\bibnamefont {Xie}}, \bibinfo {author}
  {\bibfnamefont {P.}~\bibnamefont {Dai}}, \bibinfo {author} {\bibfnamefont
  {T.}~\bibnamefont {Ito}}, \bibinfo {author} {\bibfnamefont {R.}~\bibnamefont
  {Ke}}, \bibinfo {author} {\bibfnamefont {R.~J.}\ \bibnamefont {Cava}},
  \bibinfo {author} {\bibfnamefont {S.}~\bibnamefont {Sharma}}, \bibinfo
  {author} {\bibfnamefont {M.}~\bibnamefont {Pula}}, \bibinfo {author}
  {\bibfnamefont {G.~M.}\ \bibnamefont {Luke}}, \bibinfo {author}
  {\bibfnamefont {K.~M.}\ \bibnamefont {Kojima}},\ and\ \bibinfo {author}
  {\bibfnamefont {Y.~J.}\ \bibnamefont {Uemura}},\ }\bibfield  {title}
  {\bibinfo {title} {{Magnetism of kagome metals
  $({\mathrm{Fe}}_{1\ensuremath{-}x}{\mathrm{Co}}_{x})\phantom{\rule{0.16em}{0ex}}\mathrm{Sn}$
  studied by $\ensuremath{\mu}\mathrm{SR}$}},\ }\href
  {https://doi.org/10.1103/PhysRevB.111.214412} {\bibfield  {journal} {\bibinfo
   {journal} {Phys. Rev. B}\ }\textbf {\bibinfo {volume} {111}},\ \bibinfo
  {pages} {214412} (\bibinfo {year} {2025})}\BibitemShut {NoStop}%
\bibitem [{\citenamefont {Kogan}\ and\ \citenamefont
  {Gumbs}(2021)}]{Kogan2021-fy}%
  \BibitemOpen
  \bibfield  {author} {\bibinfo {author} {\bibfnamefont {E.}~\bibnamefont
  {Kogan}}\ and\ \bibinfo {author} {\bibfnamefont {G.}~\bibnamefont {Gumbs}},\
  }\bibfield  {title} {\bibinfo {title} {Green's functions and {DOS} for some
  {2D} lattices},\ }\href {https://doi.org/10.4236/graphene.2021.101001}
  {\bibfield  {journal} {\bibinfo  {journal} {Graphene}\ }\textbf {\bibinfo
  {volume} {10}},\ \bibinfo {pages} {1} (\bibinfo {year} {2021})}\BibitemShut
  {NoStop}%
\bibitem [{\citenamefont {Gardiner}(1974)}]{GARDINER1974255}%
  \BibitemOpen
  \bibfield  {author} {\bibinfo {author} {\bibfnamefont {A.}~\bibnamefont
  {Gardiner}},\ }\bibfield  {title} {\bibinfo {title} {Antipodal covering
  graphs},\ }\href
  {https://doi.org/https://doi.org/10.1016/0095-8956(74)90072-0} {\bibfield
  {journal} {\bibinfo  {journal} {Journal of Combinatorial Theory, Series B}\
  }\textbf {\bibinfo {volume} {16}},\ \bibinfo {pages} {255} (\bibinfo {year}
  {1974})}\BibitemShut {NoStop}%
\bibitem [{\citenamefont {A.~Bader}\ \emph {et~al.}(2013)\citenamefont
  {A.~Bader}, \citenamefont {Meyerhenke}, \citenamefont {Sanders},\ and\
  \citenamefont {Wagner}}]{Bader2013}%
  \BibitemOpen
  \bibfield  {author} {\bibinfo {author} {\bibfnamefont {D.}~\bibnamefont
  {A.~Bader}}, \bibinfo {author} {\bibfnamefont {H.}~\bibnamefont
  {Meyerhenke}}, \bibinfo {author} {\bibfnamefont {P.}~\bibnamefont
  {Sanders}},\ and\ \bibinfo {author} {\bibfnamefont {D.}~\bibnamefont
  {Wagner}},\ }\href {https://doi.org/10.1090/conm/588} {\emph {\bibinfo
  {title} {{Graph Partitioning and Graph Clustering}}}},\ Vol.\ \bibinfo
  {volume} {588}\ (\bibinfo  {publisher} {Amer. Math. Soc.},\ \bibinfo
  {address} {Providence, RI},\ \bibinfo {year} {2013})\ pp.\ \bibinfo {pages}
  {1--17}\BibitemShut {NoStop}%
\bibitem [{Note1()}]{Note1}%
  \BibitemOpen
  \bibinfo {note} {{\protect \it ``A generating function is a device somewhat
  similar to a bag. Instead of carrying many little objects detachedly, which
  could be embarrassing, we put them all in a bag, and then we have only one
  object to carry, the bag.''}, G. P\'olya, Mathematics and plausible reasoning
  (1954)}\BibitemShut {NoStop}%
\bibitem [{\citenamefont {{OEIS Foundation Inc. (2025)}}()}]{oeis}%
  \BibitemOpen
  \bibfield  {author} {\bibinfo {author} {\bibnamefont {{OEIS Foundation Inc.
  (2025)}}},\ }\href {https://oeis.org} {\bibinfo {title} {{The On-Line
  Encyclopedia of Integer Sequences}}}\BibitemShut {NoStop}%
\bibitem [{\citenamefont {Laves}(1931)}]{Laves}%
  \BibitemOpen
  \bibfield  {author} {\bibinfo {author} {\bibfnamefont {F.}~\bibnamefont
  {Laves}},\ }\bibfield  {title} {\bibinfo {title} {{Ebenenteilung und
  Koordinationszahl}},\ }\href {https://doi.org/doi:10.1524/zkri.1931.78.1.208}
  {\bibfield  {journal} {\bibinfo  {journal} {Zeitschrift f\"{u}r
  Kristallographie - Crystalline Materials}\ }\textbf {\bibinfo {volume}
  {78}},\ \bibinfo {pages} {208} (\bibinfo {year} {1931})}\BibitemShut
  {NoStop}%
\bibitem [{\citenamefont {Kac}\ and\ \citenamefont {Ward}(1952)}]{Kac1952-ip}%
  \BibitemOpen
  \bibfield  {author} {\bibinfo {author} {\bibfnamefont {M.}~\bibnamefont
  {Kac}}\ and\ \bibinfo {author} {\bibfnamefont {J.~C.}\ \bibnamefont {Ward}},\
  }\bibfield  {title} {\bibinfo {title} {{A Combinatorial Solution of the
  Two-dimensional Ising Model}},\ }\href
  {https://doi.org/10.1103/PhysRev.88.1332} {\bibfield  {journal} {\bibinfo
  {journal} {Phys. Rev.}\ }\textbf {\bibinfo {volume} {88}},\ \bibinfo {pages}
  {1332} (\bibinfo {year} {1952})}\BibitemShut {NoStop}%
\bibitem [{\citenamefont {Feynman}(1972)}]{F0}%
  \BibitemOpen
  \bibfield  {author} {\bibinfo {author} {\bibfnamefont {R.~P.}\ \bibnamefont
  {Feynman}},\ }\href@noop {} {\emph {\bibinfo {title} {{Statistical Mechanics:
  A Set of Lectures}}}},\ Frontiers in Physics\ (\bibinfo  {publisher} {W. A.
  Benjamin},\ \bibinfo {year} {1972})\ p.\ \bibinfo {pages} {144}\BibitemShut
  {NoStop}%
\bibitem [{\citenamefont {Vdovichenko}(1965)}]{vdovichenko1965calculation}%
  \BibitemOpen
  \bibfield  {author} {\bibinfo {author} {\bibfnamefont {N.~V.}\ \bibnamefont
  {Vdovichenko}},\ }\bibfield  {title} {\bibinfo {title} {{A Calculation of the
  Partition Function for a Plane Dipole Lattice}},\ }\href@noop {} {\bibfield
  {journal} {\bibinfo  {journal} {Soviet Physics JETP}\ }\textbf {\bibinfo
  {volume} {20}},\ \bibinfo {pages} {715} (\bibinfo {year} {1965})},\ \bibinfo
  {note} {translated from J. Exptl. Theoret. Phys. (U.S.S.R.) 47, 715-719
  (August, 1964)}\BibitemShut {NoStop}%
\end{thebibliography}%

\vfill

\end{document}